\documentclass{aa}  
\usepackage{natbib,epsfig,txfonts,graphicx}
\bibpunct{(}{)}{;}{a}{}{,}

\newcommand{\msun}{\ensuremath{M_{\odot}}}
\newcommand{\lsun}{\ensuremath{L_{\odot}}}
\newcommand{\rsun}{\ensuremath{R_{\odot}}}
\newcommand{\Zsun}{\ensuremath{Z_{\odot}}}
\newcommand{\Teff}{\ensuremath{T_{\rm eff}}}
\newcommand{\vinf}{\ensuremath{v_{\infty}}}
\newcommand{\mdot}{\ensuremath{\dot{M}}}

\newcommand{\msunyr}{\ensuremath{M_{\odot} {\rm yr}^{-1}}}
\newcommand{\Mdu}{\ensuremath{\cdot 10^{-6}\,M_{\odot} {\rm yr}^{-1}}}
\newcommand{\mdu}{\ensuremath{10^{-6}\,M_{\odot} {\rm yr}^{-1}}}

\newcommand{\beq}{\begin{equation}}
\newcommand{\eeq}{\end{equation}}
\newcommand{\beqa}{\begin{eqnarray}}
\newcommand{\eeqa}{\end{eqnarray}}

\newcommand{\nbeq}{\begin{equation*}}
\newcommand{\neeq}{\end{equation*}}

\newcommand{\kms}{\ensuremath{{\rm km}\,{\rm s}^{-1}}}

\newcommand{\rarrow}{\rightarrow}

\newcommand{\HeI} {He\,{\sc i}}
\newcommand{\HeII}{He\,{\sc ii}}

\newcommand{\CII}{C\,{\sc ii}}
\newcommand{\CIII}{C\,{\sc iii}}
\newcommand{\CIV}{C\,{\sc iv}}

\newcommand\NII{N\,{\sc ii}}
\newcommand\NIII{N\,{\sc iii}}
\newcommand\NIV{N\,{\sc iv}}
\newcommand\NV{N\,{\sc v}}

\newcommand\OII{O\,{\sc ii}}
\newcommand\OIII{O\,{\sc iii}}
\newcommand\OIV{O\,{\sc iv}}

\newcommand\OVI{O\,{\sc vi}}

\newcommand\NeII{Ne\,{\sc ii}}

\newcommand\SiII{Si\,{\sc ii}}
\newcommand\SiIII{Si\,{\sc iii}}
\newcommand\SiIV{Si\,{\sc iv}}

\newcommand{\Hd} {H$_{\rm \delta}$}
\newcommand{\Hg} {H$_{\rm \gamma}$}

\newcommand{\Ha} {H$_{\rm \alpha}$}

\newcommand{\MV}{\ensuremath{M_V}}

\newcommand{\Rstar}{\ensuremath{R_{\ast}}}

\newcommand{\Lstar}{\ensuremath{L_{\ast}}}

\newcommand{\logg}{\ensuremath{\log g}}
\newcommand{\YHe}{\ensuremath{Y_{\rm He}}}

\newcommand{\vesc}{\ensuremath{v_{\rm esc}}}

\newcommand{\vsini}{\ensuremath{v{\thinspace}\sin{\thinspace}i}}

\newcommand{\vmic}{\ensuremath{v_{\rm mic}}}
\newcommand{\vmac}{\ensuremath{v_{\rm mac}}}

\newcommand{\fcl}{\ensuremath{f_{\rm cl}}}

\newcommand{\taur}{\ensuremath{\tau_{\rm Ross}}}

\newcommand{\trip}{\ensuremath{\lambda\lambda4634-4640-4642}}
\newcommand{\qua}{\ensuremath{\lambda\lambda4510-4514-4518}}

\newcommand{\nivem}{\ensuremath{\lambda4058}}
\newcommand{\nivab}{\ensuremath{\lambda6380}}

\newcommand\eg{\hbox{e.g.,}}
\begin{document}
\title{Nitrogen line spectroscopy in O-stars}
\subtitle{II. Surface nitrogen abundances for O-stars in the Large Magellanic
Cloud
\thanks{Based on observations collected at the European Southern Observatory Very
Large Telescope, under programmes 68.D-0369, 171.D-0237 (FLAMES) and
67.D-0238, 70.D-0164, 074.D-0109 (UVES).}
\fnmsep
\thanks{Appendix A, B and C are only available in electronic form at http://www.edpsciences.org} }

\author{J.G. Rivero Gonz\'alez\inst{1}, J. Puls\inst{1},
        F. Najarro\inst{2}, \and I. Brott\inst{3}}

\institute{Universit\"atssternwarte M\"unchen, Scheinerstr. 1, 81679 M\"unchen, 
           Germany, \email{jorge@usm.uni-muenchen.de} 
           \and
           Centro de Astrobiolog\'{\i}a, (CSIC-INTA), 
	   Ctra. Torrej\'on a Ajalvir km 4,
	   28850 Torrej\'on de Ardoz, Spain
	   \and 
	   University of Vienna, Department of Astronomy,
	   T\"urkenschanzstr. 17, 1180 Vienna, Austria}

\date{Received; Accepted}

\abstract
{Nitrogen is a key element to test the impact of rotational mixing on
evolutionary models of massive stars.  Recent studies of the nitrogen
surface abundance in B-type stars within the {\it VLT-FLAMES survey of
massive stars}\, have challenged part of the corresponding
predictions. To obtain a more complete picture of massive star
evolution, and to allow for further constraints, these studies need to be
extended to O-stars.}
{
%This is the second paper in a series aiming at the analysis of
%nitrogen abundances in O-type stars, to enable further constraints on
%the {\it early evolution}\, of massive stars. 
In this paper, we
investigate the \NIV$\lambda4058$ emission line formation, provide
nitrogen abundances for a substantial O-star sample in the Large
Magellanic Cloud, and compare our (preliminary) findings with recent
predictions from stellar evolutionary models.}
{Stellar and wind parameters of our sample stars are determined by
line profile fitting of hydrogen, helium and nitrogen lines, exploiting
the corresponding ionization equilibria. Synthetic
spectra are calculated by means of the NLTE atmosphere/spectrum
synthesis code {\sc fastwind}, using a new nitrogen model atom. We
derive nitrogen abundances for 20 O- and 5 B-stars, by analyzing 
all nitrogen lines (from different ionization stages) present in the
available optical spectra. 
}
{The dominating process responsible for emission at \NIV$\lambda$4058
in O-stars is the strong depopulation of the lower level of the
transition, which increases as a function of \mdot. Unlike the \NIII\
triplet emission, resonance lines do not play a role for typical
mass-loss rates and below. We find (almost) no problem in fitting the
nitrogen lines, in particular the `f' features. Only for some objects,
where lines from \NIII/\NIV/\NV\ are visible in parallel, we need to
opt for a compromise solution. 

For five objects in the early B-/late O-star domain which have been
previously analyzed by different methods and model atmospheres, we
derive consistent nitrogen abundances. The bulk of our sample O-stars 
seems to be strongly nitrogen-enriched, and a clear correlation of
nitrogen and helium enrichment is found. By comparing
the nitrogen abundances as a function of \vsini\ ('Hunter-plot') with
tailored evolutionary calculations, we identify  a considerable number of
{\it highly}\ enriched objects at low rotation.}
{Our findings seem to support the basic outcome of previous B-star
studies within the VLT-FLAMES survey. Due to the low initial 
abundance, the detection of strong Nitrogen enrichment in the bulk of
O-stars indicates that efficient mixing takes place already during the
very early phases of stellar evolution of LMC O-stars. For tighter
constraints, however, upcoming results from the {\it VLT-FLAMES
Tarantula survey}\ need to be waited for, comprising a much larger
number of O-stars that will be analyzed based on similar methods as
presented here.}


\titlerunning{Surface nitrogen abundances for O-stars in the Large Magellanic
Cloud}
\authorrunning{J.G. Rivero Gonz\'alez et al.}

\maketitle
%
%________________________________________________________________

\section{Introduction} 
\label{Introduction} 
One of the main products of the VLT-FLAMES survey of massive
stars\footnote{see \citealt{evans06} for an introductory publication
and \citealt{evans08} for a brief summary on the outcome of this
project} (hereafter FLAMES~I) was the determination of light element
abundances from statistically significant samples of
Galactic, Large and Small Magellanic Cloud (LMC, SMC) B-stars,
covering a broad range of rotational velocities.

The inclusion of rotational mixing into massive star evolution
(e.g., \citealt{Heger00, MeynetMaeder00, Brott11a}) brought better
agreement with spectroscopic analyses that provide evidence for a
characteristic enrichment of helium and nitrogen in many early-type
stars (reviewed by, e.g., \citealt{herrero03}, \citealt{herrero04} and
\citealt{morel09}). Several recent studies based on the nitrogen
diagnostics performed within the FLAMES~I survey have severely
challenged the predicted effects though. In this context, nitrogen is a key
element to test the predictions of rotational mixing, since it should
become strongly enriched at the stellar surface of rapidly rotating
stars already in fairly early evolutionary phases, whilst for slow
rotation almost no enhancement should occur before the red supergiant
phase. Actually, a significant number of both un-enriched
fast rotators and highly enriched slow rotators have been found within
the population of LMC core-hydrogen burning objects \citep{hunter08,
hunter09, Brott11b}.\footnote{as well as slowly rotating, highly
enriched supergiants, discussed by \citet{Vink10}} These results imply
that standard rotational mixing might be not dominant, and/or that
other enrichment processes might be decisive as well \citep{Brott11b}.

To further constrain these findings and to provide a general picture
of massive star evolution, these studies need to be extended to O-type
stars. Because of their shorter lifetimes, the time-range where
this enrichment takes place can be narrowed down, and one might be able to
constrain the mixing scenario even better than it is possible from 
B-stars alone. In this respect, the LMC is an ideal testbed, since 
the nitrogen baseline abundance is low and even a strong enrichment is
easier to measure/confirm than, e.g., in the Milky way.

Interestingly, most previous abundance studies of massive stars
are strongly biased towards intermediate and early type B-stars.
Indeed, when inspecting the available literature, metallic abundances,
in particular of nitrogen, are scarcely found for O-stars. The
situation for LMC objects is even worse, and data for only a few
supergiants (\citealt{pauldrach94c, Crowther02, evans04b}) and giants
(\citealt{walborn04}) are available. One of the reasons for this lack
of information is that the determination of nitrogen abundances is a
non-trivial task, due to the complexity of \NIII/\NIV\ line formation
related to the impact of various processes that are absent or
negligible at cooler spectral types where \NII\ is the dominant ion. 

To provide more insight into this matter, we started a series of
publications dealing with nitrogen spectroscopy in O-type stars. In
the first paper of this series \citep[hereafter Paper~I]{rivero11}, we concentrated on
the formation of the optical \NIII\ emission lines at \trip, which are
fundamental for the
definition of the different morphological
`f'-classes~\citep{walborn71b}. It turned out that the canonical
explanation in terms of dielectronic recombination \citep{mihalas73}
no longer or only partly applies when modern atmosphere codes
including line-blocking/blanketing and winds are used. The key role is
now played by the stellar wind, which induces a (relative) overpopulation of
the upper level of the transition, via pumping from the ground state
rather than by dielectronic recombination as long as the
wind-strength is large enough to enable a significantly accelerating
velocity field already in the photospheric formation region.

The main goal of the present paper is to provide nitrogen abundances
for a considerable number of O-stars in the LMC. For this purpose, we
use the corresponding sample from \citet[hereafter Mok07]{mokiem07a},
mostly based on observations within the FLAMES~I survey. So far, this
is the largest sample of O-stars studied in the LMC by means of
quantitative hydrogen and helium line spectroscopy, and allows us to
determine nitrogen abundances for a significant number of objects.
However, its size is still not comparable with the amount of corresponding
B-stars, and does not allow us to extend the B-star results (that
challenged rotational mixing) towards the O-star domain in a
statistically sufficient way. Rather, it will yield a first impression
on potential problems. A statistically significant analysis will become
possible within the VLT-FLAMES Tarantula survey (\citealt{Evans11},
`FLAMES~II'), which provides an unprecedented sample of `normal'
O-stars and emission-line stars.

This paper is organized as follows. In Sect.~\ref{prereq}, we describe
the tools used to determine nitrogen abundances, both the atmospheric
code and the nitrogen model atom. In particular, we study the
formation of the \NIV\nivem\ emission line in parallel with the
\NIV\nivab\ absorption line. Sect.~\ref{samp_obs} presents the stellar
sample and the observations used within this study. The procedure to
determine stellar and wind parameters together with nitrogen
abundances is outlined in Sect.~\ref{anal}. In Sect.~\ref{comments}, we
comment in detail on the individual objects. Sect.~\ref{results}
provides a discussion of our results, and Sect.~\ref{conclusions}
summarizes our findings and conclusions.

\section{Prerequisites for nitrogen diagnostics}
\label{prereq}
\subsection{The code}
\label{nlte_fast}
For this work, we use a recently updated version (v10.1) of the
atmosphere/line formation code {\sc fastwind} (see \citealt{santo97}
and \citealt{puls05} for previous versions). This code has been
specifically designed for the optical and IR spectroscopic analysis of
hot stars of spectral types early A to O, and accounts for NLTE
conditions, spherical symmetry and mass-loss. The current version
incorporates a variety of updates and improvements compared with
previous versions, which are briefly summarized in the following.

At first note that {\sc fastwind} differentiates between so-called
`explicit' and `background' elements, where the former are those used
as diagnostic tools (in the present context: H, He and N) and are
treated with high precision, by detailed atomic models and by
means of comoving frame transport for the line transitions.
The background elements (i.e., the rest) are used `only' for the
line-blocking/blanketing calculations, and have been treated so far by
means of the Sobolev approximation. Though this is reasonable in the
wind regime, the Sobolev approximation becomes doubtful
in regions where the velocity field is strongly curved, which is the
case in the transition zone between photosphere and wind. As we have
convinced ourselves, the induced errors are not important for the
background radiation field, but they can have a certain influence on
the temperature structure. Applying the Sobolev approximation in
regions with a strong velocity field curvature results, on average,
in too strongly populated upper levels of line transitions (see, e.g.,
\citealt{santo97}). In turn, this leads to overestimated heating
rates, which can result in too high temperatures in the transition
region (and sometimes even below). To avoid this problem, the new {\sc
fastwind} version treats also the most important lines from the
background elements in the comoving frame.

A second modification refers to the {\it photospheric} line
acceleration. So far, this quantity (being important for the
photospheric density stratification -- higher line acceleration, lower
density) has been calculated from the Rosseland opacities, which is
strictly justified only at large optical depths. In the new version,
an additional iteration cycle to calculate the photospheric structure
is performed, now by using the {\it flux-weighted} mean from the
current NLTE opacities.

By calculating a large grid of OB-star models, and comparing with
solutions from the previous {\sc fastwind} version, it turned out that
both improvements affect mostly dwarfs/giants in the effective
temperature range 30~kK $\leq\
\Teff\ \leq$ 35~kK. In particular the optical \HeII\ lines become
stronger, mostly due to the somewhat lower (electron-) densities.
Interestingly, this is just the domain where previous {\sc fastwind}
solutions showed the largest deviations from other codes
\citep{simon-diaz08}. Comparing the new structures to results from
{\sc tlusty}, excellent agreement has been found. For dwarfs/giants with
effective temperatures outside the `problematic' region, and for all
supergiants, the differences to previous results from earlier {\sc
fastwind} versions are small. 

The last major improvement concerns the implementation of dielectronic
recombination, both for the background and the explicit elements,
and was already described and successfully tested in Paper~I.

\subsection{The nitrogen model atom}
\label{atom nitro} To perform our analysis, we implemented a new
nitrogen model atom into {\sc fastwind}, consisting of the ionization
stages \NII\ to \NV,
which has already been used for the calculations performed in
Paper~I. The level structures of both \NIV\ and \NV\ have been taken 
from the {\sc wm}-basic atomic database \citep{pauldrach94c}. 

\paragraph{\NII} has been adapted to {\sc fastwind} from a previous
model ion, developed by \citet{BB89}. We had no intention to develop a
`perfect' \NII\ model, since most of our analyses deal with
O-star spectra where \NII\ becomes invisible, and particularly because
N.~Przybilla and co-workers have already constructed such a model
(based on an earlier version, see \citealt{przybilla01}) which will be
incorporated into our code after release. However, a series of tests
were performed to ensure the goodness of the model ion, see Sect~\ref{tests_nii}.

For \NII, we consider 50 LS-coupled terms, up to principal quantum number
$n = 4$ and angular momentum $l = 3$, where all fine-structure
sub-levels have been packed. Detailed information about the selected
levels is provided in Table~\ref{atom_lev_nii} and
Fig.~\ref{nii-singlet}. We account for some hundred permitted electric
dipole radiative transitions. For the bulk of the transitions,
oscillator strengths are taken from calculations performed by
\citet{BB89}, but for some transitions related to strong \NII\ lines
oscillator strengths have been taken from
NIST.\footnote{http://www.nist.gov/physlab/data/asd.cfm, firstly
described in \citet{nist}} Radiative intercombinations are neglected.
Roughly one thousand collisional bound-bound transitions are
considered, with corresponding rates using the \citet{vanregemorter62}
approximation in the radiatively permitted case and following the
semi-empirical expression by \citet{allen73} in the forbidden one.
Radiative ionization cross sections have been derived by \citet{BB89},
and adapted to the representation suggested by
\citet{seaton58}. Collisional ionization cross-sections are calculated
using the \citet{seaton62} formula in terms of the photoionization
cross-section at threshold.

\paragraph{\NIII} has
been already described in Paper~I. In brief, it consists of 41 packed
terms up to $n$~=~6 and $l$~=~4 (doublet and quartet system).  

\paragraph{\NIV.} Also this model ion consists of 50 LS-coupled terms,
up to principal quantum number $n$~=~6 and angular momentum $l$~=~4, 
with all fine-structure sub-levels packed into one term.
Table~\ref{atom_lev_niv} provides detailed information about the
selected levels. Two spin systems (singlet and triplet) are treated
simultaneously (Fig.~\ref{niv-singlet}). All allowed electric dipole
radiative transitions between the 50 levels are considered, as well as
radiative intercombinations, with a total of 520 transitions.
Corresponding oscillator strengths have been drawn from either NIST
when available or otherwise from the {\sc wm}-basic
database.\footnote{see \citet{pauldrach94c}. In brief, the atomic
structure code {\sc superstructure} (\citealt{eissner69, eissner91})
has been used to calculate all bound state energies in LS and
intermediate coupling as well as related atomic data, particularly
oscillator strengths including those for stabilizing transitions.}
Furthermore, we consider roughly one thousand bound-bound collisional
transitions between all levels, with effective collision strengths
%for electron impact excitation 
among the 12 lowest LS-states from R-matrix computations by
\citet{ramsbottom94}. Transitions without detailed data and
collisional ionizations are treated as in \NII.

Photoionization cross-sections have been taken from calculations by
\citet{Tully90}, via
TOPbase,\footnote{http://cdsweb.u-strasbg.fr/topbase/topbase.html} the
OPACITY Project on-line database \citep{cunto92}. For excited levels
with no OPACITY Project data available (5g $^1$G, 5g $^3$G, 6s
$^1$S, and 6g $^3$G, see Table~\ref{atom_lev_niv}), resonance-free
cross-sections are used, provided in terms of the \citet{seaton58}
approximation with parameters from the {\sc wm}-basic database.
Finally, the most important dielectronic recombination and reverse
ionization processes are implicitly accounted for by means of
exploiting the OPACITY Project photo cross-sections. Only for the few
levels with no such data available, we apply the `explicit' method,
using the stabilizing transitions on top of resonance-free photo
cross-section (see, e.g., Paper~I), with corresponding data from {\sc
wm}-basic.
%(i) For collisions between the 12 lowest levels, $2s^2$, $2s\
%2p$, $2p^2$, and $2s\ 3l, \ (l=s,p,d)$ (comprising singlet and triplet
%terms), collision strengths are adopted from the R-matrix computations
%by \citet{ramsbottom94}. These authors provide analytic fits
%to the effective collision strengths, $\gamma$, for collisions between all
%fine structure levels.
%(ii) For the remaining optically allowed transitions, the
%\citet{vanregemorter62} approximation is applied.
%(iii) For the optically forbidden transitions, the semi-empirical
%formula from \citet{allen73} (with $\Omega$ = 1) is used.
\paragraph{\NV.} Our model of this lithium-like ion (one doublet spin
system) consists of 27 levels, including LS-coupled and packed terms
up to $n = 7$ and $l = 6$ (see Table~\ref{atom_lev_nv} for details and
Fig.~\ref{nv-grotrian} for a Grotrian diagram). All allowed electric
dipole radiative transitions are accounted for, with a total number of
102 radiative bound-bound transitions and oscillator strengths from
{\sc wm}-basic. Collisional excitations and ionizations are treated as
in \NII,
%There is a paper, Aggarwal10 but they only show comparisons, not
%provide all data. Anyway we are not going to change the model atom for this
%calculations I would comment the last part then.
whilst photo cross-sections (in terms of the \citealt{seaton58}
approximation) have been taken from the {\sc wm}-basic atomic
database.

\smallskip
\noindent
Our complete nitrogen model atom (from \NII\ to \NV) comprises 178
LS-coupled levels, with more than 1100 radiative and more
than 3800 collisional bound-bound transitions. To calculate the final
synthetic profiles, Voigt profiles are adopted, with central
wavelengths according to NIST, radiative damping parameters from the
Kurucz database,\footnote{
www.pmp.uni-hannover.de/cgi-bin/ssi/test/kurucz/sekur.html} and
collisional damping parameters (broadening by electron impact)
computed according to \cite{Cowley71}.

\subsubsection{Testing the \NII\ model ion}
\label{tests_nii}
To test our somewhat simple \NII\ model ion,\footnote{important,
e.g., for our comparison with B-star nitrogen abundances from
alternative analyses, see Sect.~\ref{comp_hunter}} we compared synthetic line
profiles with corresponding ones from {\sc tlusty} \citep{hubeny98}
and {\sc detail/surface} \citep{Giddings81, ButlerGiddings85}, the
latter based on the newly developed \NII\ model by Przybilla et al.
(see Sect.~\ref{atom nitro}).

A summary of the various tests is provided in Table~\ref{nii_tests}.
We started by comparing with line profiles from the BSTAR2006 model
grid \citep{Lanz07}. This grid has been calculated using the model
atmosphere code {\sc tlusty} (\citealt{hubeny88, hubeny95}), a code
that assumes plane-parallel geometry, hydrostatic and radiative
equilibrium, and calculates line-blanketed NLTE model atmospheres and
corresponding synthetic profiles. Due to its restrictions, only
objects with negligible winds can be analyzed.

We used this grid of models as reference because it covers the
parameter range at which \NII/\NIII\ are the dominant ionization stages, and
the \NII\ lines are clearly visible in the synthetic spectra. In addition, the
BSTAR2006 grid has been used in numerous studies aiming at the
determination of stellar abundances and parameters in B-stars \citep[e.g.,][]{lanz08}.

A grid of {\sc fastwind} models has been calculated, covering the
temperature range 20~kK $\leq \Teff \leq$ 30~kK, using a typical step
size of 2.5~kK, and gravities representative for dwarfs and giants.
Since {\sc tlusty} does not account for the presence of a wind, we
used negligible mass-loss rates, \mdot = $10^{-9} {\ldots}$
$10^{-10}$~\msunyr. Because {\sc fastwind} allows for employing
external photospheric structures, we created a second grid using the
{\sc tlusty} photospheric structure, smoothly connected to the wind
structure as calculated by {\sc fastwind}, with the usual $\beta$
velocity law. For consistency with the {\sc tlusty} grid, all models
have been calculated with the `older' solar nitrogen abundance, [N] =
7.92 \citep{grevesse98}, where [N] = $\log$ N/H + 12 with respect to
particle numbers. Note that all tests have been performed with the
complete nitrogen model atom involving the ions \NII\ to \NV\ as
described in Sect~\ref{atom nitro}.

\begin{table}
\caption{\NII\ test series (see text).}
\vspace{0.3cm}
\label{nii_tests}
\begin{tabular}{llll}
\hline 
\hline
Series & Atomic model  & Photospheric & NLTE/line formation \\
       &               & stratification & \\ 
\hline
FW  & see Sect.~\ref{atom nitro} & {\sc fastwind} & {\sc fastwind} \\
FW2 & see Sect.~\ref{atom nitro} & {\sc tlusty} & {\sc fastwind} \\
TL  & {\sc tlusty}             & {\sc tlusty} & {\sc tlusty} \\
Prz & Przybilla et al.         & {\sc Kurucz} & {\sc detail/surface}\\
\hline
\end{tabular}
\end{table}

It turned out that there are large differences between the synthetic \NII\
line profiles calculated by {\sc tlusty} (TL) and {\sc fastwind} (FW), even
though the hydrogen/helium lines agree very well (except for the forbidden
component of \HeI\ which is stronger in FW-models). The latter code predicts
stronger \NII\ profiles, which is also true for the {\sc fastwind} results
based on the {\sc tlusty} photospheric structure (FW2, see
Figs.~\ref{comp-tlusty-2030} to \ref{comp-tlusty-2730}).

For dwarfs, there are almost no differences between the profiles from the 
FW and FW2 models. This is readily understood, since the photospheric
stratification of electron temperature, $T_{\rm e}$, and electron
density, ${n_{\rm e}}$, are essentially the same (see Fig.~\ref{mod-structure}, panel
1, 3, 5), i.e., {\sc fastwind} and {\sc tlusty} predict the same structures.

On the other hand, models for giants at higher \Teff\ display (mostly)
weak differences. E.g., at \Teff\ = 27.5~kK and \logg\ =~3.0, there is
a small disagreement of the electron densities in photospheric regions
($\taur \leq 10^{-5}$), even though the temperatures agree quite
well (Fig.~\ref{mod-structure}, panel 6). In particular, the {\sc
tlusty} and thus the FW2 structure shows a lower electron density at
optical depths where the photospheric lines are formed, because of a
higher photospheric radiative line pressure in this model.

Because of the lower electron density, the lower recombination rates (at
\Teff\ = 27.5 kK, \NIII\ is the dominant ion) lead to somewhat weaker \NII\
profiles in the FW2 models compared to the FW ones, see
Fig.~\ref{comp-tlusty-2730}. Nevertheless, differences to the
profiles as predicted by TL itself are still large, and we conclude that the
photospheric structure is not the origin of the discrepancies. 
%At this
%stage, this leaves differences in either the NLTE treatment (e.g., different
%background radiation fields) or the atomic data as potential explanations.

As an independent check, we compared our results with spectra
calculated by N.~Przybilla (priv. comm.) for two of our grid models,
denoted by Prz in the following. These spectra are based on the
NLTE/line formation code {\sc detail/surface}, and the \NII\ model ion
recently developed by Przybilla et al.. We consider this atomic model
as the superior one in the present context, because large effort on
improving and testing the atomic data has been spent, and
corresponding synthetic spectra perfectly match high resolution/high
signal-to-noise observations from various B-stars \citep{Przyb08}.

Figures~\ref{comp-tlusty-2030} and \ref{comp-tlusty-2430} show that the
agreement between the FW and Prz profiles is excellent, which leaves us with
the conclusion that there might be problems in the \NII\ atomic
data used in the BSTAR2006 grid. 
Important studies using {\sc tlusty} have been carried out during the past
years (e.g., \citealt{dufton06, Trundle07, hunter09}), including the
determination of \NII\ abundances in LMC and SMC B-stars. 
Actually, these studies utilized a different model atom, developed by
\citet{Allende03}, which has been tested by N.~Przybilla at our request,
with a positive outcome. Thus, the aforementioned analyses should be free
from uncertainties related to a potentially insufficient atomic model.

\subsection{Diagnostic nitrogen lines in the optical}
\label{nitrogen_lines}
\begin{table*}
\center \caption{Diagnostic nitrogen lines in the optical (and
adjacent) spectrum of early B- and O-type stars, together with
potential blends.}
%\vspace{0.3cm}
\label{tab_lines_nitro}
\begin{tabular}{ccccccc}
\hline 
\hline
Ion & Transition & Multiplet & Line & Wavelength(\AA) & Blends & Used\\
\hline
\NII\ & N29 - N216  & 12  & 1  &   3994.99  & - & + \\ 
      & N211 - N219  & 15  & 2  &   4447.03  & \OII\  $\lambda$4446.81, 4447.67,
4448.19, \OIII\ $\lambda$4447.79 & +\\
      & N28 - N215  & 5   & 3  &   4601.47  & \OII\  $\lambda$4602.13, 4603.23,
\NIII\ $\lambda$4604.18 & + \\
      & N28 - N215   & 5  & 4  &   4607.16  & \OII\  $\lambda$4609.44, 4610.20,
\NIII\ $\lambda$4605.16, \NeII\ $\lambda$4606.70 & + \\
      & N28 - N215  & 5   & 5  &   4621.39  & \OII\  $\lambda$4621.27, \NIII\
$\lambda$4621.04, 4623.05, \SiII\ $\lambda$4621.72 & + \\
      & N28 - N215  & 5   & 6  &   4630.54  & \OIII\ $\lambda$4630.77, \NIII\
$\lambda$4630.61, \SiIV\ $\lambda$4631.24 & + \\
      & N28 - N215  & 5   & 7  &   4643.09  & \OII\  $\lambda$4641.81, 4643.89,
\NIII\ $\lambda$4641.85 & - \\ 
      &N212 - N218  & 19  & 8  &   5005.15  & \OIII\ $\lambda$5006.84 & - \\
      &N214 - N221  & 24  & 9  &   5007.33  & \OIII\ $\lambda$5006.84 & - \\
      & N28 - N214 & 4      & 10 &   5045.10  & \NII\ $\lambda$5046.53 & - \\
      & N28 - N212 & 3      & 11 &   5666.63  & \CII\ $\lambda$5662.47 & - \\
      & N28 - N212 & 3      & 12 &   5676.01  & \NII\ $\lambda$5679.55 & - \\
      & N28 - N212 & 3      & 13 &   5679.55  & \NII\ $\lambda$5676.01 & - \\
      & N28 - N212 & 3      & 14 &   5710.77  & \SiII\ $\lambda$5707.20& - \\
      & N215 - N220 & 28   & 15 &   5941.65  & \NII\ $\lambda$5940.24, \NIII\
$\lambda$5943.44 & - \\
\hline
\NIII\ &N320 - N333  & 17     & 1  &   4003.58  & \OII\ $\lambda$4007.46 & + \\ 
       & N38 - N310  & 1     & 2  &   4097.33  & \OII\ $\lambda$4097.26, 4098.24, \Hd\
$\lambda$4101.74 & + \\ 
       &N313 - N322 & 6      & 3  &   4195.76  & \OII\ $\lambda$4192.52, 4196.26,
\SiIII\ $\lambda$4195.59, \HeII\ $\lambda$4200.00 & + \\
       &N313 - N322 & 6     & 4  &   4200.07  & \HeII\ $\lambda$4200.00& - \\
       &N321 - N334 & 18     & 5  &   4379.11  & \OII\ $\lambda$4378.03, 4378.43,
\CIII\ $\lambda$4379.47, \NII\ $\lambda$4379.59 & + \\ 
       &N312 - N316 & 3     & 6  &   4510.88  & \NIII\ $\lambda$4510.92, \NeII\
$\lambda$4511.42 & + \\
       &N312 - N316 & 3    & 7  &   4514.86  & \OIII\ $\lambda$4513.83, \NeII\
$\lambda$4514.88, \CIII\ $\lambda$4515.81, 4516.77 & + \\
       &N312 - N316 & 3    & 8  &   4518.14  & \NeII\ $\lambda$4518.14, \OIII\
$\lambda$4519.62 & + \\
       &N310 - N311 & 2    & 9  &   4634.14  & \SiIV\ $\lambda$4631.24, \OIV\
$\lambda$4632 & + \\
       &N310 - N311 & 2    & 10 &   4640.64  & \OII\ $\lambda$4638.86, \SiIII\
$\lambda$4638.28 & + \\
       &N310 - N311 & 2    & 11 &   4641.85  & \OII\ $\lambda$4641.81, 4643.39, \NII\
$\lambda$4643.08 & + \\
       &N322 - N330 & 21    & 12 &   5320.82 & \OII\ $\lambda$5322.53 &-\\
       &N322 - N330 & 21    & 13 &   5327.18 & - & - \\
       &N319 - N325 & 14    & 14 &   6445.34 & - & - \\
       &N319 - N325 & 14  & 15 &   6450.79 & \CIV\ $\lambda6449.90$& - \\
       &N319 - N325 & 14  & 16 &   6454.08 & \OII\ $\lambda$6457.05, \NII\
$\lambda$6457.68& - \\
       &N319 - N325 & 14  & 17 &   6467.02 & -                     & - \\
\hline
\NIV\  & N47 - N410 & 1    &  1 &   3478.71 & -                   & + \\ 
       & N47 - N410 & 1    &  2 &   3482.99 & -                     & + \\ 
       & N47 - N410 & 1    &  3 &   3484.96 & -                     & + \\ 
       & N49 - N412 & 3    &  4 &   4057.76 & \CIII\ $\lambda$4056.06, 4059.56 & +
\\ 
       & N48 - N49  & 2    &  5 &   6380.77 & \OIII\ $\lambda$6378.34, 6383.30, DIBs
$\lambda\lambda$ 6376.08,6379.32 & + \\ 
       & N413 - N416 & 5   &  6 &   5200.41 & \OIV\ $\lambda5198.22$ & - \\ 
       & N413 - N416 & 5   &  7 &   5204.28 & \NIV\ $\lambda5205.15$ & - \\
       & N413 - N416 & 5   &  8 &   5205.15 & \NIV\ $\lambda5204.28$, \OII\
$\lambda5206.65$ & - \\
       & N410 - N411 & 4  &  9 &   7103.24 & -                   & + \\ 
       & N410 - N411 & 4  & 10 &   7109.35 & -                   & + \\ 
       & N410 - N411 & 4  & 11 &   7111.28 & -                   & + \\ 
       & N410 - N411 & 4  & 12 &   7122.98 & -                   & + \\ 
       & N410 - N411 & 4  & 13 &   7127.25 & -                   & + \\ 
       & N410 - N411 & 4  & 14 &   7129.18 & -                   & + \\ 
\hline 
\NV\   & N53 - N54   & 1   &  1 &   4603.73 & \NIII\ $\lambda$4604.18, 4605.16, \NIV\
$\lambda$4606.33 & + \\
       & N53 - N54   & 1   &  2 &   4619.98 & \SiIII\ $\lambda$4619.66, \NIII\
$\lambda$4621.04, 4623.05& + \\
       & N518 - N525 & 9    &  3 &   4943.17 & \OIV\ $\lambda4941.29$  & - \\
       & N519 - N526 & 10    &  4 &   4943.97 & \OIV\ $\lambda4941.29$  &  - \\
       & N520 - N527 & 10.01    &  5 &   4945.29 & - &  - \\
\hline
\end{tabular}
\tablefoot{Line numbers for \NII, \NIV\ and \NV\ refer to important
transitions as indicated in Figs.~\ref{nii-singlet}, 
\ref{niv-singlet} and \ref{nv-grotrian}, respectively, and for \NIII\
to the corresponding figure (B.1) in Paper~I. Multiplet numbers
for \NII/\NIII\  are from \citet{Moore75},  and for \NIV/\NV\ 
from \citet{Moore71}. Lines used within the present
work are labeled by `+'. Lower and upper levels of the
transitions are denoted by a combination of ion and level number
according to Tables~\ref{atom_lev_nii}, \ref{atom_lev_niv} and
\ref{atom_lev_nv}, e.g., `N29' means level \#9 of \NII.}
\end{table*}

Table~\ref{tab_lines_nitro} presents a set of 51 nitrogen lines 
visible in the optical (and adjacent) spectra of OB-stars, along with
the position of potential blends. Included are the connected levels
(for corresponding term designations, see Appendix~A) and
multiplet numbers, to provide an
impression of how much independent lines are present. 

Lines from \NII\ (visible in the spectra of B and late O-stars) have
been selected after careful comparisons with profiles calculated by N.
Przybilla (priv. comm., see Sect.~\ref{tests_nii}). Only one of the
suggested lines, \NII\ $\lambda$3995, is completely isolated, and
remains uncontaminated even at high rotation rates. Moreover, this is
one of the strongest \NII\ lines located in the optical region, making
it a good choice for deriving nitrogen abundances. Other useful lines
are \NII$\lambda$5667 and $\lambda$5679, where the former is
moderately strong and the latter has roughly the same strength as
\NII$\lambda$3995.

%From these comparisons involving a multitude of potential lines, we rejected
%those lines that deviate by more than 20\%, and also
%those lines with (very) low equivalent widths, EW$ \leq 50$~m\AA\ at \Teff\
%between 20 and 25~kK.
%
%Additionally, we provide information about adjacent lines present in the
%spectra of B and late O-stars,\footnote{Information extracted from the NIST
%database.} to avoid erroneous \NII\ analyses due to blending in 
%rapidly rotating objects (\vsini\ $\geq 100\ \kms$). The major
%source of contamination arises from \OII.
%
%Only one of the suggested lines, \NII\ $\lambda$3995, is completely
%isolated, and remains uncontaminated even at high
%rotation rates. Moreover, this is one of the strongest \NII\ lines located
%in the optical region, making it a good choice for deriving nitrogen
%abundances. The typical behavior of this line for 20~kK $\leq\ \Teff\ \leq$ 
%30~kK and \logg\ = 3.0 (i.e., for dwarfs to supergiants) 
%is shown in Fig.~\ref{ew-nii3995}.

%Other useful lines are NII $\lambda$5667 and $\lambda$5679, where the former
%is moderately strong and the latter has roughly the same strength as \NII\
%$\lambda$3995. The first line is (almost) isolated, whereas the latter is 
%blended by another \NII\ line. Other lines might be useful as well, but only
%in slowly rotating stars where blending is no problem.

The subset comprising lines from \NIII\ has been discussed in Paper~I.
Prominent lines from \NIII, \NIV\ and \NV\ are among the most
well-known features in O-stars, and can be used to infer
nitrogen abundances as well as effective temperatures for the earliest
subtypes, from the reaction of the \NIV/\NV\ ionization
equilibrium.\footnote{At the earliest O-types the standard technique
for deriving effective temperatures based on \HeI\ and \HeII\
line-strengths becomes difficult or even impossible, due to vanishing
\HeI\ and rather insensitive \HeII\ lines from \Teff~=~45~kK on.} In a
similar line of reasoning, \cite{walborn02b} used the \NIV \nivem\
emission line in combination with \NIII\trip\ to split the degenerate
O3 spectral type \citep{walborn71} into three different types O2, O3,
and O3.5, relying on the \NIV/\NIII\ emission line 
ratio.\footnote{This classification scheme has been criticized by
\cite{massey04, massey05} who found that for stars with similar
effective temperature and surface gravity the \NIV/\NIII\ emission
line ratio can vary over the full range as defined for O2 and O3.5.} 
%However,
%\cite{massey04,massey05} criticize this classification criteria because
%it is not clear that the effective temperature is a crucial factor
%within this scheme. In fact, they showed that for stars with the same
%effective temperature and similar surface gravity, the \NIV/\NIII\
%emission line ratio can vary by the full range between O2 and O3.5.
%This classification scheme also lacks for a solid theoretical
%foundation, especially with respect to \NIV \nivem. 
Thus, a detailed understanding and modelling of \NIV\nivem\ (together
with the \NIII\ triplet, see Paper~I) is 
mandatory, to safely employ this powerful diagnostics.

\NIV\nivem\ and \NIV\nivab\ connect the `neighbouring' levels of the
singlet series 1s$^2$ 2s3l with l~=~s,p,d (levels \#8, 9, and 12 in
Table~\ref{atom_lev_niv}). \NIV\nivem\ (if present) is observed in
emission in the majority of stars, and has been suggested to be formed
by photospheric NLTE processes (see below) rather than by emission in
an extended atmosphere, in analogy to the \NIII\ triplet emission.
Note that there is no detailed analysis of the line formation process.
So far, only \citet{Taresch97} and \citet{Heap06} simulated the
behavior of this line as a function of effective temperature. Heap et
al. found emission for this line at \Teff~$> 40,000~K$ for $\logg =
4.0$ and [N]~=~7.92, using the plane-parallel atmospheric code {\sc
tlusty}, supporting the idea that the emission is of photospheric
origin and that velocity fields are not required to explain the basic
effect. Other arguments for the photospheric origin of \NIV\nivem\ are the
agreement with other, absorption line profiles as a function of \vsini,
unshifted radial velocites, and lack of P~Cygni profiles (N. Walborn,
priv. comm.).

Interestingly, \NIV\nivab\ appears clearly in absorption in O-star
spectra, and seems to play a similar role as the \NIII\ absorption
lines at $\lambda\lambda4097-4103$ in the \NIII\ emission line problem
(see Paper~I). We note that \NIV\nivab\ can be significantly affected
by the presence of two Diffuse Interstellar Bands (DIBs) at 
$\lambda\lambda 6376.08, 6379.32$
(\citealt{herbig75, krelowski95}), the latter being stronger if
reddening is important. Fortunately, the stars analysed in this paper
are subject to low reddening, and these DIBs only minorly affect some
of the observed \NIV\nivab\ lines (e.g. N11-038, Fig.~\ref{N11-038}).

As for the previous pair of \NIV\ lines, \NIV\nivab\ and \NIV\ 
\nivem,
also the \NIV\ multiplets around
%\NIV$\lambda\lambda3478-3482-3484$
%and \NIV$\lambda\lambda7103-7109-7111-7122-7127-7129$ 
3480~\AA\ and 7103-7129~\AA\ are formed between levels of the same
series (here within the triplet system -- levels \#7, 10, and 11 in
Table~\ref{atom_lev_niv}), and seem to mimic the behavior of these
lines: at least in the earliest O-star regime they are prominent
features, where the former multiplet appears in absorption and the
latter in emission. Both line complexes are widely used in WR-star
analyses, and the emission in the latter multiplet is a strong feature
in most WR spectra. The lack of emission at this multiplet and also at
\NIV\nivem\ has been used for classification purposes in different
WR-star studies (e.g., \citealt{negueruela05}). Likewise, the
multiplet around 3480~\AA\ has been used by \citet{walborn04} to
infer both \Teff\ and nitrogen abundances for a set of O2 stars.

The remaining \NIV\ lines listed, \NIV$\lambda\lambda5200-5204-5205$,
belong also to the triplet system, and appear, if present, in
absorption (for O-stars). Unfortunately, most spectra used within this
study do not cover this spectral region. For the few field stars where
this range is available to us (see Sect.~\ref{obs}), this multiplet is not
visible.

\smallskip 
\noindent
Finally, \NV\ lines at $\lambda\lambda$4603-4619 are produced by transitions
between the fine-structure components of 3s $^2$S and 3p $^2$P$^0$
(levels \#3 and 4 in Table~\ref{atom_lev_nv}). These doublet lines
are strong absorption features in the earliest O-stars, showing
sometimes extended absorption in their blue wings or even pronounced
P-Cygni profiles (e.g., N11-031, Fig.~\ref{N11-031}),
%(absorption lines refilled by wind emission)
revealing that they can be formed in the wind. Thus, it is clear that
mass-loss and wind-clumping will influence the formation of these
lines. Besides, we also list \NV$\lambda\lambda4943-4945$ which become
important, (almost) isolated diagnostic lines in the spectra of very
early, nitrogen rich O- and WNL-stars. Unfortunately, the
corresponding wavelength range has not been observed for the bulk of
our sample stars (see Sect~\ref{obs}), whilst no features are visible
in the few early-type spectra (field stars) where this range is
available. Note that in order to use these lines, we would need
to extend our \NV\ model ion, including high-lying levels, to allow
for cascading processes into the corresponding upper levels at $n$ = 7
(which are our present uppermost ones).

\subsection{Understanding the \NIV\ \nivem/\nivab\ line formation}
\label{niv_em}
In the following, we discuss the most important mechanisms that
explain the presence of emission at \NIV \nivem, in particular the
decisive role of mass-loss.  Our analysis is based on the model-grid
as described in Sect.~\ref{grid-mod}, and refers to LMC background
abundances plus a solar \citep{asplund05} nitrogen abundance, [N] =
7.78,\footnote{roughly 0.9 dex above the LMC baseline
abundance, [N]$_{\rm baseline}$~=~6.9, following \citet{hunter07}.}
chosen in order to obtain pronounced effects.

\begin{figure}
\resizebox{\hsize}{!}
  {\includegraphics{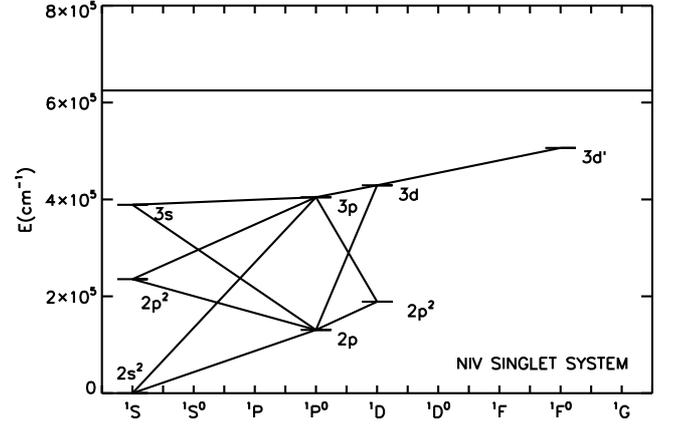}} 
\caption{
Simplified Grotrian diagram displaying the most important transitions
involved in the \NIV\ emission line problem. The horizontal line marks
the \NV\ ionization threshold. \NIV\ \nivem\ is formed by the
transition 3d $^1$D $\rightarrow$ 3p $^1$P$^0$, while the absorption line
at \nivab\ originates from the transition 3p $^1$P$^0$ $\rightarrow$ 3s
$^1$S. An efficient drain of 3p is provided by the `two-electron'
transitions 3p $^1$P$^0 \rightarrow$ 2p$^2\ ^1$S, $^1$D. Cascade processes
from 3d' $^1$F$^0$ and pumping from 2p $^1$P$^0$ are the major routes to
overpopulate the 3d $^1$D state. See text.} 
\label{grotrian_niv_em}
\end{figure}

All the important levels and the corresponding transitions involved in
the \NIV\ emission problem are summarized in
Fig.~\ref{grotrian_niv_em}. A comparison with the analogous diagram
for \NIII\ (Fig.~1 in Paper~I) shows a number of similarities but also
differences. %At first note 
In addition to what has already been outlined, the upper level of the emission line
(3d $\rarrow$ 3p) is fed by only weak dielectronic recombination, with
almost no influence on the population of 3d (contrasted to the
\NIII\ case), and there is no resonance line connected to 3d
(which turned out to be crucial for \NIII). Instead, the {\it lower}
level of \nivem, 3p, is connected with the ground-state. Similar to
\NIII, on the other hand, there are two strong `two-electron' transitions able to drain
3p, via 3p $^1$P$^0$ $\rightarrow$ 2p$^2$ $^1$S, $^1$D.
%, as well as higher levels (in particular, $3d'\ ^1F^0$) which can
%feed $3d$ by cascade processes. Moreover, level $3d$ is connected to
%$2p$ via a subordinate transition at roughly 335 \AA, where $2p$
%itself is coupled to the ground-state but also (via additional routes)
%to the draining levels $2p^2$. Finally, both diagnostic \NIV\ lines
%(\nivem\, and the absorption line at \nivab) share level $3p$ as the
%lower and the upper one, respectively, similar to the situation in
%\NIII.
%
%
% Paco comment 
% Interestingly the 3p --> 2p^2 1S transition at 591,18 is very close to
% the intercombination HeI line 1s1s1Se-1s2p3Po at 591.41, I had some 
% models in which I artificially played with the strength of the HeI 
% intercombination line but I see no effects at all.
%
\subsubsection{Basic considerations}
\label{basics}

\begin{figure}
\resizebox{\hsize}{!}
  {\includegraphics{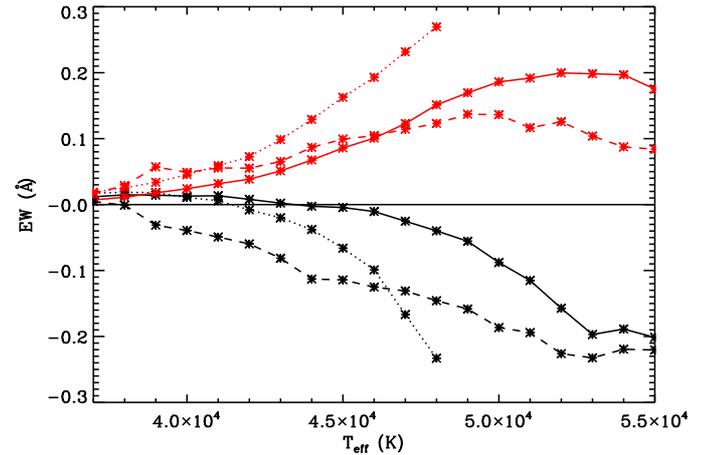}} 
\caption{Equivalent width (positive for absorption) of \NIV\nivem\
(black) and \NIV\nivab\ (red) as a function of \Teff. Solid and dotted
curves refer to low \mdot\ (model series `A') with \logg=4.0 and 3.7,
respectively, and dashed curves to supergiant mass-loss rates (model
series `E'), with \logg=4.0.
%Models with \logg=3.7 are no longer stable at hotter temperatures.
}
\label{ew4058_6380}
\end{figure}

\begin{figure}
\begin{minipage}{8cm}
\resizebox{\hsize}{!}
  {\includegraphics{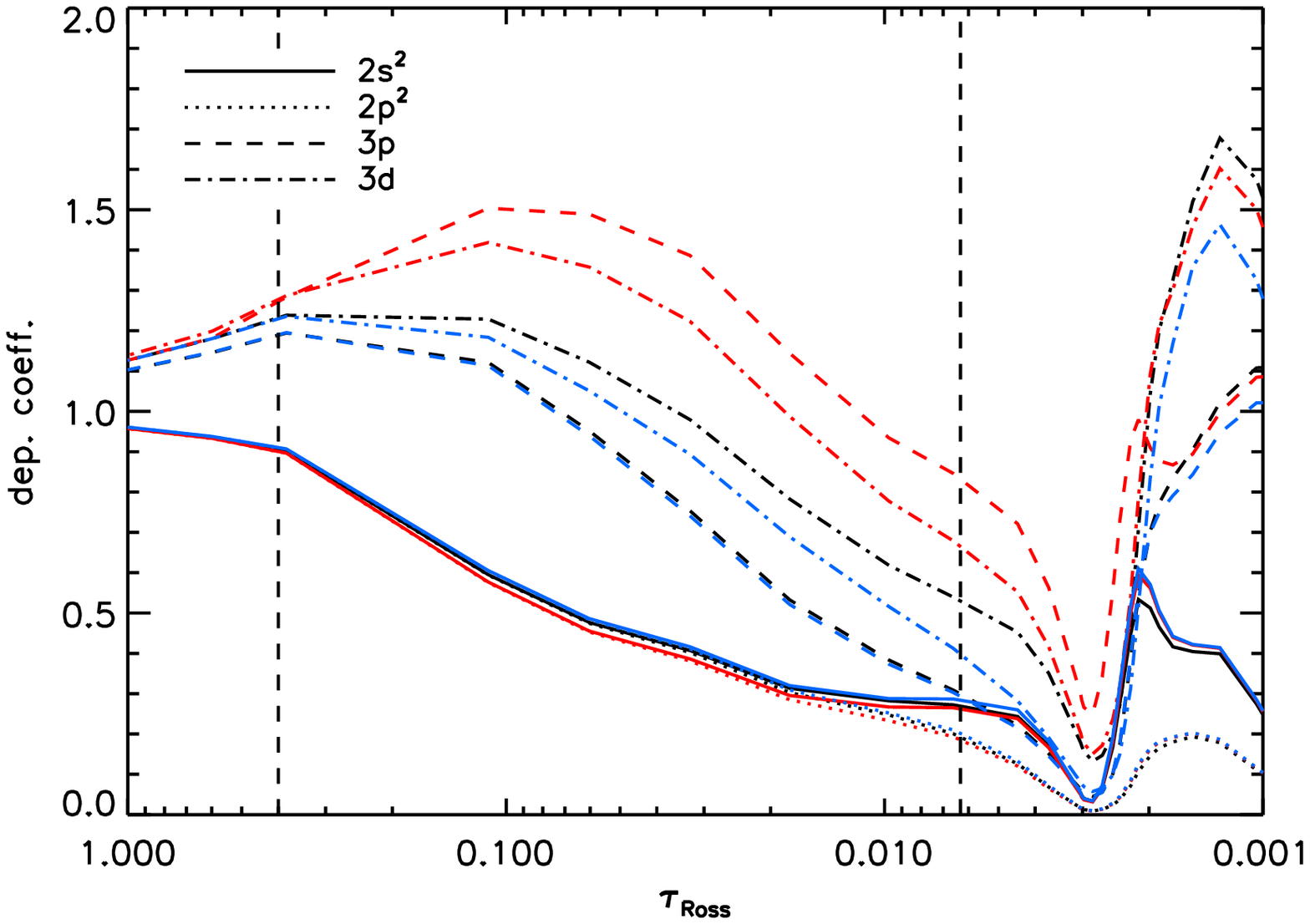}} 
\end{minipage}
\hspace{-.5cm}
\begin{minipage}{8cm}
\resizebox{\hsize}{!}
  {\includegraphics{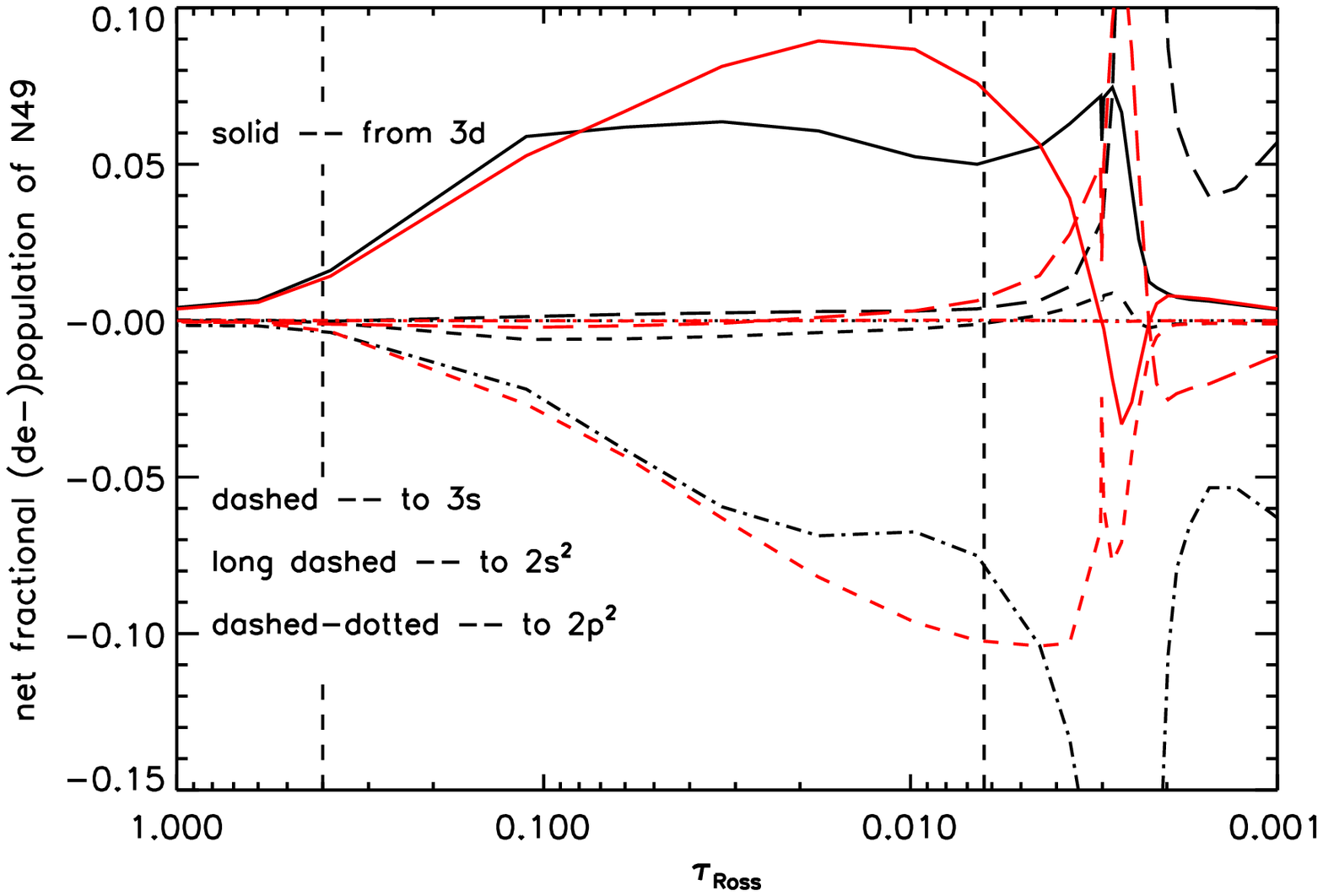}} 
\end{minipage}
\hspace{-.5cm}
\begin{minipage}{8cm}
\resizebox{\hsize}{!}
  {\includegraphics{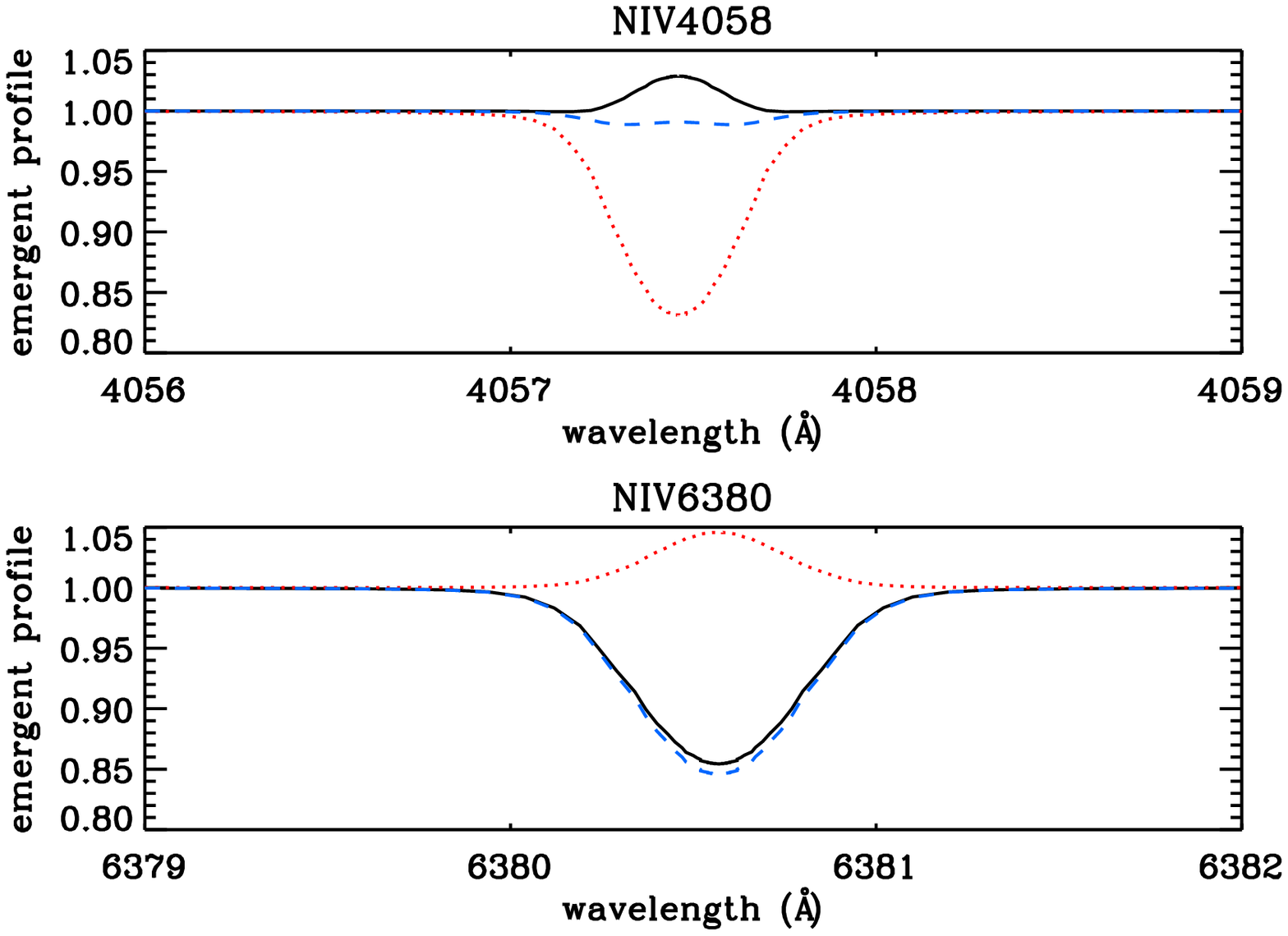}} 
\end{minipage}
\caption{Departure coefficients, fractional net rates and line
profiles for \NIV\nivem/\nivab\ and involved processes, for
model `A4540'. Designation for 2p$^2$ refers to 2p$^2$ $^2$D only
since the other fine-structure component, 2p$^2$ $^2$S, behaves
similarly. {\it Upper panel:} NLTE departure coefficients as a function
of \taur. The onset of the wind is clearly visible at \taur $\approx$
0.003. Black curves: standard model with $b_{{\rm 3d}} > b_{{\rm
3p}}$; red curves: draining transitions 3p $\rightarrow$ 2p$^2$
suppressed, leading to $b_{{\rm 3d}} < b_{{\rm 3p}}$; blue curves:
transition 2p $\rightarrow$ 3d suppressed. 
%
%Note the strong
%coupling of the ground state with levels $2p^2$ $^1D$ and $2p^2$
%$^1S$ (not indicated), which play a crucial role in depopulating level
%$3p$ via `two-electron' transitions. 
%In the standard model, $b_{3d} >
%b_{3p}$, whereas $b_{3d} < b_{3p}$ if these transitions are suppressed (red). 
%The formation region of the emission line is indicated
%by vertical dashed lines.
{\it Middle panel:} fractional net rates to and from 3p, for the standard
model (black) and the model with suppressed draining transitions
(red). 
%In the latter case, the preferred decay route switches from $3p
%\rightarrow 2p^2$ to $3p \rightarrow 3s$, though level $3p$ retains a
%much larger population. 
%
{\it Lower panel:} line profiles for \NIV\nivem\ and
\nivab, for the three models displayed in the upper panel, with
similar color coding.}
\label{A4540}
\end{figure}

In agreement with the results from \citet{Heap06}, our simulations
(see Fig.~\ref{ew4058_6380}) show that \NIV \nivem\ turns from weak
absorption (around \Teff\ $\approx$ 37~kK) into weak emission around
\Teff\ $\approx$~42~kK, for models with (very) low mass-loss rate and
\logg~=~4.0. As usual, we define equivalent
widths to be positive for absorption and to be negative for emission
lines. We find a 2~kK difference w.r.t. the turning point,
which can be attributed, to a major part, to the lower nitrogen
content of our models and different background abundances. Towards
hotter temperatures, the emission strength increases monotonically 
until a maximum around \Teff\ $\approx$ 53~kK has been reached, after which the emission
stabilizes and finally decreases. For lower gravities and/or higher
mass-loss rates, the line turns into emission at lower \Teff, so that,
for a given \Teff, the emission strength increases with decreasing
\logg\ and increasing mass-loss rate, \mdot. Since the {\it
absorption} strength of \NIV\nivab\ increases in a similar way (though
with a much weaker impact of \mdot, and only until \Teff\ $\approx$
50~kK), both lines appear (for a given \Teff) as anti-correlated, at
least for a large range of temperatures. 
%The decrease in
%equivalent width (E.W.) for higher \mdot\ and higher temperatures is mostly due to
%wind-emission. 
In contrast, the corresponding transitions of \NIII\
were found to be correlated (Paper~I).

The behaviour of both lines and the corresponding level structure
implies an efficient drain of level 3p that enhances the emission at
\nivem\ as well as prevents emission/increases absorption at \nivab,
and is provided by the two `two-electron' transitions 3p $\rarrow$
2p$^2$ $^1$D, $^1$S, similar to the case of \NIII.\footnote{though in
Paper~I we argued that in case of \NIII\ this mechanism becomes
suppressed in realistic model atmospheres with near-solar background
abundances.} 

To investigate this mechanism in more detail, and to avoid
`contamination' by wind effects, we concentrate at first on a
low-\mdot\ model with decent emission at \NIV\nivem, with $\Teff~=~45,000~K$
and $\logg\ = 4.0$.\footnote{ 
At cooler \Teff, this line is in absorption because of a lower \NIV\
ionization fraction implying deeper formation depths, which are closer
to LTE.} In the upper panel of Fig.~\ref{A4540}, we provide the
NLTE departure coefficients, $b$, of involved levels, where black
curves refer to our standard model. Obviously, level 3d is
overpopulated with respect to 3p ($b_{{\rm 3d}} > b_{{\rm 3p}}$) over the
complete line formation region. On the other hand, levels 2p$^2\ ^1$D,
2p$^2\ ^1$S, and 2p (the latter two not displayed) are (mostly
collisionally) coupled to the ground state, which in itself is
strongly depopulated, due to rather large ionizing fluxes (see below).
This situation closely resembles the situation in \NIII, where
strongly depopulated draining levels (for non-blocked models) favoured
a depopulation of the analogous level 3p.

To further clarify the impact of the different processes, we
investigate the corresponding {\it net rates} responsible for the
population and depopulation of level 3p (Fig.~\ref{A4540}, middle
panel, black curves). As in Paper~I, we display the dominating
individual net rates (i.e., $n_j R_{ji} - n_i R_{ij} > 0$ for
population, with index $i$ the considered level) as a fraction of the
{\it total} population rate. Indeed, the drain by level 2p$^2\ ^1$D
(dashed-dotted) and/or level 2p$^2\ ^1$S (not displayed) are the most
important processes that depopulate level 3p in the line formation
region. In contrast, the resonance line does not contribute to any
(de-)population of level 3p, since it is (almost) in detailed
balance (long dashed line).

To check the validity of our scenario, we calculated an alternative
model where the two draining transitions have been suppressed, by
using very low oscillator strengths. Indeed, the upper panel of
Fig.~\ref{A4540} (red curves) shows that now $b_{{\rm 3d}}$ is 
smaller than $b_{{\rm 3p}}$, and \nivem\ goes into absorption (lower panel,
red color). From the fractional net rates, we see that the preferred
decay route has switched from 3p $\rightarrow$ 2p$^2$ (standard model,
black) to 3p $\rightarrow$ 3s (red, dashed), though level 3p retains
a much larger population. 

The upper level of \nivem, 3d, is predominantly fed by cascading from
3d' $^1$F$^0$, and also by pumping from level 2p, whilst dielectronic
recombinations are negligible. Suppressing the population from level
2p leads to less emission (Fig.~\ref{A4540}, lower panel, blue
colors), due to a less populated level 3d (upper panel).

Let us now consider the behaviour of the absorption line at \nivab, 
resulting from the transition 3p $\rightarrow$ 3s, again by means of
Fig.~\ref{A4540}. As mentioned earlier, this line shows an 
anti-correlation\footnote{ when certain parameters/processes are
changed {\it for a given \Teff}, e.g., the strength of the draining
levels, the background opacities and so on. The {\it overall} increase
of these line-strengths as a function of \Teff\ is related to the
increasing ionization fractions.} with \NIV \nivem, in contrast to the
behaviour of the corresponding \NIII\ lines which appear as
correlated. In Paper~I we argued that the latter correlation results
from the proportionality of the level populations of 3p and 3s. That
is, when $b_{{\rm 3p}}$ decreases (e.g., due to increased
`two-electron' drain), $b_{{\rm 3s}}$ decreases in parallel due to
less cascading, and the absorption at $\lambda 4097$ becomes weaker in
concert with an increase in the triplet emission. Vice versa, an
increase of 3p implies less emission of the triplet lines and more
absorption at $\lambda 4097$, respectively. 

Such a reaction requires the transition 3p $\rightarrow$ 3s to be
optically thin, dominated by spontaneous decays, which is no longer
true for \NIV\nivab. Due to a mostly significant optical depth, the
radiative net rate is no longer dominated by spontaneous decays, but
depends also on absorption and induced emission processes. Now, an
increased population of 3p leads to less increase of $b_{{\rm
3s}}$,\footnote{The net radiative rate (downward) is proportional to
$A(1-\bar J/S)$, with Einstein coefficient for spontaneous decay, $A$,
scattering integral $\bar J$ and source function, $S$.} and the
absorption becomes {\it weaker} because of an increased source
function $\propto b_{{\rm 3p}}/b_{{\rm 3s}}$. Vice versa, a decrease
in the population of 3p leads to more absorption at \nivab\ in
parallel with more emission at \nivem. This behaviour becomes
particularly obvious if we investigate the reaction of the absorption
line when suppressing the draining transitions. In this case, 3p
becomes strongly overpopulated (Fig.~\ref{A4540}, upper panel, red
color), and \nivem\ goes into absorption whilst \nivab\ becomes an
emission line, due to a significantly increased source function (more
pumping than in the original scenario). We checked that if the
absorption and stimulated emission terms in the 3p $\rightarrow$ 3s
transitions are neglected, \nivab\ displays more absorption instead,
in accordance with our previous arguments. 

Note, however, that this anti-correlation is not complete. E.g., if
one changes processes which have an effect on 3d alone, the
absorption strength of \nivab\ remains unaltered. Thus, by suppressing
2p $\rightarrow$ 3d, only $b_{{\rm 3d}}$ is affected (upper panel, blue vs.
black curves), and there is less emission at \nivem\ while the
absorption at \nivab\ remains at the previous level (lower panel). 

Summarizing, we interpret the different correlations between emission
and absorption line-strength in \NIII\ and \NIV\ as due to optical
depth effects in the 3p $\rightarrow$ 3s transition. As long as this 
is optically thin, cascade effects dominate, and both lines appear as
correlated (\NIII). Larger optical depths introduce a
counteracting `source-function effect', and the lines become
anti-correlated (\NIV).  The rather large degree of such
anti-correlation supports the importance of the draining transitions,
since these are able to influence both the absolute population of the
involved levels as well as their ratios in a very efficient way, by
providing additional decay channels for level 3p.

\subsubsection{The impact of wind effects}
\label{wind}

\begin{figure}
\begin{minipage}{8cm}
\resizebox{\hsize}{!}
  {\includegraphics{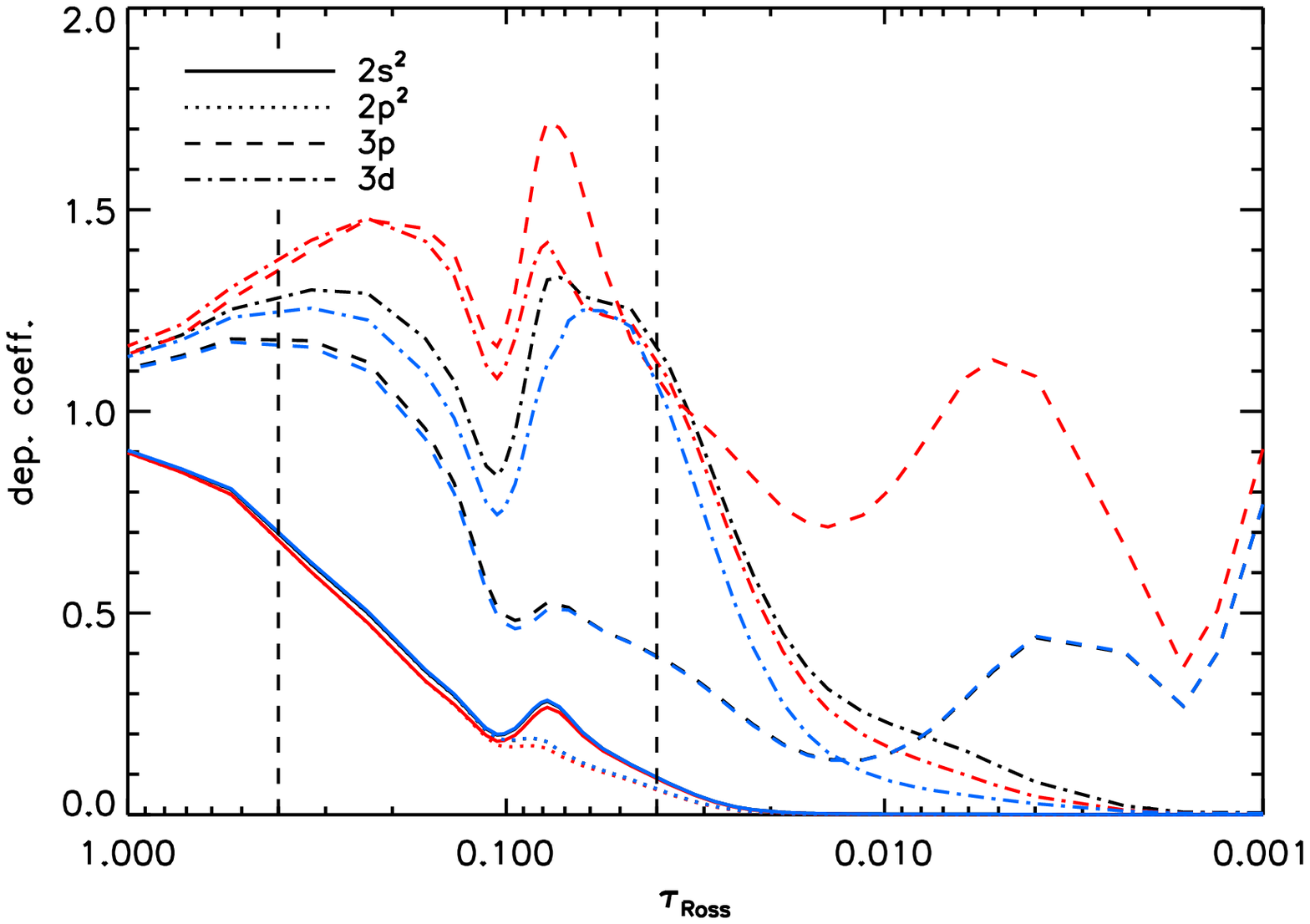}} 
\end{minipage}
\hspace{-.5cm}
\begin{minipage}{8cm}
\resizebox{\hsize}{!}
  {\includegraphics{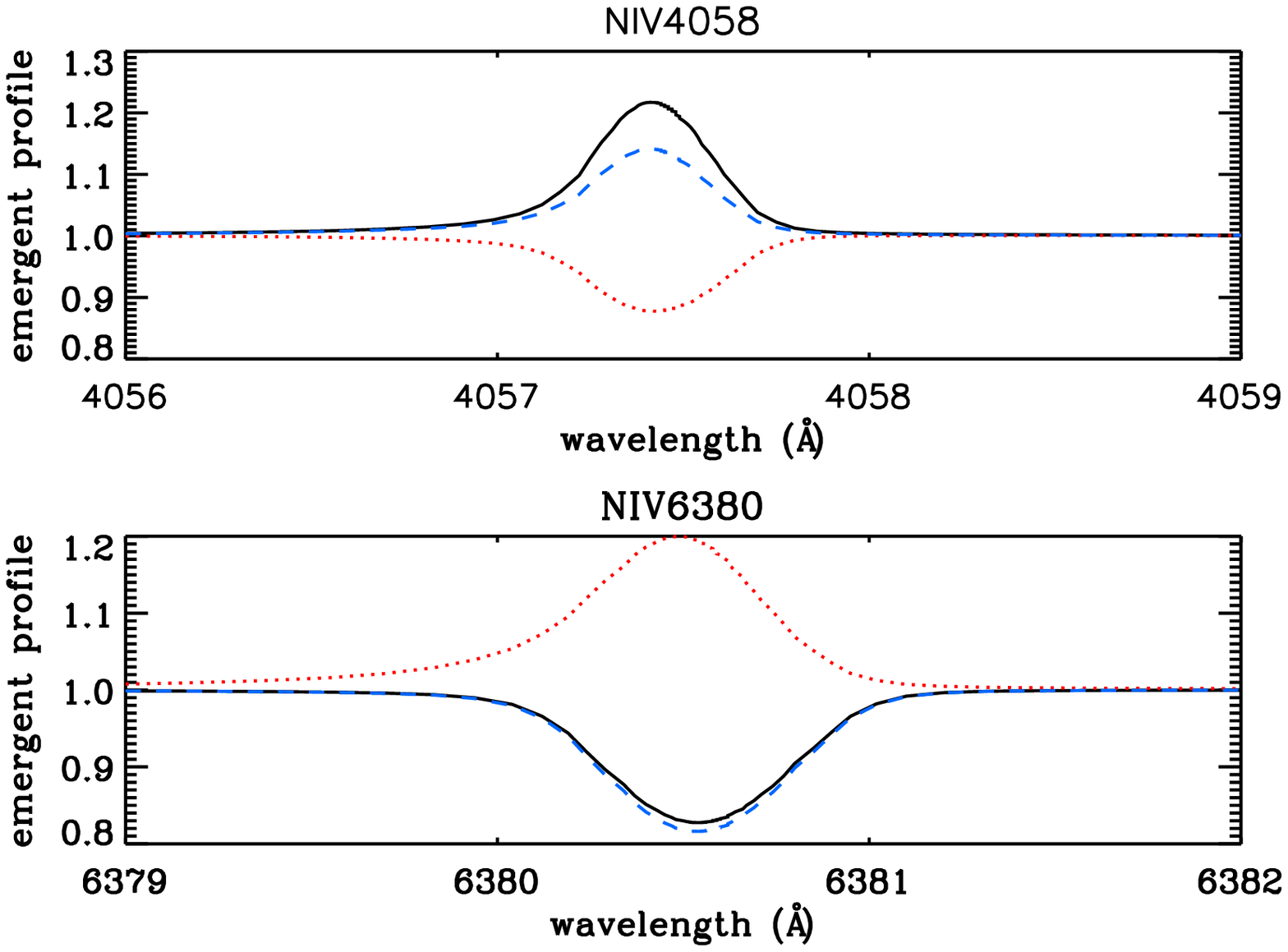}} 
\end{minipage}
\caption{Departure coefficients (top) and line profiles for 
\NIV\nivem/ \nivab\ (bottom) for the dense-wind model `E4540'. 
Level designations, lines, and color coding as in Fig.~\ref{A4540}.}
\label{E4540}
\end{figure}

So far, we discussed the possibility of obtaining emission at \NIV\nivem\
via solely photospheric NLTE processes.
In Paper~I, the presence of a wind (actually, a steep rise of the
velocity field in the outer photosphere) turned out as crucial to
explain the observed \NIII\ triplet emission in Of-stars, enabling an
efficient pumping of the {\it upper} level 
by the corresponding resonance line. 
To investigate how the presence of a wind affects the emission at
\NIV\nivem, we compare our previous model `A4540' with model
`E4540', which has the same stellar parameters but a considerably
larger, supergiant-like mass-loss rate.
%, to ensure that the wind
%originates deep enough 
%(close to the photospheric region where the line is formed) 
%to induce a significant effect. 
%This model `E4540' has
%the same effective temperature and surface gravity as `A4540', but a
%typical supergiant mass-loss rate, \mdot\ $\approx$ 3.4 \Mdu,
%contrasted to the `A' model with \mdot\ $\approx$ 0.1 \Mdu.
It turns out that the inclusion of such a strong wind has a pronounced
effect. Comparing Fig.~\ref{E4540} (lower panel) with
Fig.~\ref{A4540}, model `E4540' (black) results in much more emission
than `A4540', increasing the equivalent width of \nivem\ from $-$7 to
$-$114 m\AA. Also for higher \Teff\ the impact of \mdot\ remains
significant. The absorption line \NIV\nivab\ is affected by the wind
as well, though less pronounced. The absorption becomes slightly
stronger (by 18\% in the equivalent width), i.e., the anti-correlation
discussed above is still present. This is valid not only for model
`E4540', but also for hotter models (Fig.~\ref{ew4058_6380}), until
the wind-emission begins to contaminate \NIV\nivab.

The origin of such stronger emission at \nivem\ becomes clear if one
inspects the involved departure coefficients (Fig.~\ref{E4540}, upper
panel, black curves). Again, the onset of the wind is clearly visible
(at $\taur \approx 0.1$), now much deeper than in the `A' model, and
the line formation region is located in between $\taur\sim 0.40 {\ldots} 0.04$.
%maybe comment that line forms in the wind region for
%the E-models also.
Compared to the `A' model, `E4540' displays a larger ground-state
depopulation, where the ground-state remains coupled with the draining
levels 2p$^2$ as well as with level 2p (not displayed). This leads,
particularly in the transition region between photosphere and wind, to
an extreme depopulation of 3p. Since level 3d becomes strongly
overpopulated, mostly due to feeding by 3d' $^1$F$^0$ (which is severely
overpopulated as well), the resulting line
source function is quite large and partly even in inversion, which
explains the pronounced emission at \NIV\nivem. 
% PACO BEGIN
% Compared to the no-wind case, what we clearly see in the run of the departure
% coefficients is that 3d does not change but 2p ground state and specially
% 3p is strongly depopulated. 
% PACO END
%Jorge: I think Paco made this comment because in the version that we sent
% to Paco, the line forming region was not included
%Jorge: this was an old comment.

All these differences are caused by the onset of the wind. At first
note that the \NIV\ continuum ($\lambda <$ 160~\AA) is strongly
coupled with the \HeII\ continuum. As already realized by
\citet{Gabler89}, increasing mass-loss leads to more \HeII\
ground-state depopulation, to higher fluxes in the \HeII\
continuum and thus also to higher fluxes in the
\NIV\ continuum. Consequently, the \NIV\ ground-state becomes strongly
depopulated, and the nitrogen ionization equilibrium switches from
\NIV\ (being the dominant stage in the deeper photosphere) towards \NV\
from the transition region on. This also favours the overpopulation of
3d, by means of increased recombinations to high-lying levels with
subsequent cascades via 3d' $^1$F$^0$. 

One might now argue that the inclusion of the wind could amplify the
impact of the resonance transition(s), by producing deviations from
detailed balance leading to strong pumping, similar to the case of
\NIII. Unlike the situation in \NIII, however, such an
effect would lead to less line emission or even absorption at \NIV\nivem, 
since here the resonance line is connected to the lower level, 3p.
As it turns out, however, this effect is not present,
%, at least for
%mass-loss rates which are of typical size or below. The reason is very simple, 
since the resonance line is too
strong (\NIV\ is the dominant ion until the transition region) to
leave detailed balance before the wind has reached a significant speed.
Only then, the resonance transition becomes dominant in populating 
level 3p, but this occurs already far beyond the formation region of
\NIV\nivem. Close to the formation region, there is only a moderate
population of 3p by the resonance line, similar to the population
from 3d itself. Even this additional population is counteracted 
(actually, even slightly overcompensated) by enhanced drain towards
2p$^2$, not only in the formation region of \nivem\ but also in those
outer regions where the resonance line strongly pumps.

These arguments are supported by the fact that \NIV \nivab\ is only
slightly affected by the wind, where the increased absorption results
mostly from a diminished source function due to a less populated 3p
level. Performing the same tests as for the thin wind case, i.e.,
either suppressing the drain or suppressing the population of 3d via 2p
$\rightarrow$ 3d leads to similar results, as can be seen from the red
and blue curves and profiles in Fig.~\ref{E4540}, respectively.

We conclude that \mdot\ is a key parameter for modeling the
\NIV\ emission line, where in contrast to \NIII\ the basic mechanism 
(for typical mass-loss rates and below) is always due to the
depopulation of the lower level by the `two-electron' transitions.
This drain becomes stronger as a function of \mdot, because of
increasing ionizing fluxes leading to more ground-state depopulation.

\section{Stellar sample and observations}
\label{samp_obs}
\subsection{The stellar sample}
\label{sample} 

Table~\ref{tab_sample} lists our stellar sample which has been drawn
from the analysis of LMC O-/early B-stars by Mok07. 

Three of the 28 stars from the original sample have been discarded
from the present analysis, for two reasons. First, from our analysis
we suspect that N11-004 and N11-048 might be (SB1) binaries, where the
former object shows discrepant line shifts and for the latter we were
not able to reproduce accurately the observed He lines (shape and
strength). A possible binarity of N11-048 was also suggested by
Mok07, because of similar reasons.
%They supported this because they found a very low
%Helium abundance and they sistematically underpredicted the width
%of \HeI\ lines, whilst the \HeII\ ones were overpredicted. We found the same
%pattern in our analysis of this star. 
The other discarded star, Sk--67$^{\circ}$ 166, is the only object in the original sample which
seems to be strongly evolved (helium content \YHe~=~N(He)/N(H) =
0.20 {\ldots} 0.28, \citealt{Crowther02} and Mok07, respectively), and
has a very dense wind, with both \Ha\ and \HeII4686 in strong
emission. We confirm the stellar/wind parameters as derived by Mok07
(almost perfect fit quality of H/He lines), but did not succeed in a
reasonable fit for the nitrogen lines. A comparison with the analysis
by \citet{Crowther02} shows similar discrepancies. Because of this
problem and because of its highly evolved evolutionary status, which
does not match with all other objects in our sample, we decided to
discard this object from our present analysis and will re-consider it
in a future attempt.

The remaining sample consists of
20 O-stars, mostly giants or dwarfs, and 5 early B-type supergiants or
giants. All of the B-stars and 15 O-stars are associated with the
cluster N11, and the others are field stars. The early B-stars have
been included in our sample to allow us for a comparison with previous
analyses of such stars \citep{hunter09} in order to check the
consistency of different codes and methods in the transition region
between O- and B-types (Sect.~\ref{comp_hunter}).

Table~\ref{tab_sample} gives information about spectral type,
V-magnitude, interstellar extinction and absolute visual magnitude,
and has been taken from Mok07. Spectral types for N11
objects are based on \citet{evans06}, slightly revised by Mok07 in
collaboration with C. Evans (priv. comm.), and for the field
stars from \citet{walborn95}, and \citet{massey95,massey05}. 
For the field star Sk--70$^{\circ}$~69 we added the ((f)) designation since
the present spectra show clear emission at \NIII\trip\ and
\HeII$\lambda4686$ in absorption. 
%Photometric data for the N11 objects were obtained within the FLAMES survey
%(E06) but some of them (see Table~\ref{tab_sample}) are taken from
%\citet{Parker92}. For the field stars the data come from
%\citet{Ardeberg72}, \citet{Isserstedt79}, and \citet{Massey02}.
%Insterstellar extinctions $A_V$ and absolute visual magnitude $M_V$
%were calculated by Mok07.

\subsection{Observations}
\label{obs}
Most of the observations (for objects denoted by `N11-') have been carried
out within the FLAMES~I survey, and are described
in detail in \citet{evans06}. In brief, the data were obtained using the Fibre
Large Array Multi-Element Spectrograph (FLAMES) at the VLT,
%For
%the LMC objects, the field samples were centered towards the clusters
%N11 and NGC~2004\footnote{Mok07 only use data from cluster
%N11.}, 
for six wavelengths settings with an effective resolving power of
$R~\simeq~20,000$. The S/N ratios are in the ranges 50-200 for LMC
objects. After sky substraction, each wavelength range was co-added
and normalized by means of a cubic spline.\footnote{We
performed additional re-normalizations for different wavelength
ranges.} The final merged spectra cover two spectral ranges,
~3850-4750~\AA\ and ~6300-6700~\AA.

To improve the sampling in luminosity and temperature, Mok07 augmented
the N11 sample by LMC O-type field stars, which were observed using
the UVES spectrograph at the VLT as part of the ESO programmes
67.D-0238, 70.D-0164, and 074.D-0109 (P.I. P.~Crowther). Spectra
were obtained for four different wavelength settings, at an effective
resolving power of $R \simeq 40,000$. The final product provides
coverage between 3300-5600~\AA\ and 6300-10400~\AA\ for all stars
except for Sk--70$^{\circ}$ 69, where `only' the region
between~3300-5600~\AA\ and ~6300-6700~\AA\ had been observed. The
typical S/N ratios achieved for all spectra lie in the range 60-80.

\begin{table}
\tabcolsep1.mm
\center
\caption{Sample stars used within this study, along with spectral type,
V-magnitude, interstellar extinction $A_V$, and absolute visual magnitude
$M_V$. All quantities have been taken from Mok07.}
\vspace{0.3cm}
\label{tab_sample}
\begin{tabular}{lllccc}
\hline 
\hline
Star & Cross-IDs & Spectral Type & V  & $A_V$  & $M_V$ \\
\hline
N11-026 & - &O2 III(f$^*$) &13.51 &0.47 &$-$5.46\\
N11-031 & P3061/LH10-3061  &ON2 III(f$^*$) &13.68 &0.96 &$-$5.78\\
N11-038 & P3100 & O5 II(f$^+$) &13.81 &0.99 &$-$5.68\\
Sk--66$^{\circ}$ 100 & - & O6 II(f) &13.26 &0.34 &$-$5.58\\
N11-032 & P3168 & O7 II(f) &13.68 &0.65 &$-$5.47\\
N11-045 &    -      & O9 III &13.97 &0.50 &$-$5.03\\
BI253 &      -      & O2 V((f$^*$)) &13.76 &0.71 &$-$5.45\\
BI237 &      -      & O2 V((f$^*$)) &13.89 &0.62 &$-$5.23\\
N11-060 & P3058/LH10-3058 & O3 V((f$^*$)) &14.24 &0.81 &$-$5.07\\
Sk--70$^{\circ}$ 69 &   -  &  O5 V((f)) &13.95 &0.28 &$-$4.83\\
N11-051 &    -      & O5 Vn((f)) &14.03 &0.19 &$-$4.66\\
N11-058 &    -      & O5.5 V((f)) &14.16 &0.28 &$-$4.62\\
Sk--66$^{\circ}$ 18 &   -   & O6 V((f)) &13.50 &0.37 &$-$5.37\\
N11-065 & P1027 & O6.5 V((f)) &14.40 &0.25 &$-$4.35\\
N11-066 &    -      & O7 V((f)) &14.40 &0.25 &$-$4.35\\
N11-068 &    -      & O7 V((f)) &14.55 &0.28 &$-$4.23\\
N11-061 &    -      & O9 V &14.24 &0.78 &$-$5.04\\
N11-123 &    -      & O9.5 V &15.29 &0.16 &$-$3.37\\
N11-087 & P3042     & O9.5 Vn &14.76 &0.62 &$-$4.36\\
\hline
N11-029 &    -      & O9.7 Ib &13.63 &0.56 &$-$5.43\\
N11-036 &    -      & B0.5 Ib &13.72 &0.40 &$-$5.18\\
N11-008 & Sk--66$^{\circ}$ 15 & B0.7 Ia &12.77 &0.84 &$-$6.57\\
N11-042 & P1017     & B0 III &13.93 &0.22 &$-$4.79\\
N11-033 & P1005     & B0 IIIn &13.68 &0.43 &$-$5.25\\
N11-072 &   -       & B0.2 III &14.61 &0.09 &$-$3.98\\
\hline
\end{tabular}
\tablefoot{Primary identifications for N11
objects are from~\citet{evans06}. Identifications starting
with "Sk", "BI", "P", and "LH" are from~\citet{Sanduleak70},
\citet{Brunet75}, \citet{parker92}, and \citet{walborn02b, walborn04}, 
respectively.}
\end{table}

\section{Analysis}
\label{anal}

\subsection{Methodology}
\label{metho}

In an ideal world, we could have used the stellar and wind parameters
as obtained by Mok07 from H/He lines, and simply derived the nitrogen
abundances on top of atmospheric models with these parameters.
Unfortunately, there are reasons to re-analyze all programme stars. At
first, the present {\sc fastwind} version (see Sect.~\ref{nlte_fast})
is somewhat different from the version used by Mok07, and the
parameters need certain (mostly small) alterations in order to reach a
similar fit quality to the H/He lines. Second, we used a somewhat
different fitting strategy with respect to the `free' parameters,
which changes the optimum fit. In contrast to Mok07, we derived and
fixed the projected rotational velocity, \vsini, independently from
the actual fitting procedure (Sect.~\ref{stellar_wind_param}), whereas
Mok07 included \vsini\ as a free parameter in their fitting algorithm.
Moreover, during our fit procedure we allowed for the presence of
extra line-broadening (`macro-turbulence', \vmac), not considered by
Mok07. Differences in \vsini\ and \vmac\ can lead to certain
differences in the outcome of the fit, since the profile shapes might
change (e.g., Fig.~4 in \citealt{puls08}).  
Third, and most important, is the fact that we now aim at a
consistent fit for the H/He {\it and} nitrogen lines. Thus, and in the
sense of a compromise solution (minimization of the differences
between observed and synthetic spectra for {\it all} lines), different
stellar parameters which result in a modest change of the fit quality
of H/He alone\footnote{In particular at earliest spectral types, the
sensitivity of H/He on changes in the atmospheric parameters is rather
weak.} can lead to a significant improvement with respect to the
complete set of lines.  

In so far, we opted for an entire re-analysis, performed mostly by a
simple `fit-by-eye' method where we try to accomplish the best fit to
the strategic lines by visual inspection (for a discussion, see
\citealt{Mokiem05}). Since we start from the parameter set as provided
by Mok07 (highest `fitness'\footnote{which quantifies the
quality of the solutions resulting from a genetic algorithm
optimization, see \citet{Mokiem05} and references therein.} with
respect to their assumptions), our new solution should be located at
or close to the {\it global} maximum of the corresponding merit
function as well and not only at a local one. Note that our
derivation of nitrogen abundance and micro-turbulence for most of the
cooler sample stars relies on a more objective method
(Sect.~\ref{nitro_ab}).

\begin{table*}
\centering
\caption{Fundamental parameters for the LMC sample, assuming unclumped
mass-loss.}
\tabcolsep1.0mm
\label{tab_param}
\begin{tabular}{lllcccrccccccccc}
\hline 
\hline
\multicolumn{1}{l}{Star}
&\multicolumn{1}{l}{ST}
&\multicolumn{1}{l}{Secondary}
&\multicolumn{1}{c}{\Teff}
&\multicolumn{1}{c}{\logg}
&\multicolumn{1}{c}{$\logg_{\rm true}$}
&\multicolumn{1}{r}{\Rstar}
&\multicolumn{1}{c}{log \Lstar}
&\multicolumn{1}{c}{$Y_{\rm He}$}
&\multicolumn{1}{c}{\vmic}
&\multicolumn{1}{c}{\vsini}
&\multicolumn{1}{c}{\vmac}
&\multicolumn{1}{c}{\mdot}
&\multicolumn{1}{c}{$\beta$}
&\multicolumn{1}{c}{\vinf}\\
\multicolumn{1}{l}{}
&\multicolumn{1}{l}{}
&\multicolumn{1}{l}{\Teff\ diag.}
&\multicolumn{1}{c}{(kK)}
&\multicolumn{1}{c}{(cgs)}
&\multicolumn{1}{c}{(cgs)}
&\multicolumn{1}{r}{(\rsun)}
&\multicolumn{1}{c}{(\lsun)}
&\multicolumn{1}{c}{}
&\multicolumn{1}{l}{(\kms)}
&\multicolumn{1}{l}{(\kms)}
&\multicolumn{1}{r}{(\kms)}
&\multicolumn{1}{l}{(\mdu)}
&\multicolumn{1}{c}{}
&\multicolumn{1}{c}{(\kms)}\\
\hline
N11-026              &O2 III(f$^*$) & \NIII/\NIV/\NV   &49.0 &4.00 &4.00 &11.3 &5.82 &0.10 &10.0 & 72 &60 &1.56 &1.08 &[3120]\\
                     &              & \NIV/\NV         &52.0 &4.10 &4.10 &11.0 &5.89 &     &     &    &   &1.49 &     &\\
N11-031              &ON2 III(f$^*$)& \NIII/\NIV       &47.8 &3.95 &3.95 &13.4 &5.92 &0.11 &10.0 & 71 &60 &2.02 &1.08 &3200\\
                     &              & \NIV/\NV         &56.0 &4.00 &4.00 &12.2 &6.12 &     &     &    &   &2.20& &\\
N11-038              &O5 II(f$^+$)  & \NIII/\NIV       &40.5 &3.70 &3.71 &14.0 &5.67 &0.08 &10.0 &100 &50 &1.21 &0.98 &[2600]\\
Sk--66$^{\circ}$ 100 &O6 II(f)      & \NIII/\NIV       &39.0 &3.70 &3.71 &13.7 &5.59 &0.19 &10.0 &59 &60 &0.83 &1.27 &2075\\
N11-032              &O7 II(f)      & \NIII/\NIV       &36.0 &3.50 &3.51 &13.6 &5.44 &0.09 &10.0 & 60 &70 &0.97 &0.80 &[1920]\\
N11-045              &O9 III        & -                &32.3 &3.32 &3.33 &12.0 &5.15 &0.07 &10.0 & 64 &80 &0.69 &0.80 &[1550]\\
BI253                &O2 V((f$^*$)) & \NIV/\NV         &54.8 &4.18 &4.20 &10.7 &5.97 &0.08 &10.0 &230 &-  &1.53 &1.21 &3180\\
BI237                &O2 V((f$^*$)) & \NIV/\NV         &53.2 &4.11 &4.12 & 9.7 &5.83 &0.09 &10.0 &140 &-  &0.62 &1.26 &3400\\
N11-060              &O3 V((f$^*$)) & \NIII/\NIV/\NV   &48.0 &3.97 &3.97 & 9.5 &5.63 &0.12 &10.0 & 68 &40 &0.51 &1.26 &[2740]\\
                     &              & \NIV/\NV         &51.0 &4.10 &4.10 & 9.2 &5.71 &     &     &    &   &0.48 &     &\\
Sk--70$^{\circ}$ 69  &O5 V((f))     & \NIII/\NIV       &42.3 &3.93 &3.94 & 9.1 &5.38 &0.14 &10.0 &131 &-  &0.43 &0.80 &2750\\
N11-051              &O5 Vn((f))    & \NIII/\NIV       &41.4 &3.70 &3.83 & 8.6 &5.42 &0.08 &10.0 &350 &-  &0.41 &0.80 &[2110]\\
N11-058              &O5.5 V((f))   & \NIII/\NIV       &40.8 &3.75 &3.76 & 8.4 &5.24 &0.10 &6.0$^*$&62&60 &0.01 &1.00 &[2470]\\
Sk--66$^{\circ}$ 18  &O6 V((f))     & \NIII/\NIV       &39.7 &3.76 &3.76 &12.2 &5.52 &0.14 &10.0 & 75 &40 &1.08 &0.94 &2200\\
N11-065              &O6.5 V((f))   & \NIII/\NIV       &41.0 &3.85 &3.85 & 7.4 &5.14 &0.13 &10.0 & 60 &50 &0.05 &1.00 &[2320]\\
N11-066              &O7 V((f))     & \NIII/\NIV       &37.0 &3.70 &3.71 & 7.9 &5.02 &0.10 &10.0 & 59 &50 &0.14 &0.80 &[2315]\\
N11-068              &O7 V((f))     & \NIII/\NIV       &37.0 &3.70 &3.71 & 7.5 &4.98 &0.10 &10.0 & 30 &50 &0.13 &1.12 &[3030]\\
N11-061              &O9 V          & -                &34.0 &3.55 &3.55 &11.5 &5.20 &0.09 & 5.0 & 54 &80 &0.52 &0.80 &[1900]\\
N11-123              &O9.5 V        & -                &34.3 &4.20 &4.21 & 5.4 &4.56 &0.09 &10.0 &115 &-  &0.08 &0.80 &[2890]\\
N11-087              &O9.5 Vn       & -                &32.7 &4.04 &4.08 & 8.8 &4.90 &0.10 &10.0 &260 &-  &0.11 &0.80 &[3030]\\
\hline
N11-029              &O9.7 Ib       & \NII/\NIII       &29.0 &3.20 &3.21 &15.8 &5.20 &0.08 &10.0 & 46 &60 &0.28 &1.23 &[1580]\\
N11-036              &B0.5 Ib       & \NII/\NIII       &25.8 &3.11 &3.11 &15.5 &4.98 &0.08&12.1$^*$&39&40 &0.11 &0.80 &[1710]\\
N11-008              &B0.7 Ia       & \NII/\NIII       &26.3 &3.00 &3.00 &29.5 &5.57 &0.10&4.7$^*$& 46&60 &0.62 &1.30 &[2390]\\
N11-042              &B0 III        & \NII/\NIII       &29.2 &3.59 &3.59 &11.9 &4.97 &0.08 & 5.0 & 21 &25 &0.19 &1.19 &[2310]\\
N11-033              &B0 IIIn       & -                &26.7 &3.20 &3.34 &15.7 &5.05 &0.10 & 5.0 &256 &-  &0.25 &1.03 &[1540]\\
N11-072              &B0.2 III      & \NII/\NIII       &29.8 &3.70 &3.70 & 8.1 &4.67 &0.10&2.7$^*$& 14&10 &0.25 &1.30 &[2100]\\
\hline
\end{tabular}
\tablefoot{$\logg_{\rm true}$ is the surface gravity corrected for
centrifugal effects. Secondary \Teff\ diagnostics, used in
parallel with the \HeI/\HeII\ ionization equilibrium, is indicated for
each star. For three stars, two parameter sets are provided, due to
problems of reaching a simultaneous fit for all considered lines from three
ionization stages. For N11-026 and N11-060, the first entry is our
preferred one, and the second entry is a (disfavoured) alternative,
whilst for N11-031 we consider both solutions as possible.
(see Sect.~\ref{comments}). Brackets around \vinf: derived from
scaling via \vesc. Asterisks following \vmic: value obtained in
parallel with nitrogen abundance. For errors, see text.}
\end{table*}

Mok07 themselves used an automated fitting method (developed by
\citealt{Mokiem05}) based on a genetic algorithm optimization routine
to obtain the stellar/wind parameters by evolving a population of {\sc
fastwind} models over a course of generations, until the best fit to
H/He is found. Seven free parameters were considered to obtain the
highest `fitness': effective temperature,
\Teff, surface gravity, \logg, helium content, \YHe, projected
rotational velocity, \vsini\ (see above), micro-turbulent velocity,
\vmic, mass-loss rate, \mdot, and velocity-field exponent, $\beta$.

\subsection{Model calculations and grids}
\label{grid-mod} 

All models used within this analysis are calculated with {\sc
fastwind}, augmented by few {\sc cmfgen} models for comparison
purposes (Sect.~\ref{comments}). For these calculations, H, He and N
are treated as explicit elements. A description of our H/He model
atoms can be found in \citet{puls05}, and our nitrogen model atom has
been described in Sect.~\ref{atom nitro} and in Paper~I.

To allow us for studying the combined reaction of all diagnostic
H/He/N lines on variations of the stellar/wind parameters and nitrogen
abundances, and also for understanding the \NIV\nivem\ emission line
process (Sect~\ref{niv_em}), we generated a grid of
models.\footnote{following the basic philosophy described by
\cite{puls05}.} The grid is constructed using various nitrogen
abundances centered at the solar value [N] = 7.78 
(from [N] = 6.98 to 8.58 with step size 0.2 dex and including the LMC nitrogen baseline abundance,
[N]$_{\rm baseline}$ = 6.9), and a background metallicity, $ Z = 0.5
\Zsun$, corresponding roughly to the {\it global} metallic abundance
of the LMC (cf. \citealt{mokiem07b}). 
%{\bf Though the LMC baseline
%nitrogen abundance extends to 6.9 dex according to \citet{hunter07},
%for constructing our grid we opted to discard this abundance since for
%such a low value, nitrogen lines are barely detectable.}
The individual abundances of the background elements (in terms of mass
fractions) are scaled by the same factor with respect to the solar
abundance pattern (see \citealt{massey04}).

For a given $Z$, the grid is three-dimensional with respect to
\Teff, \logg, and $\log Q$, where $Q$ is the so-called wind-strength
parameter (or optical depth invariant), $Q~=~\mdot/(\vinf\Rstar)^{1.5}$. This
parameter allows us to condense the
dependence on \mdot, terminal velocity, \vinf, and stellar radius,
\Rstar, into one representative quantity.\footnote{relying on the fact
that the wind-emission from recombination dominated (i.e., $\rho^2$-)
processes remains unaffected as long as the wind-strength parameter
does not vary (see \citealt{puls96, puls05}).} The grids cover the
temperature range from 25 to 55~kK (with increments of 1~kK), and a
gravity range between 3.0 and 4.5 (with increments of 0.2 dex). For
$\log Q$, the different wind strengths are denoted by a letter (from
`A' to `E'), with $\log Q$ = $-$14.0, $-$13.5, $-$13.15, $-$12.8, $-$12.45,
respectively, if \mdot\ is calculated in \msunyr, \vinf\ in \kms, and
\Rstar\ in \rsun. Models with quantifier `A' correspond to thin winds,
resulting in lines that are (almost) unaffected by the wind, whereas
`E'-models correspond to a significant wind-strength typical for
O-type supergiants. Other parameters have been adopted as follows: a
solar helium abundance, \YHe\ = 0.10; \vinf\ as a function of the
photospheric escape velocity, \vesc\ (see \citealt{KP00});
the stellar radius, \Rstar, as a function of spectral type and
luminosity class, corresponding to prototypical values; the velocity
field exponent, $\beta$, from empirical values (\citealt{KP00}), with
$\beta = 0.9$ for O-stars, and higher values towards later types; and
the micro-turbulence, $\vmic$ = 10~\kms.

\subsection{Determination of stellar and wind parameters}
\label{stellar_wind_param}

The different steps performed in our analysis can be summarized as
follows, and are detailed in the next sections.

First, we determine \vsini\ for each object. Then we use the results
from Mok07 in combination with our model grid to roughly constrain the
stellar (\Teff, \logg, [N]) and wind-strength parameters ($\log
Q$), by inspecting the synthetic and observed H/He and nitrogen line
profiles (\YHe, $\beta$ and $\vmic$ already specified within the grid.)
During this step, we determine the extra line-broadening
parameter, \vmac, by reproducing the profile shape of the weaker
lines.

Subsequently, the stellar/wind parameters, now including \YHe\ and
$\beta$, are fine-tuned by calculating a grid of much higher resolution
around the initial guess and adopting \vinf\ from Mok07. After the
fundamental parameters have been fixed and, in case, \vmac\ has been
adjusted, we fit the nitrogen abundance, using two different methods
(for cooler and hotter objects, respectively), and also update \vmic,
which Mok07 solely derived from H/He lines. In certain cases, we need
to re-adapt the stellar/wind parameters to obtain the (almost) final
solution. Now, we are able to calculate the stellar radius from $M_V$
and the synthetic fluxes, and we can update the mass-loss rate to its
final value, by scaling with the new radius. A final consistency check
with the new \mdot\ and \Rstar\ values is performed to ensure the
stability of our results. 

Tables~\ref{tab_param} and \ref{tab_abun} list all quantities derived
in this way, and Table~\ref{tab_diff_mok} yields the main differences 
between our and the Mok07 results. Note that these quantities
refer to unclumped winds, whilst in Sect.~\ref{clump} we discuss the impact of
wind clumping. 

\paragraph{Projected rotational velocities and macro-turbulence.}
Before we are able to perform the actual (fine-)analysis, we need to
constrain the line broadening parameters, i.e., \vsini\ and \vmac\
(\vmic\ will be - when possible - inferred in parallel with the
nitrogen abundances, see Sect~\ref{nitro_ab}). As outlined above,
Mok07 derived \vsini\ directly from their automated fitting method,
from the H/He lines. More suitable is to use metal lines, since these
are not affected by Stark-broadening. Only in case of high rotational
velocities or high temperatures, where metallic lines are blended or
are very weak, \HeI\ lines might be used. Thus, we derived \vsini\
from scratch, employing the Fourier method \citep{gray76}, as
implemented and tested in the OB-star range by \citet{simon-diaz06}
and \citet{simon-diaz07}. This method has the advantage to easily
discriminate the rotational contribution from other broadening
mechanisms that affect the line shapes. In dependence of temperature
and rotational velocity of the star, we used lines from \OII, \NII,
\CII, and \SiIII\ for B- and late O-type stars. For earlier O-types,
higher ionization states are predominant, and mostly \NIII/\NIV\ and
\SiIV\ lines have been considered, together with \HeI\ lines for the
earliest types.

To finally reproduce the actual profile shape, in most cases some
extra line-broadening is needed, conventionally called
macro-turbulence. Though the physical origin of this broadening
remains still to be proven, there are some strong indications that it
is associated with (high order, non-radial) stellar pulsations 
(\citealt{aerts09,simon-diaz10}). To account for this effect, we used
a radial-tangential description of \vmac\ to fit the profile {\it
shapes} of nitrogen (and partly helium) lines, using our first
estimates on the stellar and wind parameters (see above) and the new
\vsini\ values. For the fastest rotators of our sample, however,
corresponding values could not been constrained, since large \vsini\
produce either too weak nitrogen lines, or these lines, together with
\HeI\ lines, loose their sensitivity to distinct changes in \vmac.

As expected, most of our \vsini\ values turn out to be systematically
lower than those provided by Mok07 (Table~\ref{tab_diff_mok}), by
typically 30-40\%. The derived range of \vmac\ values is consistent
with results from similar investigations, e.g.,~\citet{dufton06,
simon-diaz06, lefever07, MP08, simon-diaz10}.  The uncertainty of our
estimates is typically on the order of $\pm10~\kms$, being larger for stars
with relatively low rotational speeds.

\paragraph{Effective temperatures.} This parameter is mostly
constrained by the \HeI/\HeII\ ionization equilibrium. For this
purpose, we primarily use \HeI\ $\lambda\lambda$~4471, 4713, 4387 and
\HeII\ $\lambda\lambda$~4200, 4541. In most cases, we did not meet the
so-called \HeI\ singlet problem \citep{najarro06b}. As a consistency
check on \Teff\ and especially for the hotter stars, where the \HeI\
lines can no longer serve as an efficient temperature indicator, we
make additional use of the nitrogen ionization equilibrium, by means
of the lines listed in Table~\ref{tab_lines_nitro}. For B- and late
O-stars, we investigate \NII/\NIII, for mid O-stars \NIII/\NIV, and
for early O-stars \NIV/\NV\ or even - in a few cases -\NIII/\NIV/\NV\
(see Table~\ref{tab_param} for the specific diagnostics applied to
a particular object). To this
end, we either use our coarse grid or our specific `fine-grid' models,
with a similar gridding of nitrogen abundances ([N] = 6.9
{\ldots} 8.58). By exploiting this additional information, i.e.,
roughly `fitting' the nitrogen lines of different ions at a unique
abundance, we are able to fine-tune \Teff\ (and also some of the other
parameters, see below). For objects where only lines from
one ionization stage are present (N11-033, 045, 061, 087, 123), such a
concistency check only allows for rather weak constraints, if at all.

We estimate a typical uncertainty for \Teff\ according to the grid
resolution, $\Delta\Teff \approx$ 1~kK. For N11-066 and N11-068 we
can only provide rough estimates on the stellar parameters,
consistent with the nitrogen ionization equilibrium, and we adopt a
larger error, $\Delta\Teff \approx$ 2~kK.  For some problematic stars,
N11-026, N11-031, and N11-060, we consider an even larger error, about 
4~kK, in agreement with Mok07 (see Sect.~\ref{comments}). 

\paragraph{Surface gravities.} We derived \logg\ using the classical approach,
from the Stark-broadened wings of the Balmer lines, basically \Hg\ and
\Hd, which should be uncontaminated by wind-emission. As for \Teff, we
use nitrogen lines as a final consistency check.
These surface gravities need to be corrected for stellar rotation,
applying a centrifugal correction (see \citealt{Repolust04} and
references therein). The estimated error for \logg\ is 0.1 dex.

\paragraph{Helium abundances.} To ensure the reliability of the final
parameters, especially \Teff\ and and \logg, and for our discussion on
the abundance enrichment, we need to revisit the helium line fits,
since inconsistent helium abundances can influence these parameters. A
small sub-grid was constructed around the stellar parameters derived
in previous steps, for different \YHe, from 0.08 (corresponding to the
approximate LMC baseline abundance, see Sect.~\ref{results}) to 0.14, in
steps of 0.02. A rough estimate on the error is half this stepsize.

\begin{table}
 \centering
\caption{Differences between the fundamental parameters as derived
here and by Mok07.}
\label{tab_diff_mok}
\tabcolsep1.2mm
\begin{tabular}{lrrrrrr}
\hline 
\hline
\multicolumn{1}{l}{Star}
&\multicolumn{1}{c}{$\Delta\Teff$}
&\multicolumn{1}{c}{$\Delta\logg$}
&\multicolumn{1}{c}{$\Delta Y_{\rm He}$}
&\multicolumn{1}{c}{$\Delta\vsini$}
&\multicolumn{1}{c}{$\Delta\log\mdot$}
&\multicolumn{1}{c}{$\Delta\beta$}\\
\multicolumn{1}{l}{}
&\multicolumn{1}{c}{(kK)}
&\multicolumn{1}{c}{(cgs)}
&\multicolumn{1}{c}{}
&\multicolumn{1}{l}{(\kms)}
&\multicolumn{1}{l}{(\mdu)}
&\multicolumn{1}{c}{}\\
% Star & $\Delta\Teff$  & $\Delta\logg$ & $\Delta Y_{\rm He}$
% & $\Delta\vsini$  & $\Delta\log\mdot$ &
% $\Delta\beta$ \\
\hline
N11-026 & $-$4.3  &    -   & $-$0.01  & $-$37 & $-$0.06 &-\\
N11-031 &  2.8    & 0.10   &  0.01    & $-$45 & $-$0.28 & 0.19\\
N11-038 & $-$0.5  &$-$0.02 & $-$0.02  & $-$45 & $-$0.10 &-\\
Sk--66$^{\circ}$ 100 &    -    &    -   &     -    & $-$25 & $-$0.03 &-\\
N11-032 &  0.8    & 0.05   &     -    & $-$36 &  0.08   &$-$0.23\\
N11-045 &    -    &    -   &     -    & $-$41 &  0.10   &-\\
BI253   &  1.0    &    -   & $-$0.01  & 39    & $-$0.10 &-\\
BI237   &    -    &    -   & $-$0.01  & 14    & $-$0.10 &-\\
N11-060 &  2.3    & 0.05   &     -    &$-$38  & $-$0.01 &-\\
Sk--70$^{\circ}$ 69  & $-$0.9  & 0.06   & $-$0.03  &  -    & $-$0.38 & 0.02\\
N11-051 & $-$1.0  &$-$0.05 &     -    & 17    & $-$0.39 & 0.20\\
N11-058 & $-$0.5  &$-$0.14 &     -    &$-$23  & $-$1.18 &$-$0.42\\
Sk--66$^{\circ}$ 18  & $-$0.5  &    -   &     -    & $-$7  &     -   &-\\
N11-065 & $-$0.7  &$-$0.04 & $-$0.04  &$-$23  & $-$0.86 & 0.20\\
N11-066 & $-$2.3  &$-$0.17 & $-$0.01  &$-$12  & $-$0.47 &-\\
N11-068 & $-$2.9  &$-$0.43 &     -    &$-$24  & $-$0.42 &-\\
N11-061 &  0.4    & 0.04   &     -    &$-$33  &  0.39   &$-$1.00\\
N11-123 & $-$0.5  &$-$0.02 &     -    &  5    &     -   &-\\
N11-087 &    -    &    -   &     -    &$-$16  & $-$0.10 &-\\
\hline
N11-029 &$-$0.4   &$-$0.03 &  0.01    &$-$31  &  0.22   &$-$0.40\\
N11-036 &$-$0.5   &$-$0.20 &     -    &$-$15  &     -   &-\\
N11-008 & 0.3     & 0.02   &     -    &$-$35  &  0.09   &-0.57\\
N11-042 &$-$1.0   &$-$0.10 & $-$0.02  &$-$21  & $-$0.02 &-\\
N11-033 &$-$0.5   &$-$0.01 &  0.02    &   -   &  0.02   &-\\
N11-072 &$-$1.0   &$-$0.08 & $-$0.02  &   -   &  0.02   & 0.46\\
\hline
\end{tabular}
\tablefoot{Positive values indicate larger values from this study.
Dashes: no difference compared to the Mok07 analysis. For
N11-031, the comparison is made with respect to the cooler solution
(see Sect.~\ref{comments}).}
\end{table}

\paragraph{Terminal velocities} cannot be reliably derived from the
optical, and have been adopted from Mok07. For the field stars, values
have been inferred from UV P Cygni profiles, by \citet{massa03} and
\citet{massey05}. For Sk--66$^{\circ}$ 18, it was measured by Mok07
using UV \OVI\ lines.
%Crowther02 for SK67166, Massa03 SK7069 and SK66100,
%Massey2005 BI237 and BI253
For the FLAMES N11 stars, only N11-031 could be analyzed with respect
to this parameter, by \citet{walborn04}. For all other stars, \vinf\
has been estimated from \vesc, following \citet{KP00}.

\paragraph{Velocity field exponent $\beta$.} 
The sample used within this study does not contain any star with such
a dense wind that \Ha\ is in emission, therefore an accurate
determination of this parameter is difficult\footnote{if not
impossible, due to the \mdot - $\beta$ degeneracy (e.g.,
\citealt{markova04}), and effects from wind clumping (e.g.,
\citealt{Puls06}).} for optical spectroscopy. We applied the following
philosophy. If the combination \mdot-$\beta$ provided by Mok07
resulted in reasonable \Ha-fits, we kept $\beta$. Otherwise, we set
$\beta$ to prototypical values, $\beta~=~0.8 \ldots 1.30$, in
dependence of spectral type and results from earlier analyses
performed in our group. Moreover, for some of the sample stars,
N11-008, N-029, N11-058, and N11-061, the automated fitting method
used by Mok07 resulted in quite large values for $\beta$ (e.g.,
N11-061: $\beta$=1.8), whilst for N11-051 a rather low value,
$\beta=0.6$, was inferred. We consider such values as either
unphysical or indicating a substantial amount of wind clumping. In all 
these cases, we modified $\beta$ as outlined above (see also
Sect.~\ref{comments}).

\paragraph{Mass-loss rates} are derived from
fitting the synthetic \Ha\ profiles to the observations, given
$\beta$ (see above). Usually \HeII$\lambda4686$ is used as a
consistency check. It turned out that for many 
sample stars we were not able to successfully fit both
lines at the same \mdot, since \HeII$\lambda4686$ showed more
absorption than consistent with the observations when \Ha\ was
fitting. This might indicate a certain problem regarding
\HeII$\lambda4686$ in the new {\sc fastwind} version, or some impact
of wind clumping. The problem needs to be investigated in future, but
has no impact within the present study.

Another consistency check for \mdot\ is provided by the nitrogen
lines, particularly by the \NIII\ and \NIV\ emission lines (and
sometimes also by the \NV\ doublet), which are strongly affected by
the wind strength. In the case of two stars, N11-058 and N11-065,
which showed quite good line fits to H, He and N, the synthetic
\NIV\nivem\ profile displayed weak emission, whereas the observed one
was clearly in weak absorption. Consistency could be achieved by
lowering \mdot\ until this line could be acceptably fitted, leaving
the remaining nitrogen lines and \Ha\ almost unaltered. In both cases
it turned out that \Ha\ was already almost insensitive to reductions
in \mdot. Other stars which showed a similar problem, N11-051, Sk--66$^{\circ}$
18, and Sk--66$^{\circ}$ 100, could not be `cured' by this approach,
since in this case a reduced value of \mdot\ was no longer consistent
with (unclumped!, see Sect.~\ref{clump}) \Ha.

Since in all of our objects \Ha\ is in absorption, leading to
the well known \mdot-$\beta$ degeneracy, we estimate quite a large
error on \mdot, namely plus/minus a factor of two, which is typical in
this situation (e.g., \citealt{markova04}). The impact of the error in
\Rstar\ is negligible here, as outlined in the next
paragraph.

\paragraph{Stellar radii.} Since the effective temperatures derived
within this work are different from those of Mok07 (overall, these
differences are modest, except for N11-026, N11-031, and N11-068, see
Table~\ref{tab_diff_mok}), this leads to different theoretical fluxes
and thus to different stellar radii. Similar to Mok07, we followed
\citet{kudritzki80} and \citet{herrero92},
%
%\beq
%5 \log \Rstar = 29.58 + \left(\displaystyle\ V_{theo} - M_V\right),
%\label{eq_rstar}
%\eeq
%\beq
%V_{theo} = - 2.5 \log \int_{filter} 4 H_{\lambda} S_{\lambda} d\lambda,
%\label{eq_rstar}
%\eeq
%
and calculated the `new' radii from the theoretical Eddington fluxes
and the (de-reddened) absolute magnitudes from Table~\ref{tab_sample}.

Once the radii have been re-determined, 
\mdot\ needs to be modified as well, to preserve the fit quality of
\Ha, which depends on the optical depth invariant $Q$ (see above).
Contrasted to the case of Galactic objects, where the error of $M_V$
(due to unknown distances) dominates the error budget of \mdot, this
plays a secondary role in our sample, due to sufficiently well-known
distances and the rather large error introduced by the \mdot-$\beta$
degeneracy.

\subsection{Nitrogen abundances and micro-turbulences}
\label{nitro_ab} 

\begin{table}
\caption{Helium and nitrogen abundances for the LMC sample, with 
stellar parameters from Table~\ref{tab_param}.}
%\vspace{0.3cm}
\label{tab_abun}
\tabcolsep1.5mm
\begin{tabular}{lrclccc}
\hline 
\hline
Star & \vsini & \YHe & [N] & $ \Delta$[N] & $ \Delta$[N]$^{\rm cl}$&
Literature
\\
\hline
N11-026  & 72 &0.10 &7.80\tablefootmark{1} &$+$0.40     &0.25-0.30 &-\\
         &    &     &7.75\tablefootmark{2} &$+$0.40     &          & \\       
N11-031  & 71 &0.11 &7.83$^1$ &$\pm$0.15 &-         &8.00$\pm0.18$\\
         &    &     &8.30\tablefootmark{2} &$\pm$0.15   &          & \\           
             
N11-038  &100 &0.08 &7.85 &$\pm$0.15 &0.05-0.10 &-\\
Sk--66$^{\circ}$ 100  & 59 &0.19 &8.48\tablefootmark{3} &$\pm$0.15 &-         &-\\
N11-032  & 60 &0.09 &7.87 &$\pm$0.15 &0.05-0.10 &-\\
N11-045  & 64 &0.07 &6.98 &$\pm$0.20 &-         &-\\
BI253    &230 &0.08 &7.90 &$\pm$0.15 &0.15-0.20 &-\\
BI237    &140 &0.09 &7.38 &$\pm$0.15 &0.05-0.10 &-\\
N11-060  & 68 &0.12 &8.20\tablefootmark{1} &+0.30     &0.05-0.10 &-\\
         &    &     &8.15\tablefootmark{2} &+0.30     &          & \\           
             
Sk--70$^{\circ}$ 69   &131 &0.14 &8.05\tablefootmark{3} &$\pm$0.15 &0.15-0.20 &-\\
N11-051  &350 &0.08 &7.58\tablefootmark{3} &$\pm$0.20 &-         &-\\
N11-058  & 62 &0.10 &\emph{8.09} &\emph{$\pm$0.15} &- &-\\
Sk--66$^{\circ}$ 18   & 75 &0.14 &8.48\tablefootmark{3} &$\pm$0.15 &-         &-\\
N11-065  & 60 &0.13 &8.17\tablefootmark{3} &$\pm$0.15 &0.05-0.10 &-\\
N11-066  & 59 &0.10 &8.17 &$\pm$0.20 &0.05-0.10 &-\\
N11-068  & 30 &0.10 &7.85 &$\pm$0.20 &0.05-0.10 &-\\
N11-061  & 54 &0.09 &7.18 &$\pm$0.15 &-         &-\\
N11-123  &115 &0.09 &7.00 &$\pm$0.15 &-         &-\\
N11-087  &260 &0.10 &7.38 &$-$0.20\tablefootmark{4} &-         &-\\
\hline
N11-029  & 46 &0.08 &7.43 &$\pm$0.15 &-         &7.10$\pm0.35$\\
N11-036  & 39 &0.08 &\emph{7.85} &\emph{$\pm$0.17} &- &7.76$\pm0.11$\\
N11-008  & 46 &0.10 &\emph{8.08} &\emph{$\pm$0.11} &- &7.84$\pm0.11$\\
N11-042  & 21 &0.08 &7.00 &$\pm$0.15 &-         &6.92$\pm0.24$\\
N11-033  &256 &0.10 &7.28 &$-$0.20\tablefootmark{4} &-            &-\\
N11-072  & 14 &0.10 &\emph{7.68} &\emph{$\pm$0.15} &- &7.38$\pm0.06$\\
\hline
\end{tabular}
\newline
\tablefoottext{1}{From \NIII/\NIV, uncertainty from \NV.}\\
\tablefoottext{2}{From \NIV/\NV, uncertainty from \NIII.}\\
\tablefoottext{3}{Improvement of \NIV\nivem\ fit with weak clumping.}\\
\tablefoottext{4}{Uncertainty of upper limit.}
\tablefoot{[N] is our best-fitting
value, $\Delta$[N] the corresponding uncertainty, and
$\Delta$[N]$^{\rm cl}$ the change in [N] (always positive) if
clumping with $f_{\infty} = 0.1$ is included in the models
(Sect.~\ref{clump}). 
%Note that this uncertainty would mostly cancel with
%the corresponding change in the derived [N] if a background abundance of
%$z = 0.4$ would have been used (Sect.~\ref{zeta}).
Values displayed in italics have been derived by the `curve-of-growth'
method. Literature values are from \citet{walborn04} for N11-031 and
from \citet{hunter09} for the B-stars and N11-029. When two entries
are provided, the first one corresponds to our preferred solution, 
except for N11-031 where we consider both solutions as possible.}
\end{table}

After having determined the stellar and wind
parameters and their uncertainties (some fine-tuning might still be
necessary), we are now in a position to derive the nitrogen abundances
and the corresponding micro-turbulent velocities, \vmic. 
Because the latter parameter significantly affects the
strength of both He and metal lines and thus the implied abundances,
it is useful to determine both quantities in parallel. 
%Moreover, \vmic\
%might be interpreted as an indicator for sub-surface convection
%\citep{Cantiello09}, and thus needs special attention.

To carry out this analysis, we calculate a fine grid of typically 25
models, by combining different abundances centered at the rough
estimates derived in Sect.~\ref{stellar_wind_param}, with five different
values for \vmic\ = 0,5,10,15,20 \kms.

When possible (see below), we use a `curve of growth' method based on
the equivalent widths of the lines, which has been applied in the past
years to different sets of B-stars for obtaining various metallic 
abundances (e.g., \citealt{urbaneja04, simon-diaz06, MP08}).
%
%For each of the considered lines and for all values of \vmic, we
%construct a `curve of growth' by plotting the theoretical equivalent
%width as a function of abundance. By means of the {\it observed} equivalent
%widths (including appropriate errors) then, we are able to derive
%individual abundances/uncertainties, per line and \vmic. Subsequently
%and again for each micro-turbulence, we create a diagram which displays
%the derived abundances as a function of equivalent width, combining 
%the results for all considered lines.
%These diagrams might be also used as a tool for identifying
%`problematic' lines, \ie, lines which deviate considerably from the
%general trend, as well as for checking the ionization equilibrium if
%lines from different ionization states are present.
%
%The final pair of abundance and \vmic\ (incl. errors) is found from
%the condition that all lines should display the same abundance, i.e.,
%by interpolating w.r.t. \vmic\ in such a way as to obtain a zero
%slope in the equivalent width-abundance diagram. 
In brief, this method uses synthetic and observed equivalent widths
including uncertainties from {\it all} considered lines, to derive a unique
pair of abundance and \vmic\ (incl. errors). 
%from the condition that all lines should display the same abundance.
Results from such analyses are indicated by an asterisk in the
\vmic-column of Table~\ref{tab_param}.

Unfortunately, this procedure was not applicable to the bulk of the 
sample stars because of various reasons, \eg\ blending and diluted 
lines due to fast rotation, almost invisible lines due to low nitrogen
content, and peculiarities in the observed lines from some of the
stars with highly ionized nitrogen. All these problems will be
commented on in Sect.~\ref{comments}, and we opted for a determination
of the nitrogen abundance/micro-turbulence pair by means of a visual
inspection of the fit quality, using the same fine grid as described
above. 

We estimate the errors in \vmic\ as 3-5 \kms, both for the equivalent
width and the visual method. To provide an impression on the impact of
such errors on the derived nitrogen abundances, we note that a
decrease in \vmic\ by 5~\kms\ leads to an increase of [N] by 0.05-0.07
dex. For the error associated to [N] when derived by `visual' fitting,
we decided to be quite conservative. Even though we are able to obtain
quite good fits for the bulk of the stars implying an uncertainty of
0.1 dex, we rather adopt a larger value of 0.15 dex to roughly account
for the additional dependence on the stellar and wind parameters. For
two stars, N11-033 and N11-087, we can only provide an upper limit on
the abundance, with an estimated uncertainty of 0.20 dex. This value
is also adopted for N11-066, N11-068, and N11-045, N11-051. For the first two
stars only rough estimates on their stellar parameters could be
obtained, and for the latter two only one nitrogen multiplet could be
used. Even larger uncertainties have been
derived for two `problematic' stars, N11-026 and N11-060, see
Sect.~\ref{comments}.

Table~\ref{tab_abun} lists the obtained nitrogen abundances together
with their estimated errors. When available, corresponding literature
values have been added.

\subsection{Additional considerations}
\subsubsection{Wind clumping}
\label{clump} 
So far, we neglected wind clumping in our analysis. Since most of the 
nitrogen lines are formed in the intermediate or outer
photosphere,\footnote{Note that this might no longer be true for the
\NV\ doublet, e.g., N11-031.} at least for not too extreme winds as
considered here, they should remain rather unaffected by {\it direct}
clumping effects, though indirect effects could be important (see
Paper~I). Clumped winds have lower mass-loss rates compared to their
unclumped counterparts, by a factor of $1/\sqrt\fcl$ if the clumping
factor \fcl\ were radially constant, which could influence both the \NIII\
and the \NIV\ emission lines, due to their sensitivity on \mdot.
Given the multitude of evidence for wind-clumping 
(e.g., \citealt{puls08} and references therein), it is necessary
to examine the impact of clumping on our abundance determinations.

We adopt the parametrization as used by \citet{hilliermiller99} and
\citet{hillier03},
\beq
f(r) = f_{\infty}+(1-f_{\infty})\exp\bigl(-v(r)/v_{\rm cl}\bigr)
\eeq
where $f(r)$ is the volume filling factor,\footnote{which is, within
the standard assumption of micro-clumping and a void inter-clump
medium, the inverse of the clumping factor \fcl.} $f_{\infty}$ its
asymptotic value (if $\vinf \gg v_{\rm cl}$) and $v_{\rm cl}$ the
velocity where the volume filling factor reaches values close to
e$^{-1}$, if $f_{\infty} \ll 1$.  To allow for maximum effects, we
set $v_{\rm cl} = 30~\kms$, close to the sonic speed for O-stars,
%We
%considered models with $f_{\infty}$ = 0.5, 0.25 and 0.1, corresponding
%to \mdot\ reductions by factors between roughly 0.7-0.3 (consistent with
%recent investigations allowing for macro-clumping, see
%\citealt{sundqvist11}), which preserve the fit quality for \Ha. 
%
and concentrated on models with $f_{\infty}$ = 0.1, 
corresponding to \mdot\ reductions by a factor of roughly 0.3
(consistent with recent investigations allowing for macro-clumping, see
\citealt{sundqvist11}).

Since most of our sample stars display thin winds, no major reaction
due to clumping is to be expected.
%Overall, low clumping factors were derived ($f_{cl}\sim~2$) for the bulk of the
%stars, only for Sk-66100 a larger value was prefered ($f_{cl}\sim~5$).
%This is associated to small changes on \mdot\ and therefore on the nitrogen
%profiles, marginal in most cases. Only for a few stars, (N11-051, N11-065,
%Sk66100, SK-6618 and SK-7069), the inclusion of clumping favours to a better
%fit to \NIV\nivem, see Sect~\ref{comments}.
Indeed, a value of $f_{\infty}$~=~0.1 did not induce any noticeable
change in the spectrum for the bulk of the stars. 

For stars N11-038, 032, BI237, N11-060, 065, 066, and 068,
mostly the \NIII\ triplet is affected (for BI237, also \NV),
requiring 0.05 to 0.1 dex more nitrogen to preserve the {\it previous} fit
quality of these lines. Since, on the other hand, the other nitrogen
lines remain unmodified, the inclusion of clumping did not change our
[N]-values for these stars, but only improves or deteriorates the
particular representation of the triplet lines.

For BI253 and Sk--70$^{\circ}$ 69, clumping of the considered amount has a
larger effect, particularly on the \NIII\ triplet, \NIV\nivem\ and the \NV\
doublet. Here, a clumping factor of $f_{\infty}$ = 0.1 requires that
[N] needs to be increased by 0.15{\ldots}0.20 dex.

Finally, for N11-026 and N11-031, the inclusion of clumping affects
the stellar parameters as well. Due to the lower
\mdot, a hotter temperature (by 1 to 2 kK) is needed to preserve the
\HeI\ fit. In case of N11-026 then, [N] needs to be considerably
increased, by 0.25 to 0.3 dex. This is the only case where we
encountered a significant effect. For N11-031, on the other hand, at
least the `cool solution' (see Sect.~\ref{comments}) remained at the
previous nitrogen abundance.

Besides these general effects, for a few stars, Sk--66$^{\circ}$ 100,
Sk--70$^{\circ}$ 69, N11-051, Sk--66$^{\circ}$ 18 and N11-065, the inclusion of
clumping (of a lesser degree than $f_\infty$ = 0.1) favours a better fit to
\NIV\nivem, see Sect~\ref{comments}.

\subsubsection{Background abundances}
\label{zeta} 
One of the central results from Paper~I was that the \NIII\ triplet
emission increases with decreasing background metallicity, $Z$, due to
reduced line-blocking. In this investigation we assume, following
\citet{mokiem07b}, a {\it global} $Z$ of the LMC, $Z =
0.5~\Zsun$. Since this value is somewhat controversial, and a
$Z = 0.4~\Zsun$ might be appropriate as well
(e.g., \citealt{Dufour84}), we need to test the impact of this
difference on the derived nitrogen abundances.

Overall, such lower background $Z$ does not produce any
extreme changes. As expected, the triplet emission increases,
requiring a roughly 0.05 dex lower abundance to recover the previous
fits. Interestingly, we also found that \NIV\nivem\ and the \NIII\
quartet lines tend to more emission and weaker absorption, respectively, but to a
lesser extent. The effect on the remaining nitrogen lines is marginal.
Note that a lower [N] value due to lower background abundances would partly
cancel with the corresponding increase in the derived [N] due to 
moderately clumped winds.

\section{Comments on the individual objects}
\label{comments}
In the following, we give specific comments on the individual objects,
regarding peculiarities and problems found during our analysis. We
separate between B-/late O-stars and (hotter) O-stars, and sort by
luminosity class and spectral type, starting at the hotter side. All
nitrogen line fits (including corresponding limits according to
Table~\ref{tab_abun}) are displayed in
Appendix.~\ref{fits_ap}, except for the objects N11-072, N11-032, and
BI237 which have been included in the main paper, since they are
exemplary for objects with different features. All spectra are
corrected for radial velocity shifts. 
%For each star, three different
%profiles are plotted together with the observed spectra (green): a
%best solution (black), an upper (red) and a lower (blue) limit for the
%derived nitrogen abundance, refering to the results from
%Table~\ref{tab_abun} assuming unclumped winds. 

We selected those lines that were clearly visible, at least in
most of the cases. In particular, for \Teff\ $\le$ 35~kK, we used
\NII$\lambda\lambda$~3995, 4447, 4601, 4607, 4621, 4630 and
\NIII$\lambda\lambda$~4003, 4097, 4195, 4379, \trip, and \qua.  For
\Teff\ $>$ 35~kK, we analyzed the lines from \NIII/\NIV/\NV: \NIII\ as
before, \NIV$\lambda\lambda$ 4058, 6380, and
\NV$\lambda\lambda4603-4619$ (see Table~\ref{tab_lines_nitro}).
Additionally, the \NIV\ multiplets around 3480~\AA\ and 7103-7129~\AA\
have been used for the field stars observed with UVES (except for
Sk--70$^{\circ}$~69 where only \NIV$\lambda$3480 is available).

\subsection{Late O- and B-supergiants/giants}
\label{bstars}
\subsubsection{Supergiants}

\paragraph{\bf N11-029 -- O9.7 Ib.} This is the only O-supergiant
within our sample. Since it is of late nature, we discuss it here
together with the B-supergiants. No major problem has been
encountered, and the largest difference with respect to Mok07 concerns
the large $\beta$ = 1.63 derived in their analysis. We opted for a lower
value, $\beta$ = 1.23, still at the limit of prototypical values. To
compensate for this modification, \mdot\ needs to be somewhat increased.

Figure~\ref{N11-029} shows the best fit for the nitrogen lines. An
abundance of [N] = 7.43$\pm0.15$ has been inferred, mainly from
\NII$\lambda3995$ and the \NIII\ quartet lines. The bulk of the \NII\
lines are not helpful since there are blended by \OII\
(Table~\ref{tab_lines_nitro}). This is also the case for most of the
\NIII\ lines.

The most interesting feature, however, is the discrepancy for
\NIII$\lambda4634$,\footnote{similar to \NIII$\lambda4640$ which is
strongly blended by \OII.} which shows an almost completely refilled
profile but is predicted to be in absorption. To synthesize a profile
with E.W. $\approx$ 0 would require a higher \Teff\ or a lower \logg,
inconsistent with the He ionization equilibrium. This seems to be the 
first observational evidence for one of the problems discussed in 
Paper~I. For a certain temperature range (around 30 to 35~kK), {\sc
fastwind} spectra predict too few emission in the \NIII\ triplet,
compared to results from {\sc cmfgen}, due to (still) missing overlap
effects between the \NIII\ and \OIII\ resonance lines around 374~\AA\ 
(which are treated consistently in {\sc cmfgen}; for details, see Paper~I).

For further tests, we calculated a {\sc cmfgen} model at the same 
stellar/wind parameters and abundances as derived in the present
analysis. As expected, the corresponding synthetic profiles are closer
to the observations (though still not as refilled as observed). To check
whether the oxygen abundance plays a significant role, two different
abundances were considered, [O] = 8.66 and 8.30 dex (solar and factor
two lower). In agreement with our theoretical argumentation from
Paper~I, such a difference did not affect the predicted \NIII\ triplet emission
strength. 

\begin{figure*}
\center
%\resizebox{\hsize}{!}
  {\includegraphics[totalheight=0.45\textheight]{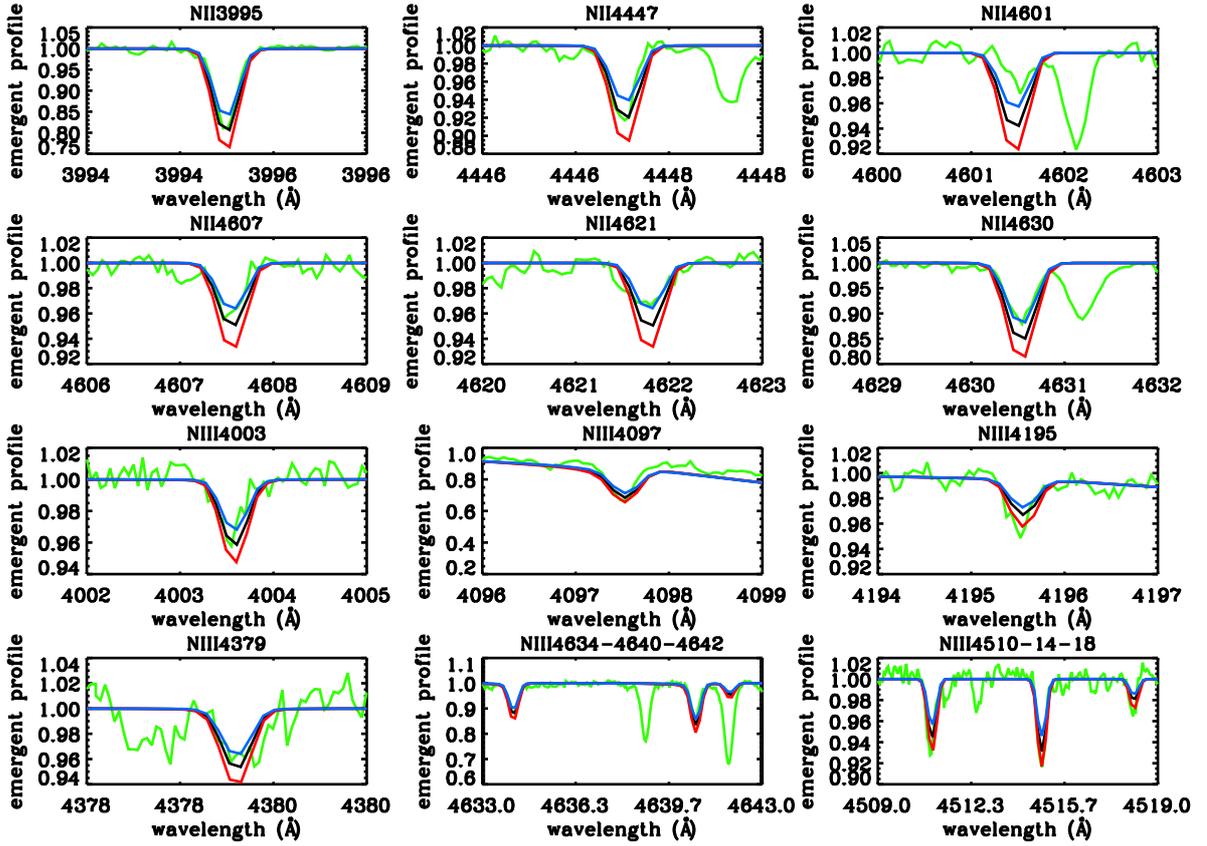}}
%\vspace{-0.5cm}
\caption{N11-072 - B0.2 III. Observed (green) and best-fitting optical
nitrogen spectrum (black). Blue and red spectra correspond to
synthetic line profiles with [N] at the lower and upper limit,
respectively.  For details, see Sect.~\ref{comments}.}
\label{N11-072}
\end{figure*} 

\paragraph{\bf N11-036 -- B0.5 Ib.} The largest difference to Mok07 is
that we derive a lower value for \logg\ (by 0.2 dex) as well as a
lower \Teff\ (by 500~K), by exploiting \HeI/\HeII\ in parallel with
the \NII/\NIII\ ionization equilibrium.

The nitrogen lines are well fitted, both for \NII\ and \NIII\
(Fig.~\ref{N11-036}). The only discrepancy found relates to an
underprediction of absorption strength in \NII$\lambda4607$, 
whereas the discrepancy at \NII$\lambda4601$ is due to an
\OII\ blend. Quite a large enrichment is found,
[N] = 7.85$\pm0.17$.
%, corresponding to $\sim$ nine times the LMC baseline
%abundance. 
%Note that the upper solution for the abundance circumvent
%the problem for \NII$\lambda4607$.

\paragraph{\bf N11-008 -- B0.7 Ia.} Again, we adapted the rather large
velocity field exponent derived by Mok07, $\beta=1.87$, to a more
typical value of $\beta=1.30$. We derive \vsini\ = 46 \kms,
which is approximately half the value as obtained by Mok07, and
compensate by invoking a \vmac\ = 60 \kms.

The best fitting abundance, [N]~=~8.08$\pm0.11$ (Fig.~\ref{N11-008}),
has been obtained by the `curve of growth method', and yields 
reasonable fits except for \NII$\lambda\lambda$4447, 4621, and
\NIII$\lambda4003$ which are slightly overpredicted.

\subsubsection{Giants}
\paragraph{\bf N11-042 -- B0 III.} We derive a lower projected
velocity than Mok07, \vsini\ = 21 \kms, as well as a lower \Teff,
together with a corresponding decrease of \logg, 
for consistency with the \NII/\NIII\ ionization equilibrium. 
For this cooler solution, the helium abundance needs to be
lowered as well, \YHe = 0.08, close to the LMC He baseline
abundance (see Sect.~\ref{results}).

Due to the low rotational speed, we are able to clearly inspect all
\NII\ and \NIII\ lines (Fig.~\ref{N11-042}) which are fitted almost 
perfectly. The only discrepancy occurs at \NIII$\lambda4097$, caused
by a coincident \OII\ line. We are able to see two strong \OII\
absorption lines at both sides of \NIII$\lambda4640$, and to the left
of \NIII$\lambda4379$ (where the former cannot be used for the diagnostics of 
similar stars with rapid rotation, N11-008, N11-029, N11-036, and
N11-045).

A low nitrogen content has been derived, [N]~=~7.00$\pm0.15$, 
consistent with the low helium abundance.

\paragraph{\bf N11-033 -- B0 IIIn.} We find a slightly cooler \Teff\ 
compared to Mok07, and our helium line fits suggest \YHe = 0.10.

Due to its large rotation, all nitrogen lines are severely diluted and
almost `vanish' from the spectrum (Fig.~\ref{N11-033}), implying an
upper limit of [N] = 7.28 from the \NII\ lines.
% Jorge: removed 7.28-0.20 as indicated by the editor

\paragraph{\bf N11-072 -- B0.2 III.} This object shows a very sharp-lined 
spectrum, with the lowest \vsini\ value within our sample. 
A consistent solution for nitrogen and helium suggests a slightly
cooler \Teff\ (by 1~kK) and lower \logg\ (by 0.1 dex). Our
best fit indicates \YHe\ =0.10.

As for the similar object N11-042, we are able to obtain a good
fit from the `curve of growth' method, with [N]~=~7.68$\pm0.15$, 
and a rather low \vmic\ = 2.7 \kms. Thus, and in contrast to N11-042,
this object is clearly enriched. Note that N11-072 and N11-042 belong
to different associations, LH-10 and LH-9, respectively.

\subsection{O-stars}
\label{ostars}

\subsubsection{Giants}
\paragraph{\bf N11-026 -- O2 III(f$^*$).} This star is one of the four
O2 stars in our sample, together with N11-031, BI237, and BI253,
comprising the hottest objects. Unlike Mok07, we favour a cooler
solution, \Teff\ = 49 kK (Mok07: 53~kK), and a somewhat lower \mdot.
This comparably large difference for \Teff\ relies both on the better
reproduction of \HeI$\lambda4471$ and on the fit to the nitrogen
lines, with three ionization stages visible (Fig.~\ref{N11-026}).
Mok07 considered such cooler solution as well (almost included in
their error bars), which would have improved their fit to
\HeI$\lambda4471$, but argued in favour of the hotter one, accounting
for the global fit quality. By considering nitrogen now, we find
support for a lower \Teff, since for \Teff\ $\geq$ 50~kK the \NIII\
lines vanish from the spectrum (cf. BI237 and BI253).

The derived nitrogen abundance results from a compromise solution,  
[N] = 7.80, where this value is also the lower limit.
Moreover, we estimate quite a large uncertainty (upper limit) of 0.4
dex, arising from our difficulties to fit all three ionization stages
in parallel.
%\footnote{Similar problems have been found for N11-031 and N11-060.} 
We favour a solution that provides a good fit for the
\NIII\ quartet and the \NIV\ lines, whereas a larger abundance, [N] =
8.20, is needed to fit the \NV\ doublet. From Fig.~~\ref{N11-026} it
is clear that such a large abundance (red) is in disagreement with the
remaining nitrogen lines.  Of course, we also tried a fit at hotter
temperatures. At \Teff\ = 52~kK (close to the result by Mok07), \logg\
= 4.1 (which is still consistent with the Balmer line wings) and
\mdot\ = 1.5\Mdu, it is possible to fit both \NIV\ and \NV, for quite
a similar abundance, [N] = 7.75. At this temperature, however, all
\NIII\ lines have vanished though. To recover them we would need to
increase the abundance again, also by 0.4 dex. Because of the poorer
prediction of \HeI$\lambda4471$ 
%and the somewhat large \logg\ value 
we opt for the cooler solution. Since in both cases the implied nitrogen
abundances are similar, this does not lead to severe problems for our
further analysis, but requires larger error bars than typical.

\begin{figure*}
\center
%\resizebox{\hsize}{!}
{\includegraphics[totalheight=0.45\textheight]{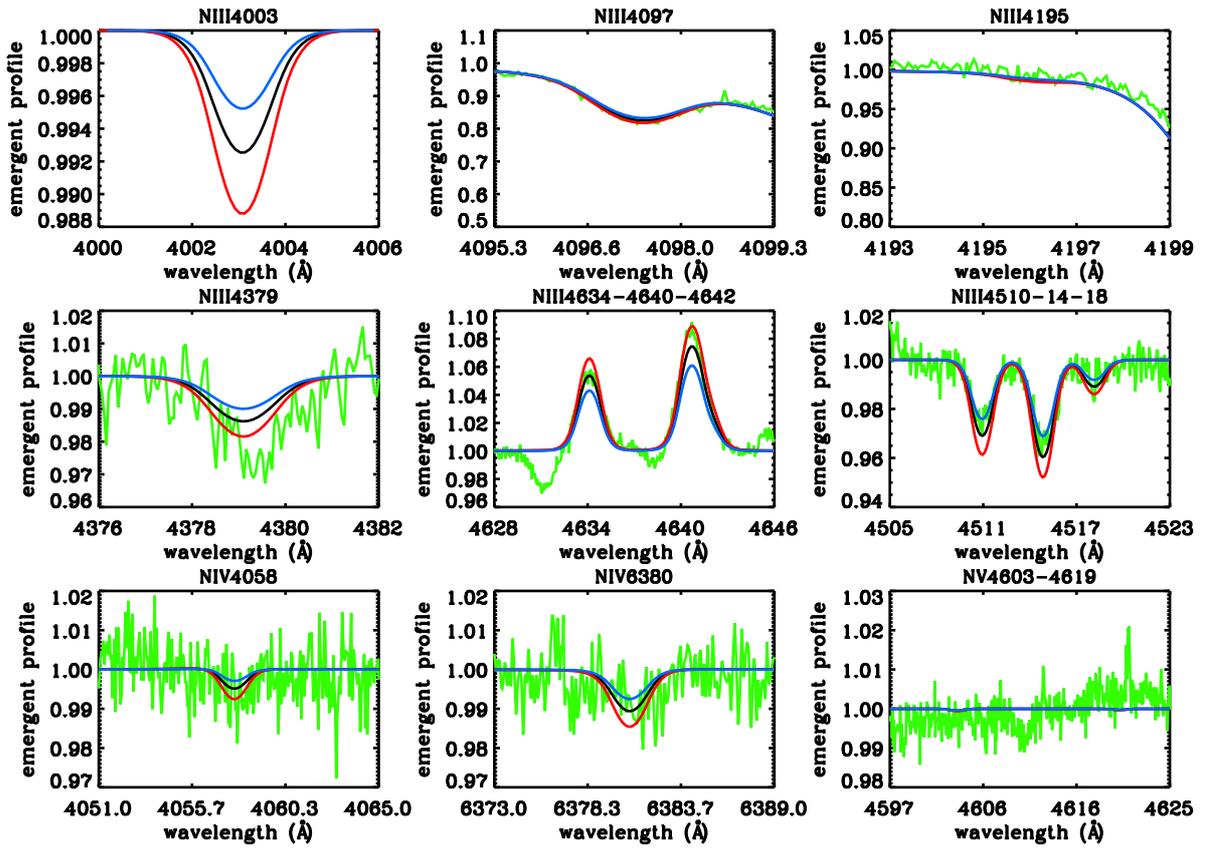}}
%\vspace{-0.5cm}
\caption{N11-032 - O7 II(f). Same color coding as Fig.~\ref{N11-072}. 
For this star, \NIII$\lambda$4003 has not been observed.}
\label{N11-032}
\end{figure*}

\begin{figure*}
\center
%\resizebox{\hsize}{!}
{\includegraphics[totalheight=0.45\textheight]{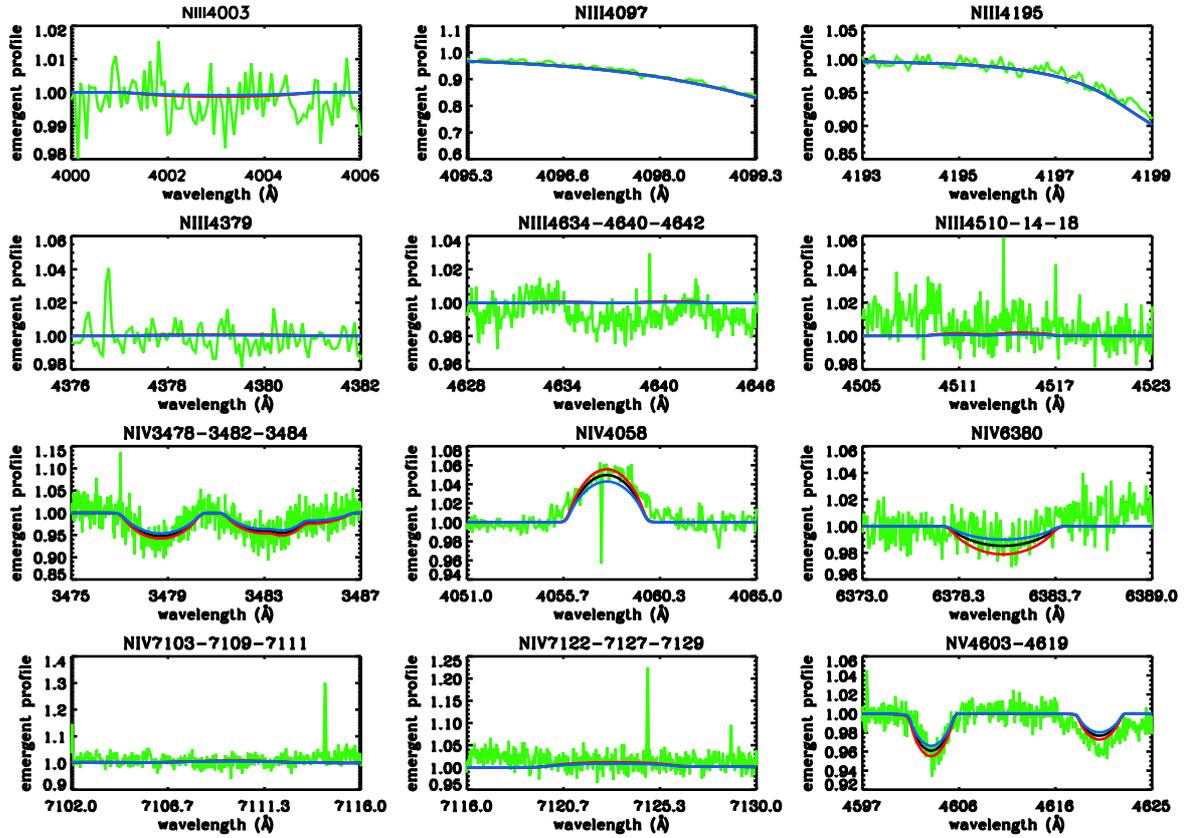}}
%\vspace{-0.5cm}
\caption{BI237 - O2 V((f$^*$)).} 

\label{BI237}
\end{figure*}

\paragraph{\bf N11-031 -- ON2 III(f$^*$).} This star raised the most
severe difficulties in our sample when trying to fit the nitrogen
lines (Fig.~\ref{N11-031}). Already its `ON' designation 
indicates strong nitrogen features in its spectra,
in this case \NIV$\lambda\lambda4058, 3480$ (for the latter, see
\citealt{walborn04}), and the \NV\ doublet lines. We were not
able to consistently fit these strong features together with the
remaining nitrogen lines. In contrast, \NIV$\lambda6380$ has almost
the same strength as in N11-026. 
%Since in the remaining objects this
%line turned out to be quite reliable, even for the hottest objects, we
%adopt a conservative solution using the \NIII/\NIV\ ionization
%equilibrium and discard the problematic lines.

If we try to reproduce the (rather weak, but clearly visible)
\HeI$\lambda$4471 in parallel with \NIII\ and \NIV$\lambda6380$,  
we obtain \Teff~=~47.8~kK,
a bit cooler than N11-026, whilst Mok07 derived
\Teff~=~45~kK, excluding \Teff\ values larger than 47~kK based
on the helium ionization analysis. 

If, on the other hand, we try to fit the problematic lines, we need a
much higher \Teff. A consistent fit for all \NIV\ lines (including
\NIV$\lambda6380$) together with those from \NV\ requires \Teff~=~56~kK,
\logg\ = 4.00, and \mdot\ = 2.2\Mdu, together with a very high
abundance, [N]~=~8.30. Clearly, this set of stellar parameters neither
reproduces the \NIII\ lines nor the weak \HeI$\lambda4471$.  N11-031
has been previously analyzed by \citet{walborn04} and~\citet{doran11}
using the \NIV/\NV\ lines (without discussion of \HeI\ and \NIII). The
former authors obtained quite similar parameters, \Teff\ = 55~kK,
\logg\ = 4.00, and a somewhat lower $\mdot\ = 1.0 \Mdu$, presumably
because of a clumped wind (though clumping has not been mentioned). At
these parameters, they derived [N]~=~8.00$\pm0.18$, which for our
models would still be too low. \citet{doran11} only provided \Teff\
within their analysis, deriving \Teff~=~54.7~kK for this object.

Because of the similarities with N11-026 (except for the two strong
features), the fact that \NIV$\lambda6380$\footnote{which turned out
to be quite `reliable' in the remaining objects, even for the hottest
ones.} behaves `normally' and that \HeI$\lambda4471$ is clearly
visible, the cooler solution with [N]~=~7.83$\pm0.15$ cannot be
discarded so far. We checked our synthetic spectra by independent
{\sc cmfgen} calculations. The results are quite similar, in
particularly the predicted \NIII\ emission lines are even a bit weaker
than those produced by {\sc fastwind}, again pointing to a cooler
solution.

We also tried to attribute the problematic feature to the presence
of X-rays, by means of {\sc cmfgen} calculations including typical
X-ray strengths and distribution, but almost no effect on these lines
has been found (basically because the line forming region is still
inside or close to the photosphere). 
%It could also be related to mass transfer on binary system.
%references on Walborn04!!!

Of course, it would be helpful to consider \NIV$\lambda$3480 as
well, which unfortunately is not included in our dataset. A by-eye
comparison with the corresponding profile displayed in \citet[their
Fig. 1]{walborn04} showed that both our cooler and hotter solutions
are not incompatible with this spectrum.\footnote{Interestingly, our
cooler solution predicts somewhat stronger absorption than the hotter
one, even though the
corresponding nitrogen abundance is significantly lower.}
%Jorge: I checked more in detail and the cooler solution presents stronger
% absorption (even with much less N abundance) profiles (peak heigth down to
% 0.86) than the hotter solution for this multiplet (peak heigth down to 0.92).
% There is  no indication of how deep the profile is in Walborn figure but it
% seems that LH10-3061 (N11-031) is stronger than LH64-16, which if you go to
% Fig 4 the absorption goes down to approx. 0.93 for this multiplet for LH64-16
% and therefore the solution for N11-031 would be between our two possible
% solutions if we consider also this line. I guess that writing "not
% incompatible" is sufficient is still true and would keep the referee happy
% because we include a rough comparison for this multiplet.

Thus, the nature of the strong \NIV$\lambda4058$ and \NV\ doublet
features remains open. Of course, binarity could be a plausible 
solution, where a cooler component could be responsible for \HeI\ and
\NIII, and a hotter one for the intense \NIV\ and \NV\ lines. 
We note that this is the brightest (\MV=$-$5.78) object of the
O-star sample and that other, previously thought single early O-stars,
displaying both strong \NV\ lines as well as the presence of \HeI\ in
their optical spectra such as CygOB2-22 and HD\,93129A, were subsequently
resolved as binaries (\citealt{walborn02b}, \citealt{nelan04}). Thus,
the binarity scenario for N11-031 needs to be clarified in future
investigations. 

Note, however, that the other ON-stars discussed by
\citealt{walborn04}, LH64-16 and NGC~346-3, seem to display very
similar features, though the presence of \HeI\ is not as clearly
visible as in our spectrum for N11-031.
Similar, but less dramatic, problems are also found
for those other sample stars where we were able to analyze
\NIII/\NIV/\NV\ in parallel, namely N11-026 and N11-060 (see below).
While the presence of such discrepancies in all these objects may
point towards a less likely binary scenario for N11-031, we note that
the differences in \Teff\ for the two alternative solutions reach
8,000~K in N11-031, and remain at moderate 3,000~K for the other two objects
analyzed here.

We will reconsider N11-031 and other ON-stars in future investigations,
to clarify the question how it is possible to have weak \HeI\ and
\NIII\ in parallel with strong \NIV\ and \NV. For the remainder of
this paper, however, we discuss the results for N11-031 in terms of both the
cool and the hot solution, without preferring either of them.

\paragraph{\bf N11-038 -- O5 II(f$^+$).} The parameter set derived for
this stars is quite similar to Mok07, with only slightly lower
\YHe~=~0.08 and \vsini~=~100 \kms. This star displays a peculiar
\HeI$\lambda4471$ profile of triangular shape that could not be
reproduced, even if invoking macro-turbulent broadening (as speculated by
Mok07).

We obtain [N]~=~7.85$\pm0.15$, similar to the case of N11-032, by
means of quite a good fit to all lines from three ionization stages.
Besides the peculiar \HeI$\lambda4471$ profile, \NIV\nivab\ shows
contamination by the DIBs at $\lambda\lambda$ 6376.08, 6379.32, and the
\NV$\lambda4603$ absorption is much stronger than predicted,
contrasted to the other component.

%Though both problems could be related to blends (which are not present
%in other stars), one might also speculate on binarity.
%The former feature could be related to an \OIII\ blend to
%the \NIV\ line. However, we do not find any evidence of this on the
%other stars of the sample and additionally at such rotation both lines
%should be blended. ~The feature associated to the \NV\ line might be
%related to a moderately strong \NIII\ line (gf~=~0.512 from NIST
%database) for high-lying levels that belong to the quartet system that
%not included in our present \NIII\ model atom. 

\paragraph{\bf Sk--66$^{\circ}$ 100 -- O6 II(f)} is one of the field stars
within our sample. The inspection of the H/He fits suggests no major
revision of the values provided by Mok07.

Figure~\ref{SK66100} shows the best fit to the nitrogen lines. An 
abundance of [N]~=~8.48$\pm0.15$ was needed to obtain a consistent
fit. This large value agrees well with the large He content found for
this object, \YHe~=~0.19. However, we also found some problems
regarding the \NIV\ fits, except for the multiplet around
3480~\AA\ where the fit is perfect. Interestingly, we were able to
`cure' these problems by invoking a clumped wind with $f_{\infty}$ =
0.2, with no significant changes in the remaining nitrogen lines. 

\paragraph{\bf N11-032 -- O7 II(f).} No major problems were found for
this star. Our solution is slightly hotter than in Mok07, and we
preferred a typical $\beta$ value for O-stars, $\beta = 0.80$.

An excellent fit is obtained for this star of the (f) category
(Fig.~\ref{N11-032}), resulting in [N]~=~7.87$\pm0.15$ 
for both the \NIII\ triplet and the quartet lines. 
The fit quality of \NIII$\lambda4097$ is remarkable as well. 
Note that for this star \NIII$\lambda4003$ has not been observed.
Since the \NIV\ lines are weak and rather noisy, we can only state that
our simulations are consistent with the observations.

\paragraph{\bf N11-045 -- O9 III.} Our analysis is in good agreement 
with Mok07. We confirm that a low He abundance (lower than the 
estimated LMC He baseline abundance) matches the observations.

The only clear N-abundance indicators are the \NIII\ quartet lines,
since most other lines are weak and the spectrum is noisy
(Fig.~\ref{N11-045}). \NIII$\lambda4097$ is also weakly visible, and
consistent with the quartet lines. Around this \Teff, the \NIII\
triplet turns from absorption to emission, and thus
this object does not belong to the `f' category. The absence of lines,
the fact that the star does not display a fast (projected) rotation,
and the very low helium content is consistent with the very low 
nitrogen abundance, [N]~=~6.98$\pm0.20$. These findings
suggest that this object is of unevolved nature.

\subsubsection{Dwarfs}

\paragraph{\bf BI253 -- O2 V((f$^*$)).} This was one of the
stars that were used by \citet{walborn02b} to define the O2 spectral
type. A slightly hotter solution (by 1~kK, \Teff~=~54.8 kK) than in
Mok07 was obtained, by using the \NIV/\NV\ ionization equilibrium.
This star was also analyzed by~\citet{massey05} and~\citet{doran11}. The former
authors provide only a lower limit on \Teff\ and a consistently lower \logg\
(\Teff\ $\ge$ 48~kK and \logg~=~3.9), both in agreement with the error bars by
Mok07, whilst \citet{doran11} derive a somewhat cooler solution (by 2~kK)
compared to our findings. 

%By means of the Nitrogen lines, we are able to confirm a hotter solution
%for this star. 
%As argued by Mok07, a different \mdot\ from the latter authors
%relies on the different $\beta$ used.

Figure~\ref{BI253} shows that \NIV\nivab, \NIV$\lambda3480$, the
\NIV\ multiplet around 7120~\AA (where we reproduce the observed
emission), and the \NV\ lines are nicely fitted, with
[N]~=~7.90$\pm0.15$. At this \Teff\ and \mdot, no \NIII\ is visible in
the spectrum (see also our discussion on N11-031). The feature located
around \NIII$\lambda4634$ corresponds to \OIV$\lambda4632$. On the
other hand, we were not able to reproduce the rather broad emission of
\NIV\nivem\ (a larger \vsini\ is inconsistent with the remaining
lines). For this star, we compared again with a {\sc cmfgen} model at
similar parameters, in particular with the same [N]. Contrasted to our
solution, the width of the \NIV\ emission line could be fitted, whilst
the other lines indicated either a hotter temperature or a higher
abundance.

\paragraph{\bf BI237 -- O2 V((f$^*$)).} As for BI253, the nitrogen
ionization equilibrium favours the hotter solution proposed by Mok07
(\Teff~=~53.2~kK) rather than the cooler limit derived
by~\citet{massey05}. Our result is also consistent with the work
by~\citet{doran11}. This object is quite similar to BI253 with 
a somewhat thinner wind. 

Figure~\ref{BI237} displays a good fit for the nitrogen lines. 
\NIV\nivem\ does not show a broad profile as in BI253, and we are able to
perform an excellent fit to this line. We derive a lesser enrichment
than for BI253, with [N]~=~7.38$\pm0.15$.

\paragraph{\bf N11-060 -- O3 V((f$^*$)).} With \Teff~=~48~kK, which is
about 2~kK hotter than in Mok07, we found a consistent description of
H/He and nitrogen. As argued by Mok07, \HeI$\lambda4471$ 
is similar to N11-031, and they derived a similar \Teff~around 45~kK
for both stars. Our findings indicate a larger value for both
stars, indicating the internal consistency.

Again, we encountered the problem already met for N11-026 and
N11-031, i.e., it is rather difficult to fit the lines from three
different nitrogen ionization stages in parallel, see
Fig.~\ref{N11-060}. We opted for a compromise solution with 
[N]~=~8.20$+0.30$, which is quite large but in line 
with the helium enrichment. Note that we also estimate quite a large
upper limit, motivated by the following reasoning.

As for N11-026, we considered the impact of a hotter solution.  With
\Teff~=~51~kK and \logg~=~4.1, we are able to fit both \NIV\ and \NV\ 
at a similar abundance, [N]~=~8.15. To recover the \NIII\ triplet
lines at such hot temperatures, however, requires an increase of [N]
by roughly 0.3 dex. Thus, we are able to derive a quite similar
nitrogen abundance for different \Teff, either using the \NIII/\NIV\
or \NIV/\NV\ ionization equilibrium. The upper limit results from the
condition to match either \NV\ or \NIII, respectively. 

\paragraph{\bf Sk--70$^{\circ}$ 69 -- O5 V((f)).} For this field star, we had
some problems to reconcile H/He and N at the parameters provided by Mok07.
Our final solution is cooler by 1~kK, and we derived \YHe~=~0.14 which
is lower than the \YHe~=~0.17 estimated by Mok07.

Figure~\ref{SK7069} presents the best fit for all nitrogen lines, for 
[N]~=~8.05$\pm0.15$. There are only two disagreements: the
`right' wing of \NIII$\lambda4634$ is predicted a bit too narrow, and
\NIV\nivem\ is predicted to be in very weak emission, contrasted to
the observations. We consider the fit as acceptable, particularly
since in a clumped wind this weak emission becomes almost suppressed.

\paragraph{\bf N11-051 -- O5 Vn((f)).} This star is the fastest rotator
in our sample, and we derived a slightly higher \vsini\ and cooler \Teff.
As well, we used a prototypical 
value of $\beta = 0.8$ instead of the rather low $\beta = 0.6$ 
derived by Mok07, implying also a lower \mdot.

Unlike N11-033, where the fast rotation removes almost all
information, this object shows the \NIII\ triplet in emission, 
%according to its ((f)) designation, 
which is quite well reproduced by our model,
with [N]~=~7.58$\pm0.20$ (Fig.~\ref{N11-051}). At such high \vsini,
these lines are blended with \CIII$\lambda\lambda4647-4650-4652$.
Since carbon is not included in our calculations, it is not possible
to predict the right wing of the blended profile. The fact that the
carbon triplet has the same strength as the \NIII\ one indicates that
this star could belong to the newly defined Ofc category
\citep{walborn10,sota11},
%inspecting Galactic stars. 
which seems to be strongly peaked at spectral type O5 for all
luminosity classes. 

The only problem of our fitting procedure is found for \NIV\nivem, 
predicted to be in slight emission and actually not present in the
observed spectra. Contrasted to the case of N11-058 and N11-065, it
was not possible to circumvent this discrepancy by lowering \mdot,
since the fit to \Ha\ becomes inacceptable then. By means of a clumped
wind with lower \mdot, on the other hand, we can fit both \Ha\ and
obtain a better result for \NIV\nivem, whilst not compromising the
remaining nitrogen lines.

\paragraph{\bf N11-058 -- O5.5 V((f)).} To find a consistent solution
for all H, He, and N lines, a very low \mdot~=~0.01\Mdu\ 
is required (similar to the case of N11-065),
and also a low \logg\ (0.14 dex lower than Mok07), which is still consistent
with the Balmer line wings but appears to be rather low for a dwarf.

This was the only O-star that could be analyzed by the `curve growth
method' with respect to [N] and \vmic. The derived value of
\vmic~=~6$\pm3$~\kms is lower than for the other O-stars, though such
a difference does not drastically affect the derived abundance, 
as argued in Sect.\ref{nitro_ab}. Quite a large
abundance, [N]~=~8.09$\pm0.15$, was determined which fits all the 
lines (Fig.~\ref{N11-058}). 

\paragraph{\bf Sk--66$^{\circ}$ 18 -- O6 V((f)).} Our estimates agree well with
those from Mok07. As for Sk--66$^{\circ}$ 100, we found a very large nitrogen
abundance, [N]~=~8.48$\pm0.15$ (Fig.~\ref{SK6618}). Note that the
\NIV\nivem\ line is predicted in weak emission but appears in
absorption. Again, lowering \mdot\ was not sufficient to cure this
problem. A somewhat better fit to the this line was obtained for a
weakly clumped wind with \fcl\ = 2.3, included in the figure.

\paragraph{\bf N11-065 -- O6.5 V((f)).} Our best fit to the He lines
indicates \YHe~=~0.13, lower than the value derived by Mok07, \YHe\ =
0.17. As already discussed in Sect.~\ref{stellar_wind_param}, a satisfactory
reproduction of \NIV\nivem\ requires a very low \mdot~=~0.05\Mdu. 
Since this line clearly appears in absorption, whereas our model with  
\mdot\ from Mok07 predicts much too less absorption, we lowered \mdot,
but were able to preserve the fit to \Ha\ and the remaining nitrogen
lines. By the inclusion of clumping we obtained an even better
fit quality, with [N]~=~8.17$\pm0.15$ from \NIII\ and \NIV.

\paragraph{\bf N11-066 -- O7 V((f)).} 
\Teff\ and \logg\ as derived by Mok07 turned out to be inconsistent
with the \NIII/\NIV\ ionization equilibrium. We had considerable
problems to fit both \NIII\ and \NIV\ lines at the same abundance,
and particularly to reproduce the absorption within \NIV\nivem and the
\NIII\ quartet lines. To this end, a lower \Teff\ was mandatory, but
only rough estimates using our coarse grid could be obtained,
resulting in \Teff~=~37~kK and \logg~=~3.7 dex, which are 2.3~kK and
0.17 dex lower than the values provided by Mok07, respectively.  The
gravity is somewhat low for a dwarf but still inside the error bars
assigned by Mok07.  With these values, we determined
[N]~=~8.17$\pm0.20$, see Fig.~\ref{N11-066}, which seems rather large
for \YHe\ = 0.1.

\paragraph{\bf N11-068 -- O7 V((f)).} 
As for N11-066, we were only able
to fit the H/He and N lines by means of the coarse grid. The
differences to the results by Mok07 are significant, but still
consistent with the observations and identical to those of N11-066
which has the same spectral type. 

In contrast to N11-066, however, particularly the \NIII\ triplet shows
weaker emission, and thus a lower [N]~=~7.85$\pm0.20$ has been found,
see Fig.~\ref{N11-068}. 

\paragraph{\bf N11-061 -- O9 V.} Contrasted to Mok07, we chose a
prototypical value for the velocity field exponent, $\beta = 0.8$,
together with a larger value for \mdot. All H/He lines could be
reproduced without difficulties.

The nitrogen analysis is quite similar to the case of N11-045, 
and we derived [N]~=~7.18$\pm0.15$ from \NIII\ alone (Fig.~\ref{N11-061}). 

\paragraph{\bf N11-123 -- O9.5 V((f)).} 
Our parameters are very similar
to Mok07, and Fig.~\ref{N11-123} shows our solution for \NII/\NIII.
Even though the rotation is not extreme, almost no nitrogen
is visible, and the features overlapping with the `non-existent' 
\NII$\lambda4630$ and \NIII$\lambda4640$
are blends by \OIII\ and \SiIII,
respectively. Thus, we infer a very low nitrogen content,
[N]~=~7.00$\pm0.15$, roughly corresponding to the LMC baseline abundance.

\paragraph{\bf N11-087 -- O9.5 Vn.} Also for this rapid rotator, we found good
agreement with Mok07. Due to rotation, all nitrogen lines are diluted
into the continuum (Fig.~\ref{N11-087}), 
and only an upper limit for [N] could be estimated, [N]~=~7.38.

\section{Discussion}
\label{results}

\subsection{Comparison with results from Mok07}
\label{comp_mok07} 
In our discussion of the derived results, let us first concentrate on
a brief comparison with the findings by Mok07 (see
Table~\ref{tab_diff_mok}). Except for few cases, we derive somewhat
cooler \Teff. To a certain extent, this might be attributed to the
improved temperature structure in the new {\sc fastwind} version, and
also to the possibility to exploit the nitrogen ionization
equilibrium. The largest changes concern N11-026 (4.3~kK cooler), and
N11-031 (already the cooler solution is 2.8~kK hotter). For the
earliest stars in our sample (two O2 dwarfs and two O2 giants), we
determine a temperature range of $47.8 \leq \Teff\ \leq 54.8$, quite
similar to Mok07, but now including N11-031, since we infer hotter
solutions for this object. \citet{massey05} found a similar range,
using two dwarfs and two giants.
%Further
%discussion on this matter will be provided in Paper~III.
Gravities changed in parallel for most of the cases, with 0.10 and $-$0.43 dex as
largest positive and negative difference, respectively. Stellar radii
show a very good agreement, even for the two stars with the most
extreme changes in \Teff\ (less than 5\% difference in \Rstar).

Regarding \mdot, we derive values which are typically lower by less
than a factor of two ($\Delta \log \mdot \approx$ $-$0.1{\ldots} $-$0.4
dex), with a maximum change of $-$1.2 dex for N11-058 based on our
analysis of the nitrogen lines. The agreement of the resulting \YHe\
values is good, except for two stars with differences considerably
larger than the adopted errors. For both stars (Sk--70$^{\circ}$ 69 and
N11-065), we find a lower helium content. 

The largest differences relate to \vsini\ and \vmic. Differences
around 30-40\% in \vsini\ stress the importance of obtaining this
parameter in a separate step, when using an automated fitting method.
The substantial differences in \vmic, on the other hand, should not be
regarded as worrisome though. To a major part, \vmic\ has not been
literally derived during this work, but was only adopted (as \vmic =
10~\kms), where the resulting fit quality did not indicate any
problems with this value, within $\pm$ 5~\kms. Only for four mostly
cooler stars we were actually able to infer more robust estimates,
indicating quite a low \vmic\ $\approx$ 5~\kms. Note, however, that
the latter value refers to nitrogen lines only, and inconsistencies in
\vmic\ from H/He (used by Mok07) and metal lines have been found
already in various studies.

Since the topic of a potential relation between \vmic\ and stellar
type (\logg!) is of recent interest,\footnote{e.g., Mok07 found a weak
correlation for objects with $\logg \leq 3.6$; see also
\citet{kilian91}, \citet{GiesLambert92}, \citet{Daflon04} for similar
results for Galactic B-stars and \citet{hunter07} for LMC B-stars.}
because it might indicate (together with other evidence) the presence
of sub-surface convection \citep{Cantiello09}, a more thorough
investigation is certainly required. A derivation from light elements 
will become difficult for the hotter O-stars though, due to the
restricted number of visible lines and the complex formation mechanism of the
ubiquitous (photospheric) emission lines. Here, it will become
advantageous to exploit the information contained in the numerous UV
Fe and Ni lines (e.g., \citealt{haser98}).

\subsection{Overlap with B-star nitrogen analyses}
\label{comp_hunter}
To ensure the consistency between O-star nitrogen abundances from this
(and upcoming) work and previous results from B-stars (using different
codes, model atoms and analysis methods), a more thorough inspection
of the cooler objects is certainly required. Indeed, we are able to
compare with alternative nitrogen abundances from some overlapping
objects (compiled by \citealt{hunter09},
%, except for the fast rotator N11-033 (
see Table~\ref{tab_abun}), which base on stellar parameters obtained by
means of {\sc tlusty} and using the \SiIII/\SiIV\ ionization
equilibrium \citep{hunter07}. 

Already in the latter work, \citet{hunter07} realized
that their \Teff-values were somewhat lower than corresponding results
from Mok07 (who used exclusively H and He), but also pointed out
that \Teff\ estimates based on \HeII$\lambda4541$ would be in much
closer agreement. This is even more true regarding our `new'
values, which lie in between the Hunter et al. estimates from Si and those
from Mok07.
%Corresponding differences of \Teff\ from metals and from H/He have
%been found also in other investigations (e.g., SimonDiazxx)
Such differences in \Teff\ derived either from metals or from H/He are somewhat
disturbing, due to their influence on the metallic abundances when
using lines from one ionization stage only.

Anyhow, \citet{hunter07} decided to keep their cooler solution, to
preserve the internal consistency of their analysis. Consequently,
the nitrogen abundances derived in the present study are
systematically larger, due to our higher \Teff\ (leading to
intrinsically weaker \NII\
lines), with $\Delta\Teff \approx$~1kK for N11-008 and N11-072,
and $\approx$~2kK for N11-036. For N11-042, the \Teff\ is rather
similar, and for this object the derived [N]-values indeed overlap by
better than 0.1~dex. Also for N11-029, our \Teff\ is rather similar, but we derive
a 0.1~dex lower \logg, which leads to weaker \NII\ lines.

To check our model atom and our diagnostic approach, we performed an
additional analysis by reproducing the conditions adopted by
\citet{hunter07}, i.e., we used their stellar parameters (with
negligible \mdot\ to mimic {\sc tlusty} models) and \NII\
lines only, within the `curve of growth' method and adopting their \NII\
equivalent widths and uncertainties. For N11-029 and N11-042, we could only
perform a `by eye' estimate because Hunter et al. considered one
\NII\ line alone. Corresponding results are compared in
Table~\ref{tab_hunter}, and the agreement is almost excellent.

Thus, we conclude that the consistency of nitrogen abundances in the
overlapping B- and O-star regime is satisfactory, and that the
different codes and methods produce a reasonable agreement. However, 
there {\it is} a slight offset on the order of 0.1 to 0.2 dex, which
we attribute mostly to different effective temperatures. Since our analysis
is based on both \NII\ and \NIII\ lines (in contrast to Hunter et al.),
and has been performed in parallel with the analysis of H/He, we
prefer our values though.

\begin{table}
\center
\caption{Comparison between nitrogen abundances derived by means of {\sc
fastwind} and by \citet{hunter09}, using their stellar parameters and 
\NII\ diagnostics alone.}
\label{tab_hunter}
\begin{tabular}{llcc}
\hline 
\hline
Star & Sp. Type & [N] t.w. & [N] Hunter \\
\hline
N11-029 & O9.7 Ib & 7.20 &7.10 \\
N11-036 & B0.5 Ib & 7.86 &7.76 \\
N11-008 & B0.7 Ib & 7.89 &7.84 \\
N11-042 & B0 III  & 6.99 &6.92 \\
N11-072 & B0.2 III& 7.43 &7.38 \\
\hline
\end{tabular}
\end{table}

\subsection{Nitrogen abundances}
\label{comment nitro}
\begin{figure}
\begin{minipage}{9cm}
\resizebox{\hsize}{!}
  {\includegraphics{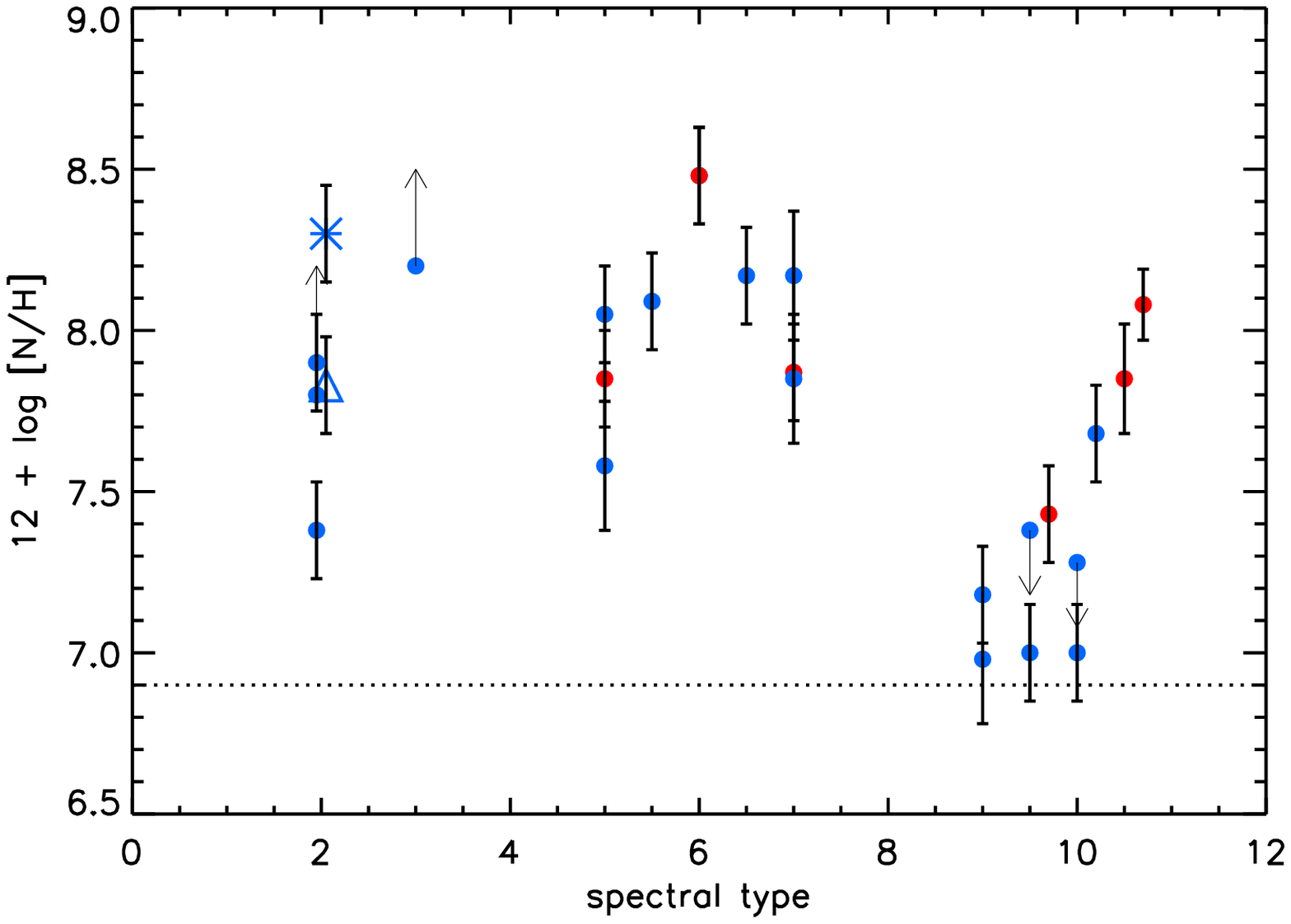}} 
\end{minipage}
\hspace{-.5cm}
\begin{minipage}{9cm}
\resizebox{\hsize}{!}
  {\includegraphics{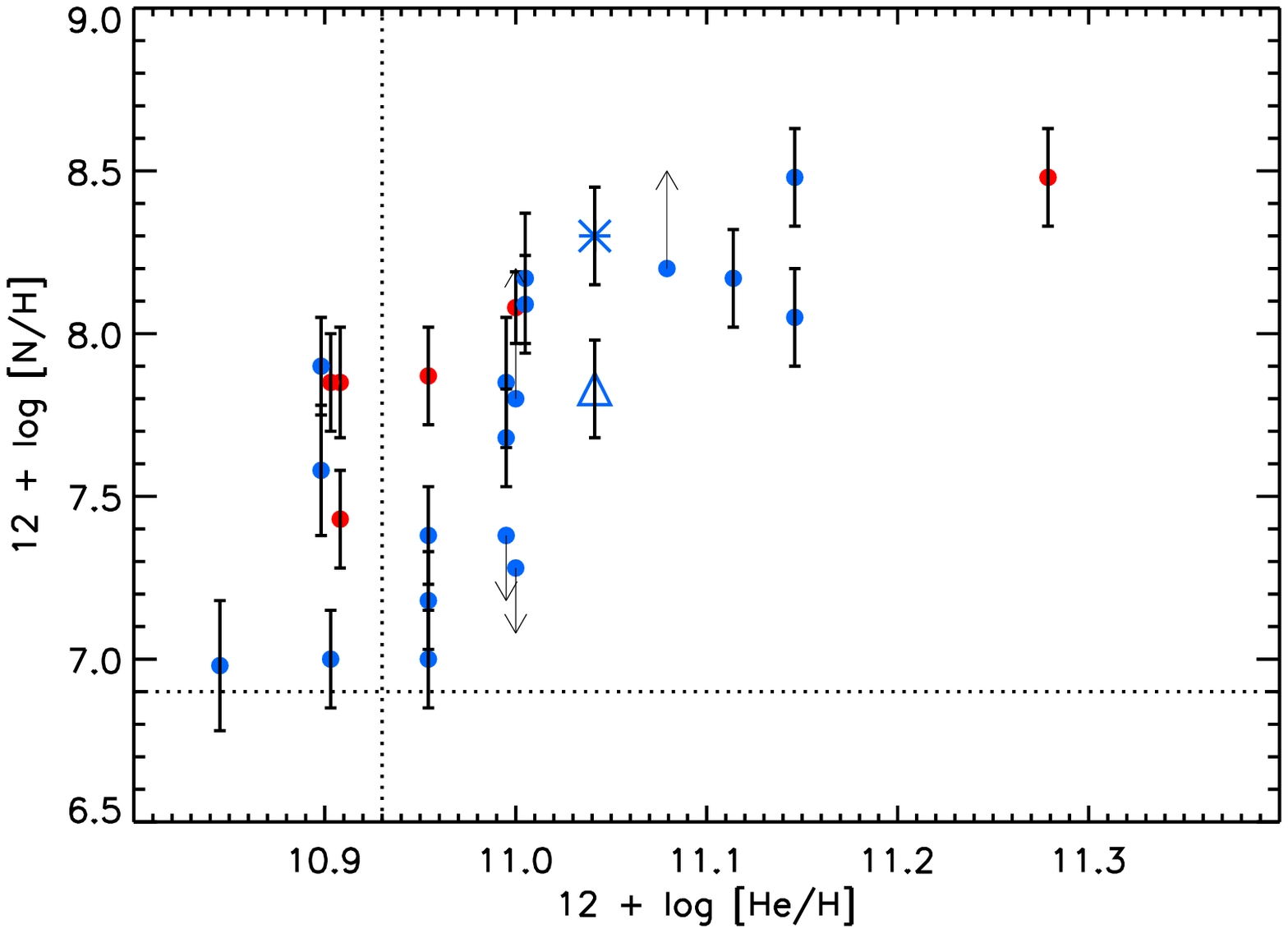}} 
\end{minipage}
\caption{Nitrogen abundances derived for our LMC sample. Upper panel: 
As a function of O-star spectral type (`10-12' correspond to B0-B2;
for \Teff, see Table~\ref{tab_abun}). Lower panel: As a function of
helium content, 12 + $\log$(He/H) = 12 + $\log$(\YHe). Red: luminosity
class I-II; blue: III-V. Arrows indicate upper or lower limits. 
Alternative solutions for N11-031 (see
Sect.~\ref{comments}) are indicated by a triangle and an asterisk for
the cooler and hotter solution, respectively. 
The estimated (1-$\sigma$) error for $\log$(\YHe) is 0.05 dex. The dotted
lines indicate the LMC nitrogen \citep{hunter07} and the average
helium (see text) baseline abundances. Some objects have been
slightly shifted horizontally, because of better visibility.} 
\label{nyhe}
\end{figure}

Figure~\ref{nyhe} summarizes the basic outcome of our analysis, by
displaying the derived nitrogen abundances as a function of spectral
type and helium content, together with the LMC baseline abundance from
\citet{hunter07}. Evidently, there are only few cooler objects located
close to the baseline, whereas the majority of the objects
(independent of luminosity class!) is strongly enriched, with [N] in
between 7.5 and 8.1.\footnote{Note that most stars have `normal' or
only moderately enriched helium abundances (see below) so that the
high [N] cannot be an effect of decreasing H content.} Five objects
display extreme enrichment, with [N] from 8.17 to 8.5, which is close
to the maximum nitrogen content given by the CNO equilibrium value
for nitrogen, [N]$_{\rm max} \approx$ 8.5. However, this is well above
the enrichment reached for a 40~\msun\ star with an initial rotational
velocity of 275~\kms \citep{Brott11a}.

The lower panel of Fig.~\ref{nyhe} is more promising though. There
seems to be a strong correlation between the nitrogen and the helium
enrichment, here displayed logarithmically. 
The LMC helium abundance should be located, in terms of number
fraction, around \YHe\ = 0.08-0.094 corresponding to [He]~=~10.90-10.97
(\citealt{RusselDopita90, MM01, Vermeij02, Peimbert03, Tsamis03}), and
agrees quite well with our minimum values for the derived helium abundance. We
found only five stars with considerable enriched nitrogen  close to this
value, three (super-)giants and two dwarfs, but note also the
attributed uncertainty in helium content. Except for these objects,
the correlation is almost perfect, and there is a certain clustering
around the pair [He]=11.0/[N]=8.0.

\begin{figure}
\resizebox{\hsize}{!}
{\includegraphics[angle=270]{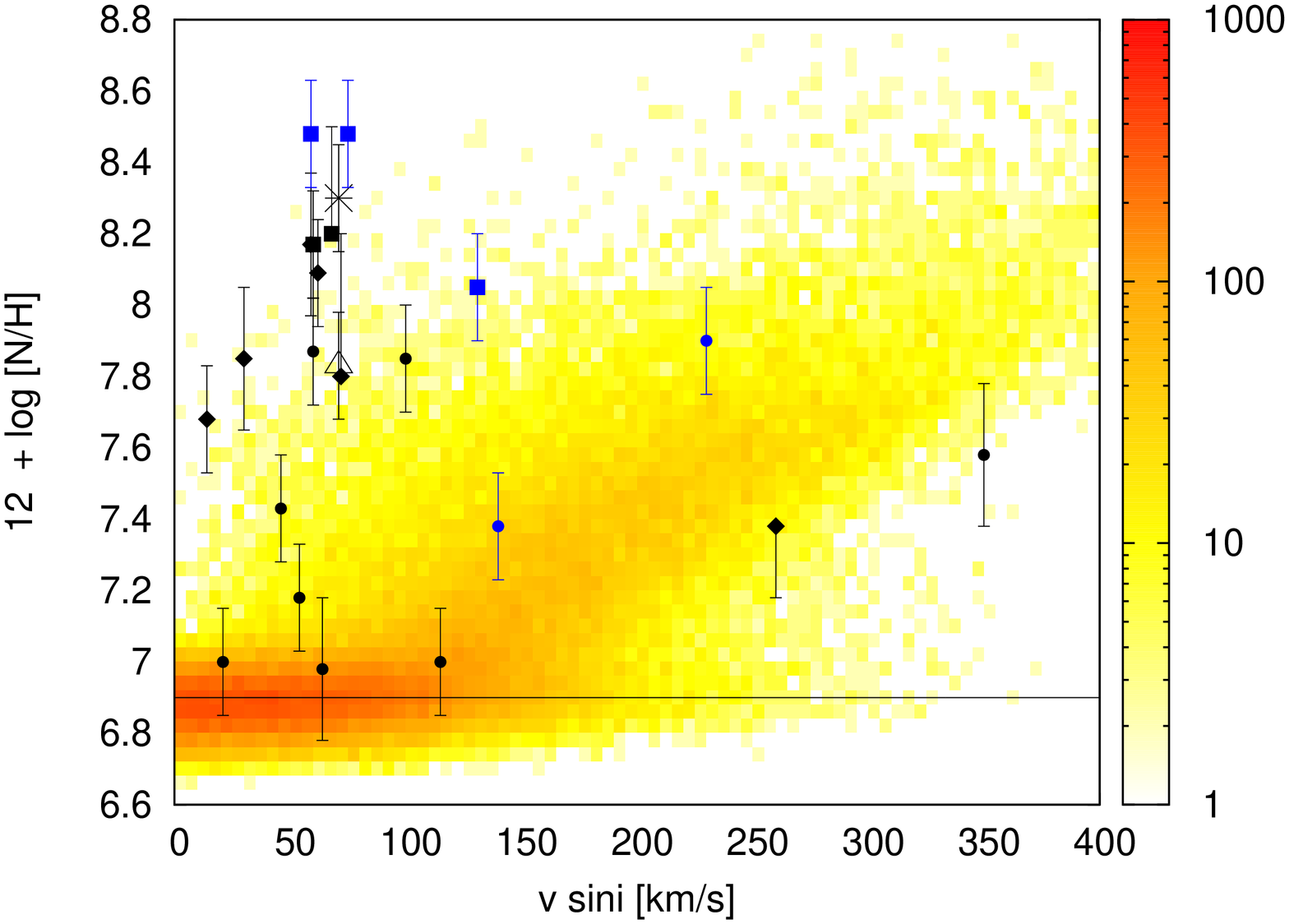}}
\caption{`Hunter-plot' displaying the nitrogen abundance vs.
projected rotational speed. Population synthesis from
\citet{Brott11b}, for \Teff\ $\ge$ 29kK and a magnitude limit $V \le
15.29$, shown as a density plot in the background. The color coding
corresponds to the number of stars per bin, with binsize 5~\kms $\times$ 
0.04 dex. Overplotted are data from
this study. Black: N11 stars; blue: field stars. Circles, diamonds and
squares correspond to objects with low, intermediate and strong helium
enrichment, respectively (see text). Alternative solutions for
N11-031 (intermediate He enrichment) as in Fig.~\ref{nyhe}.}
\label{hunter_plot}
\end{figure}

A somewhat different view is provided in Fig.~\ref{hunter_plot}, which
displays the so-called `Hunter-plot', nitrogen-abundance vs. projected
rotational speed, for all our sample stars with \Teff\ $\ge$ 29~kK.
N-11 stars are indicated in black, field stars in blue. Circles,
diamonds and squares correspond to objects with low (\YHe~$<$~0.1),
intermediate (\YHe~=~0.1), and strong (\YHe~$>$~0.1) helium enrichment,
respectively.

The background of this figure consists of results from the recent
population synthesis by \citet{Brott11b}, for all objects with \Teff\
$\ge$ 29~kK, and a magnitude limit (corresponding to our sample) of $V
\le 15.29$, shown as a density plot. The underlying simulation assumes
a rather broad Gaussian rotational velocity distribution as derived
for LMC early-type stars, peaking at 100~\kms with a standard
deviation of $\sigma$ = 140~\kms\ \citep{hunter08b,
hunter09},\footnote{Supergiants and stars above 25~\msun\ have been
discarded from their analysis, to avoid effects from mass loss induced
spin-down.} and random inclinations.

Such diagrams ([N] vs. \vsini, compared with evolutionary
calculations) have been presented the first time by \citet{hunter08},
to summarize the outcome of the B-star analyses within the FLAMES-I
survey, and to investigate the predicted effects of rotational mixing.
One of their major findings was the unexpected presence of a
significant number of objects with slow rotation and large enrichment, not
predicted by (single-star) theory, so-called `group 2' objects.

In the O-star case now, this problem becomes even more severe. We refrain
here from a detailed statistical analysis, since the number of
investigated objects is too low, and postpone this objective until the
results from the FLAMES Tarantula survey (with more than 200 `useful'
O-stars) have become available. 

Nevertheless, the trend is obvious. Roughly one third of the objects
are located at positions where they should be expected (those at the
baseline and the `diagonal'), another one-third is located at the
predicted upper limit, and the last one-third (beyond [N] = 8.0)
extends to very large values where the predicted population density is
almost zero. Let
us note that the two objects with the largest [N] enrichment ($\approx$
8.5 dex, which is (incidentally?) just the maximum nitrogen content 
given by the CNO equilibrium value) are two
field stars, Sk--66$^{\circ}$ 100 and Sk--66$^{\circ}$ 18, one O6 giant and
another O6 dwarf. Both stars did not present any difficulties in the nitrogen
analysis, thus indicating a reasonable quality. 

In terms of the original Hunter et al. sample, roughly two-third of
our objects would be denoted by `group 2'.\footnote{In contrast to
the B-star `group 2' objects, however, the deviations between
predictions and `observations' for some of
the objects are much more extreme.} The
corresponding number of objects is so large that inclination effects
w.r.t. \vsini\ should be irrelevant.  Interestingly, however, the
corresponding He-abundances are in line with our findings. The first
group has a low abundance (Fig.~\ref{hunter_plot}, circles), the second group mostly consists of
enriched objects (diamonds), and the third one comprises objects with
considerable He enrichment (squares). Thus, in parallel with the derived
correlation between {\it observed} nitrogen and helium content, the
discrepancy between observations and theory becomes the stronger the
larger the He-abundance is.

In evolutionary models is the amount of He transported to the surface 
strongly controlled by the parameter $f_\mu$, which describes the
inhibiting effect of mean molecular weight gradients (in this case,
the H-He gradient) on the transport of elements (see
\citealt{Heger00}).
%However, measurements of He abundances are%
%sparse in the literature, and limited to low \vsini. 
The models of \citet{Brott11a} have adopted $f_\mu$ = 0.1, from an
earlier calibration of \citet{yoon06}.
%which an update of the calibration of Heger00a to low vsini O-star
%abundances of He from Herreo(92,98).  
Lowering the sensitivity to the mean molecular weight barrier would
increase mixing of both, nitrogen and helium, to the surface, but also
reduce the minimum mass and velocity required for chemical
homogeneous evolution in the models (see also the discussion in
\citealt{Heger00}). Given the present values of [N] and \YHe, it
might be possible to derive further constraints on $f_\mu$ in future
work.

\section{Summary}
\label{conclusions}
In this paper, we investigated the \NIV\nivem\ emission line formation,
determined the nitrogen abundance of a sample of 25 LMC O- and early
B-stars, and performed a first comparison with corresponding predictions from
stellar evolution including rotational mixing. The results of this
work can be summarized as follows.
\begin{enumerate}
\item In O-stars, the dominating process responsible for the \NIV\ line
emission is the strong depopulation of the lower level by the
`two-electron' transitions 3p $\rarrow$ 2p$^2$, of (mostly)
photospheric origin. This drain increases as a function of \mdot,
because of increasing ionizing fluxes (which are coupled to the \HeII\
continuum), leading to more depopulation of the ground and the coupled
2p$^2$ states. Resonance lines (as for the \NIII\ emission triplet) do
not play a role for typical O-star mass-loss rates and below.

Since in addition to nitrogen there are many other elements which
display optical line emission in the hot star regime (C, O, Si), it might be
suspected that similar processes might be invoked, because of similar
electronic configurations/transitions.

\item To infer the nitrogen abundances, we re-determined the
stellar and wind parameters, by means of `by eye' fits, starting with
the values provided by Mok07, but exploiting in parallel the nitrogen
ionization equilibrium and deriving \vsini\ in a first, separate step.
Moreover, we accounted for extra line-broadening expressed in terms of
\vmac. In addition to systematically lower \vsini, we also derived
mostly lower \Teff\ (partly because of using an improved {\sc fastwind}
version) and thus \logg, but differences to Mok07 are generally small,
except for few objects. 
\item Based on these parameters, we derived nitrogen abundances,
mostly by varying the abundance and comparing with all nitrogen lines
present in the spectrum. In a few cases, we were able to estimate [N]
and \vmic\ in parallel, by means of a curve-of-growth method. 
\item Again in most cases, we found no problems in fitting the
nitrogen lines, and reproduced the `f' features quite well. Only for
some of the (hotter) objects where lines from all three stages, 
\NIII, \NIV\ and \NV, are visible, we needed to aim at a compromise
solution. Real problems were encountered for one star, N11-031 (ON2
III(f$^\ast$)), where only either \HeI, \NIII\ and \NIV\nivab\ (at
cooler \Teff) or \NIV\ and \NV\ (at higher \Teff) could be fitted in
parallel. The difference in the derived \Teff\ amounts to 8,000~K,
which is far from satisfactory, and requires future effort to resolve
the problem. A solution in terms of binarity, though somewhat
unlikely, cannot be ruled out so far.
\item For some cooler objects already analyzed by \citet{hunter07}
by means of {\sc tlusty} using \NII\ lines alone, we found
differences in [N] on the order of 0.1 to 0.3 dex, with larger values
from our analysis. These differences could be exclusively attributed
to different stellar parameters, mostly \Teff. Overall, however, are the
nitrogen abundances in the overlapping B- and O-star domain
consistent within a reasonable error. 
\item Within our sample, we found only three cooler objects close to
the LMC nitrogen baseline abundance, [N]$_{\rm baseline}$ = 6.9. The
majority of the analyzed O-stars (independent of luminosity class)
seems to be strongly enriched, with [N] = 7.5 to 8.1. Five objects indicate an
extreme enrichment, with [N]~=~8.17 to 8.5.
\item There is a rather good correlation between the derived nitrogen
and helium surface abundances.
\item Comparing the nitrogen abundances as a function of \vsini\ with
tailored evolutionary calculations, we found a significant number of
highly enriched, low \vsini\ (`group 2') objects. Interestingly, the
correlation between He and N becomes also visible in this comparison:
Whilst most objects with unenriched He are located just in the region
where the predicted population density is largest (accounting for
selection effects), objects with enriched He are located at the upper
limit of this distribution and above, and particularly those with
the largest He enrichment lie well above this limit.
\end{enumerate}
Due to the low initial (baseline) nitrogen abundance, the detection of
strong nitrogen enrichment in the bulk of O-stars might indicate that
efficient mixing takes place already during the very early phases of
stellar evolution of LMC O-stars. Nevertheless, it would be premature
to draw firm conclusions from our results, since the sample size is
still low. Upcoming results from the VLT-FLAMES Tarantula survey
(which will be derived in a similar way as presented here, drawing
from our experience) will enable a more complete view. In particular,
the determination of O-star nitrogen abundances in the LMC will place
very tight constraints on the early evolutionary phases of O-stars and
thus on the theory of massive star evolution.

\begin{acknowledgements}
{We like to thank our anonymous referee and N.~Walborn for valuable
comments and suggestions. Many thanks to P.~Crowther for providing us
with the UVES spectra for the LMC field stars, and N.~Przybilla for
synthetic spectra based on his \NII\ model ion and for performing test
calculations with the \NII\ data set developed by C. Allende Prieto.
Many thanks also to John Hillier for providing the {\sc cmfgen} code, and
particularly to Keith Butler for his advice and help on the nitrogen
atomic data.

J.G.R.G. gratefully acknowledges financial support from the German
DFG, under grant 418 SPA 112/1/08 (agreement between the DFG and the
Instituto de Astrof\'isica de Canarias). J.P and F.N. acknowledge
financial support from the Spanish Ministerio de Ciencia e
Innovaci\'on under projects AYA2008-06166-C03-02 and
AYA2010-21697-C05-01.}
\end{acknowledgements}

\bibliographystyle{aa}
\bibliography{bib_niii}

\Online
\appendix

\section{Details of the nitrogen model atom}
\label{details_nit}

\begin{figure}[b]
\begin{minipage}{9cm}
  \resizebox{\hsize}{!}
  {\includegraphics{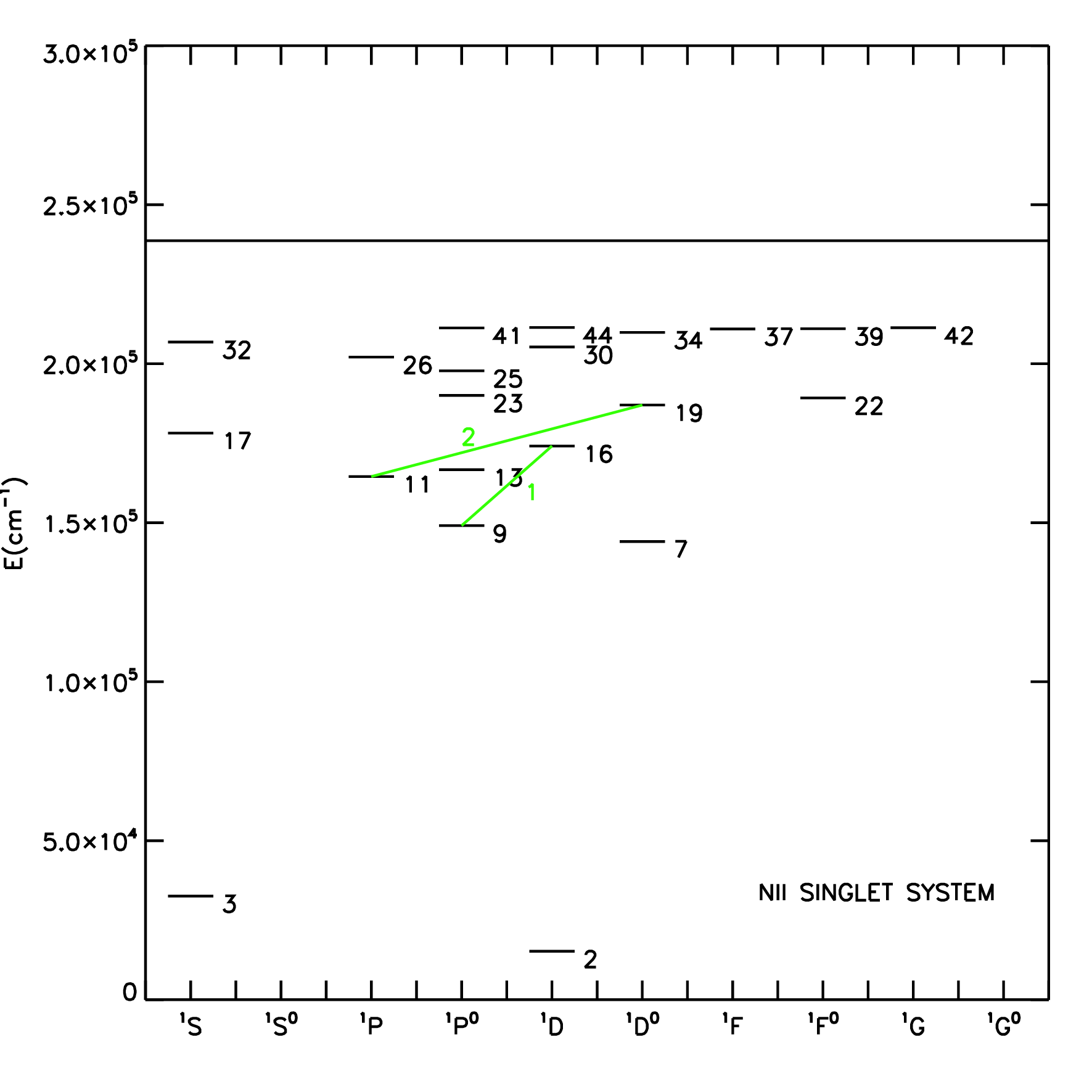}}
\end{minipage}
\vspace{-0.5cm}
\begin{minipage}{9cm}
  \resizebox{\hsize}{!}
  {\includegraphics{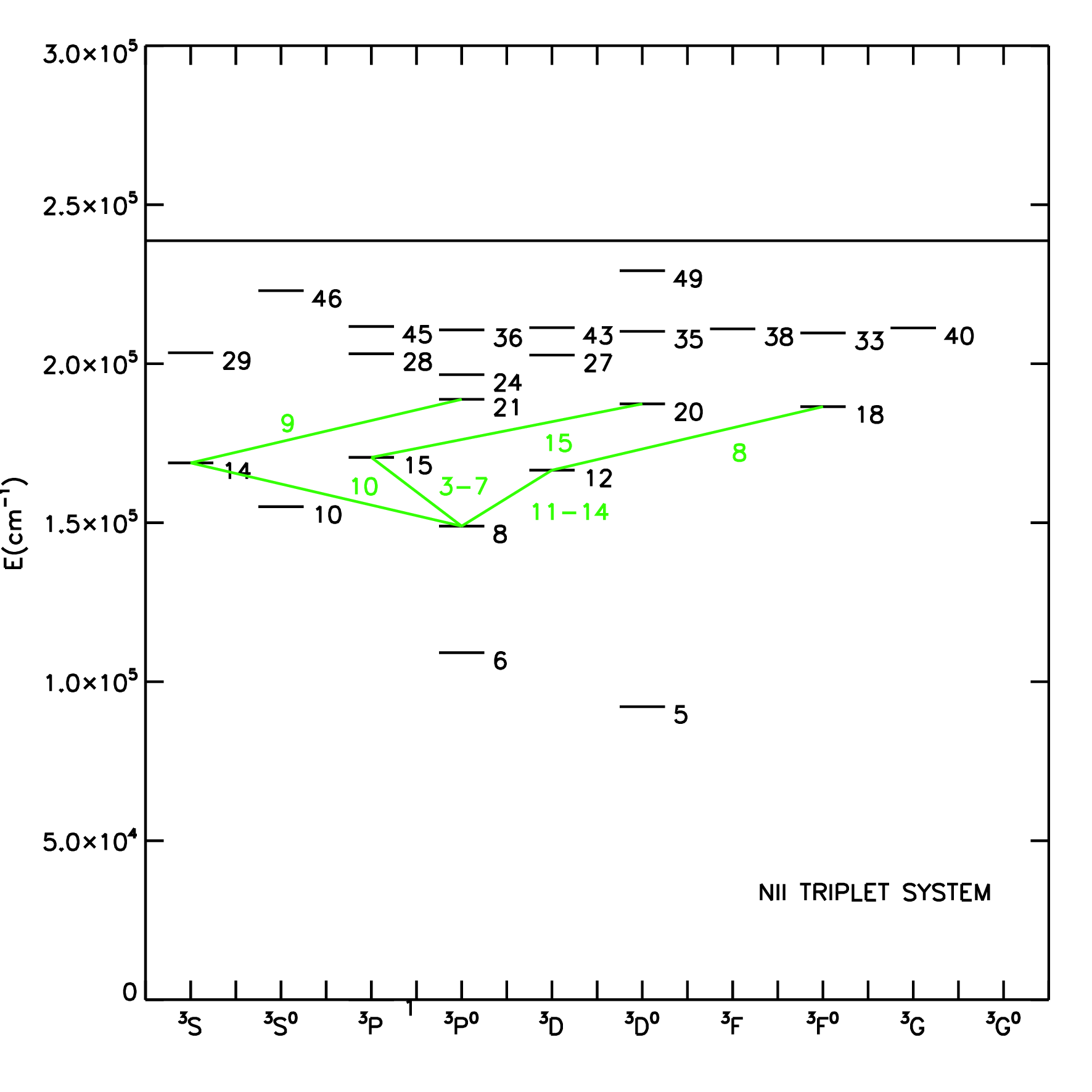}}
\end{minipage}
\caption{Grotrian diagrams for the \NII\ singlet (upper panel) and triplet
(lower panel) systems. Level designations refer
to Table~\ref{atom_lev_nii}. Important optical transitions are indicated by
green lines and numbers referring to entries in Table~\ref{tab_lines_nitro}.
} 
\label{nii-singlet}
\end{figure}

\begin{figure*}
\begin{minipage}{9cm}
\resizebox{\hsize}{!} 
{\includegraphics{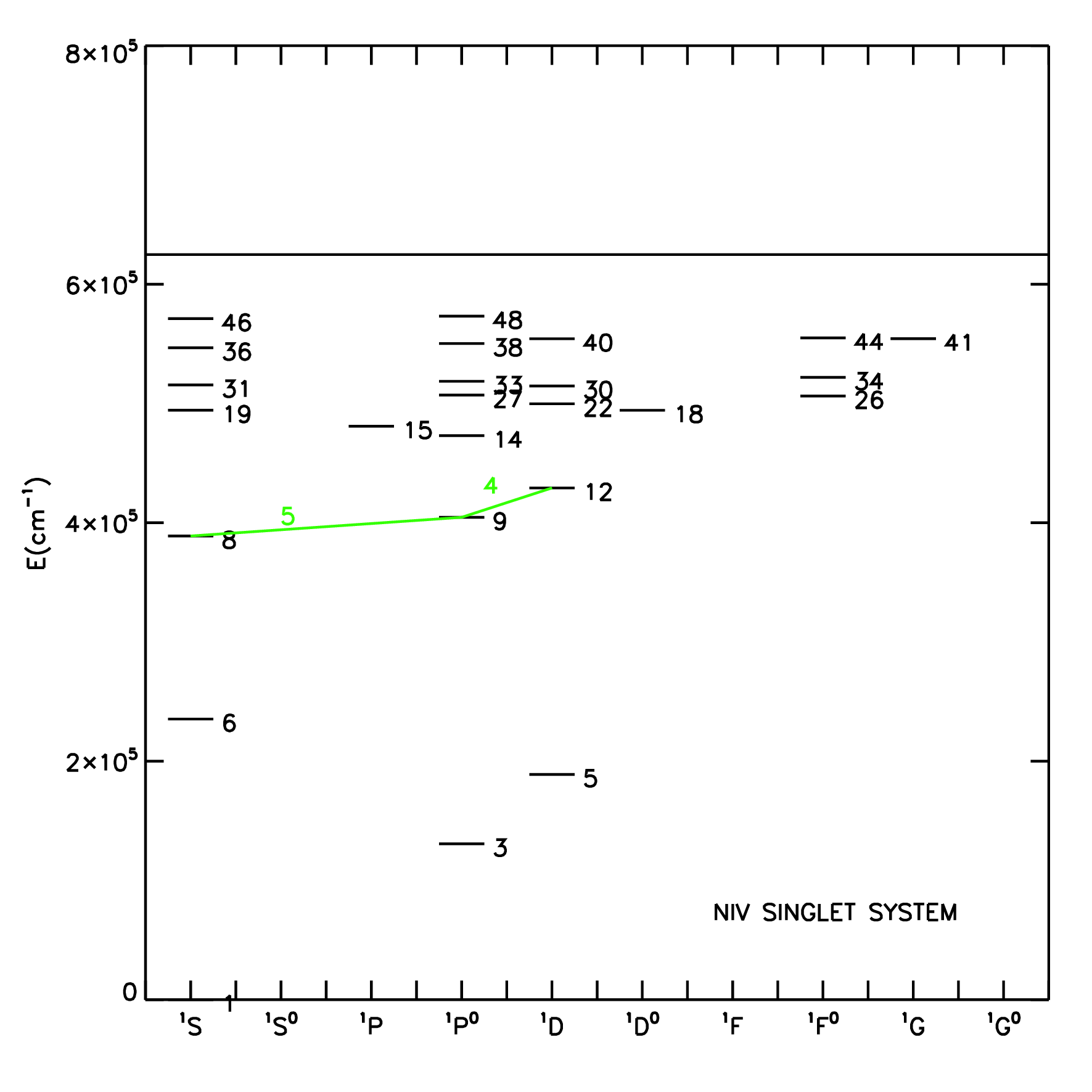}}
\end{minipage}
\begin{minipage}{9cm}
\resizebox{\hsize}{!} 
{\includegraphics{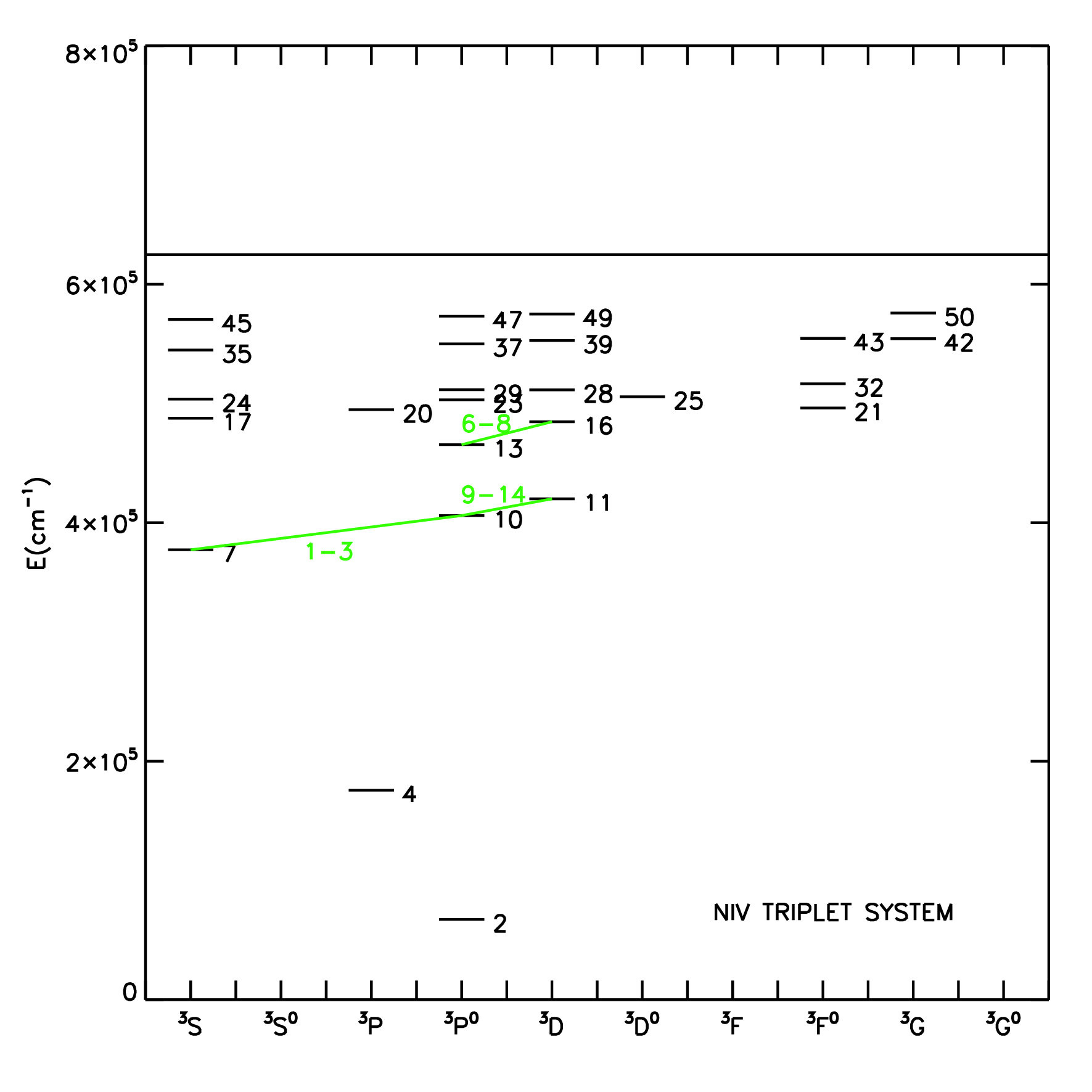}}
\end{minipage}
\caption{As Fig.~\ref{nii-singlet}, but for the \NIV\ singlet (left) 
and triplet (right) system. Level numbers refer to
Table~\ref{atom_lev_niv}.} 
\label{niv-singlet}
\end{figure*}

\begin{figure}
\resizebox{\hsize}{!}
  {\includegraphics{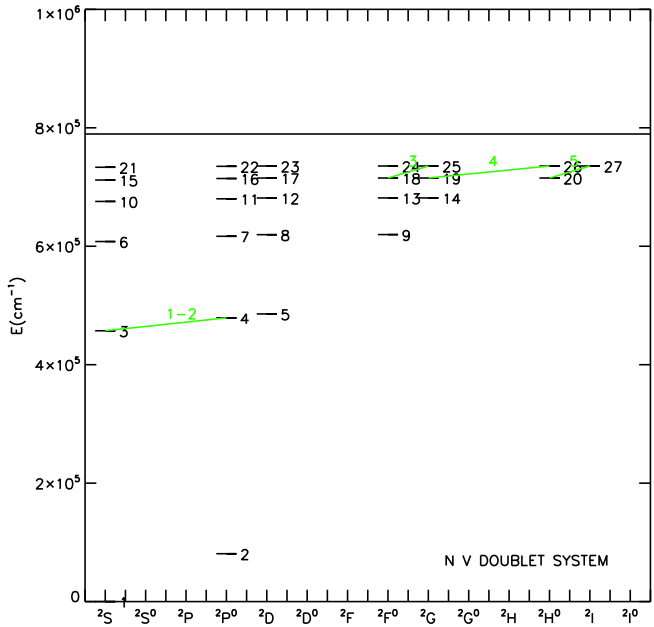}}
%\vspace{-0.5cm}
\caption{As Fig.~\ref{nii-singlet}, but for the \NV\ model ion.
Level numbers refer to Table~\ref{atom_lev_nv}.}
\label{nv-grotrian}
\end{figure}

This section provides some details of our \NII, \NIV\ and \NV\ model
ions (corresponding material for \NIII\ has been already presented in
Paper~I). Configurations and term designations are outlined in 
Tables~\ref{atom_lev_nii}, \ref{atom_lev_niv} and \ref{atom_lev_nv}, 
whilst Figs.~\ref{nii-singlet}, \ref{niv-singlet} and
\ref{nv-grotrian} display the Grotrian diagrams for the \NII\ singlet
and triplet system (the quintet system comprises five levels only),
the \NIV\ singlet and triplet system, and the \NV\ doublet system,
respectively. In these figures, important optical transitions as given
in Table~\ref{tab_lines_nitro} are indicated as well.

\begin{table}
\caption{Electronic configurations and term designations of our \NII\ model
ion. The level numbers correspond to the entries in the Grotrian diagrams in
Fig.~\ref{nii-singlet}, for the singlet and the triplet terms.}
\label{atom_lev_nii}
%\vspace{0.3cm}
\tabcolsep1.3mm
\begin{tabular}{rll | rll}
\hline 
\hline
\# & Configuration & Desig. & \#  & Configuration  & Desig. \\
\hline
  1      & 1s$^2$ 2s$^2$ 2p$^2$              &    2p$^2$ $^3$P    &     
 26      & 1s$^2$ 2s$^2$ 2p($^2$P$^0$) 4p    &    4p   $^1$P        \\
  2      & 1s$^2$ 2s$^2$ 2p$^2$              &    2p$^2$ $^1$D    &      
 27      & 1s$^2$ 2s$^2$ 2p($^2$P$^0$) 4p    &    4p   $^3$D        \\
  3      & 1s$^2$ 2s$^2$ 2p$^2$              &    2p$^2$ $^1$S    &      
 28      & 1s$^2$ 2s$^2$ 2p($^2$P$^0$) 4p    &    4p   $^3$P        \\
  4      & 1s$^2$ 2s     2p$^3$              &    2p$^3$ $^5$S$^0$  &       
 29      & 1s$^2$ 2s$^2$ 2p($^2$P$^0$) 4p    &    4p   $^3$S        \\
  5      & 1s$^2$ 2s   2p$^3$                &    2p$^3$ $^3$D$^0$  &      
 30      & 1s$^2$ 2s$^2$ 2p($^2$P$^0$) 4p    &    4p   $^1$D        \\
  6      & 1s$^2$ 2s   2p$^3$                &    2p$^3$ $^3$P$^0$  &      
 31      & 1s$^2$ 2s 2p$^2$($^4$P) 3s        &    3s'   $^5$S        \\
  7      & 1s$^2$ 2s   2p$^3$                &    2p$^3$ $^1$D$^0$  &      
 32      & 1s$^2$ 2s$^2$ 2p($^2$P$^0$) 4p    &    4p   $^1$S        \\
  8      & 1s$^2$ 2s$^2$ 2p($^2$P$^0$) 3s    &    3s   $^3$P$^0$  &   
 33      & 1s$^2$ 2s$^2$ 2p($^2$P$^0$) 4d    &    4d   $^3$F$^0$        \\
  9      & 1s$^2$ 2s$^2$ 2p($^2$P$^0$) 3s    &    3s   $^1$P$^0$  &      
 34      & 1s$^2$ 2s$^2$ 2p($^2$P$^0$) 4d    &    4d   $^1$D$^0$        \\
 10      & 1s$^2$ 2s  2p$^3$                 &    2p$^3$ $^3$S$^0$  &      
 35      & 1s$^2$ 2s$^2$ 2p($^2$P$^0$) 4d    &    4d   $^3$D$^0$        \\
 11      & 1s$^2$ 2s$^2$ 2p($^2$P$^0$) 3p    &    3p   $^1$P    &    
 36      & 1s$^2$ 2s$^2$ 2p($^2$P$^0$) 4d    &    4d   $^3$P$^0$        \\
 12      & 1s$^2$ 2s$^2$ 2p($^2$P$^0$) 3p    &    3p   $^3$D    &    
 37      & 1s$^2$ 2s$^2$ 2p($^2$P$^0$) 4f    &    4f   $^1$F        \\
 13      & 1s$^2$ 2s   2p$^3$                &    2p$^3$ $^1$P$^0$  &       
 38      & 1s$^2$ 2s$^2$ 2p($^2$P$^0$) 4f    &    4f   $^3$F        \\
 14      & 1s$^2$ 2s$^2$ 2p($^2$P$^0$) 3p    &    3p   $^3$S    &    
 39      & 1s$^2$ 2s$^2$ 2p($^2$P$^0$) 4d    &    4d   $^1$F$^0$        \\
 15      & 1s$^2$ 2s$^2$ 2p($^2$P$^0$) 3p    &    3p   $^3$P    &    
 40      & 1s$^2$ 2s$^2$ 2p($^2$P$^0$) 4f    &    4f'   $^3$G        \\
 16      & 1s$^2$ 2s$^2$ 2p($^2$P$^0$) 3p    &    3p   $^1$D    &    
 41      & 1s$^2$ 2s$^2$ 2p($^2$P$^0$) 4d    &    4d   $^1$P$^0$        \\
 17      & 1s$^2$ 2s$^2$ 2p($^2$P$^0$) 3p    &    3p   $^1$S    &    
 42      & 1s$^2$ 2s$^2$ 2p($^2$P$^0$) 4f    &    4f'   $^1$G        \\
 18      & 1s$^2$ 2s$^2$ 2p($^2$P$^0$) 3d    &    3d   $^3$F$^0$  &    
 43      & 1s$^2$ 2s$^2$ 2p($^2$P$^0$) 4f    &    4f'   $^3$D        \\
 19      & 1s$^2$ 2s$^2$ 2p($^2$P$^0$) 3d    &    3d   $^1$D$^0$  &    
 44      & 1s$^2$ 2s$^2$ 2p($^2$P$^0$) 4f    &    4f'   $^1$D        \\
 20      & 1s$^2$ 2s$^2$ 2p($^2$P$^0$) 3d    &    3d   $^3$D$^0$  &    
 45      & 1s$^2$ 2s 2p$^2$($^4$P) 3s      &    3s'   $^3$P        \\
 21      & 1s$^2$ 2s$^2$ 2p($^2$P$^0$) 3d    &    3d   $^3$P$^0$  &    
 46      & 1s$^2$ 2s 2p$^2$($^4$P) 3p      &    3p'   $^3$S$^0$        \\
 22      & 1s$^2$ 2s$^2$ 2p($^2$P$^0$) 3d    &    3d   $^1$F$^0$  &    
 47      & 1s$^2$ 2s 2p$^2$($^4$P) 3p      &    3p'   $^5$D$^0$        \\
 23      & 1s$^2$ 2s$^2$ 2p($^2$P$^0$) 3d    &    3d   $^1$P$^0$  &    
 48      & 1s$^2$ 2s 2p$^2$($^4$P) 3p      &    3p'   $^5$P$^0$        \\
 24      & 1s$^2$ 2s$^2$ 2p($^2$P$^0$) 4s    &    4s   $^3$P$^0$  &    
 49      & 1s$^2$ 2s 2p$^2$($^4$P) 3p      &    3p'   $^3$D$^0$        \\
 25      & 1s$^2$ 2s$^2$ 2p($^2$P$^0$) 4s    &    4s   $^1$P$^0$  &    
 50      & 1s$^2$ 2s 2p$^2$($^4$P) 3p      &    3p'   $^5$S$^0$        \\
\hline
\end{tabular}
\end{table}

\begin{table}
\caption{Electronic configurations and term designations of our \NIV\ model
ion. The level numbers correspond to the entries in the Grotrian diagrams in
Fig.~\ref{niv-singlet}.}
\label{atom_lev_niv}
%\vspace{0.3cm}
\tabcolsep1.3mm
\begin{tabular}{rll | rll}
\hline 
\hline
\# & Configuration & Desig. & \#  & Configuration  & Desig. \\
\hline
  1      & 1s$^2$ 2s$^2$                      &    2s$^2$ $^1$S    &     
 26      & 1s$^2$ 2p($^2$P$^{0}_{3/2}$) 3d    &    3d'   $^1$F$^0$        \\
  2      & 1s$^2$ 2s 2p                       &    2p $^3$P$^0$    &      
 27      & 1s$^2$ 2s 4p                       &    4p   $^1$P$^0$        \\
  3      & 1s$^2$ 2s 2p                       &    2p $^1$P$^0$    &      
 28      & 1s$^2$ 2s 4d                       &    4d   $^3$D        \\
  4      & 1s$^2$ 2s 2p                       &    2p $^3$P  &       
 29      & 1s$^2$ 2p($^2$P$^{0}_{3/2}$) 3d    &    3d'   $^3$P$^0$        \\
  5      & 1s$^2$ 2p$^2$                      &    2p$^2$ $^1$D  &      
 30      & 1s$^2$ 2s 4d                       &    4d   $^1$D        \\
  6      & 1s$^2$ 2p$^2$                      &    2p$^2$ $^1$S  &      
 31      & 1s$^2$ 2p($^2$P$^{0}_{3/2}$) 3p    &    3p'  $ ^1$S        \\
  7      & 1s$^2$ 2s 3s                       &    3s $^3$S  &      
 32      & 1s$^2$ 2s 4f                       &    4f   $^3$F$^0$        \\
  8      & 1s$^2$ 2s 3s                       &    3s  $ ^1$S  &      
 33      & 1s$^2$ 2p($^2$P$^{0}_{3/2}$) 3d    &    3d'   $^1$P$^0$        \\
  9      & 1s$^2$ 2s 3p                       &    3p   $^1$P$^0$  &      
 34      & 1s$^2$ 2s 4f                       &    4f   $^1$F$^0$        \\
 10      & 1s$^2$ 2s 3p                       &    3p $^3$P$^0$  &      
 35      & 1s$^2$ 2s 5s                       &    5s   $^3$S        \\
 11      & 1s$^2$ 2s 3d                       &    3d   $^3$D    &    
 36      & 1s$^2$ 2s 5s                       &    5s  $ ^1$S        \\
 12      & 1s$^2$ 2s 3d                       &    3d   $^1$D    &    
 37      & 1s$^2$ 2s 5p                       &    5p   $^3$P$^0$        \\
 13      & 1s$^2$ 2p($^2$P$^0$)   3s          &    3s' $^3$P$^0$  &       
 38      & 1s$^2$ 2s 5p                       &    5p   $^1$P$^0$        \\
 14      & 1s$^2$ 2p($^2$P$^{0}_{3/2}$) 3s    &    3s'   $^1$P$^0$    &    
 39      & 1s$^2$ 2s 5d                       &    5d   $^3$D        \\
 15      & 1s$^2$ 2p($^2$P$^0_{1/2}$) 3p      &    3p'   $^1$P    &    
 40      & 1s$^2$ 2s 5d                       &    5d   $^1$D        \\
 16      & 1s$^2$ 2p($^2$P$^0$) 3p            &    3p'   $^3$D    &    
 41      & 1s$^2$ 2s 5g                       &    5g   $^1$G        \\
 17      & 1s$^2$ 2p($^2$P$^{0}_{3/2}$) 3d    &    3d'   $^3$S    &    
 42      & 1s$^2$ 2s 5g                       &    5g   $^3$G        \\
 18      & 1s$^2$ 2p($^2$P$^0_{1/2}$) 3d      &    3d'   $^1$D$^0$  &    
 43      & 1s$^2$ 2s 5f                       &    5f   $^3$F$^0$        \\
 19      & 1s$^2$ 2s 4s                       &    4s  $ ^1$S  &    
 44      & 1s$^2$ 2s 5f                       &    5f   $^1$F$^0$        \\
 20      & 1s$^2$ 2p($^2$P$^0$) 3p            &    3p'   $^3$P  &    
 45      & 1s$^2$ 2s 6s                       &    6s   $^3$S        \\
 21      & 1s$^2$ 2p($^2$P$^0$) 3d            &    3d'   $^3$F$^0$  &    
 46      & 1s$^2$ 2s 6s                       &    6s  $ ^1$S        \\
 22      & 1s$^2$ 2p($^2$P$^{0}_{3/2}$) 3p    &    3p'   $^1$D  &    
 47      & 1s$^2$ 2s 6p                       &    6p   $^3$P$^0$        \\
 23      & 1s$^2$ 2s 4p                       &    4p   $^3$P$^0$  &    
 48      & 1s$^2$ 2s 6p                       &    6p   $^1$P$^0$        \\
 24      & 1s$^2$ 2s 4s                       &    4s   $^3$S  &    
 49      & 1s$^2$ 2s 6d                       &    6d   $^3$D        \\
 25      & 1s$^2$ 2p($^2$P$^0$) 3d            &    3d'   $^3$D$^0$  &    
 50      & 1s$^2$ 2s 6g                       &    6g   $^3$G        \\
\hline
\end{tabular}
\end{table}

\begin{table}
\caption{Electronic configurations and term designations of our \NV\ model
ion. The level numbers correspond to the entries in the Grotrian diagram
Fig.~\ref{nv-grotrian}.}
\label{atom_lev_nv}
%\vspace{0.3cm}
\tabcolsep1.3mm
\begin{tabular}{rll | rll}
\hline 
\hline
\# & Configuration & Desig. & \#  & Configuration  & Desig. \\
\hline
  1      & 1s$^2$ 2s    &    2s $^2$S    &     
   15    & 1s$^2$ 6s    &    6s   $^2$S             \\
  2      & 1s$^2$ 2p    &    2p $^2$P$^0$    &      
 16      & 1s$^2$ 6p    &    6p   $^2$P$^0$        \\
  3      & 1s$^2$ 3s    &    3s $^2$S    &      
 17      & 1s$^2$ 6d    &    6d   $^2$D        \\
  4      & 1s$^2$ 3p    &    3p $^2$P$^0$  &       
18       & 1s$^2$ 6f    &    6f   $^2$F$^0$        \\
  5      & 1s$^2$ 3d    &    3d $^2$D  &      
 19      & 1s$^2$ 6g    &    6g   $^2$G        \\
  6      & 1s$^2$ 4s    &    4s $^2$S  &      
 20      & 1s$^2$ 6h    &    6h   $^2$H$^0$        \\
  7      & 1s$^2$ 4p    &    4p $^2$P$^0$  &      
 21      & 1s$^2$ 7s    &    7s   $^2$S        \\
  8      & 1s$^2$ 4d    &    4d   $^2$D  &      
 22      & 1s$^2$ 7p    &    7p   $^2$P$^0$        \\
  9      & 1s$^2$ 4f    &    4f   $^2$F$^0$  &      
23       & 1s$^2$ 7d    &    7d   $^2$D        \\
 10      & 1s$^2$ 5s    &    5s $^2$S  &      
  24     & 1s$^2$ 7f    &    7f   $^2$F$^0$        \\
 11      & 1s$^2$ 5p    &    5p   $^2$P$^0$    &    
 25      & 1s$^2$ 7g    &    7g   $^2$G        \\
 12      & 1s$^2$ 5d    &    5d   $^2$D    &    
26       & 1s$^2$ 7h    &    7h   $^2$H$^0$        \\
 13      & 1s$^2$ 5f    &    5f $^2$F$^0$  &      
27       & 1s$^2$ 7i    &    7i   $^2$I        \\
 14      & 1s$^2$ 5g    &    5g   $^2$G   \\
\hline
\end{tabular}
\end{table}

\section{Tests of the \NII\ model ion}

\begin{figure*}
\centerline
{\includegraphics{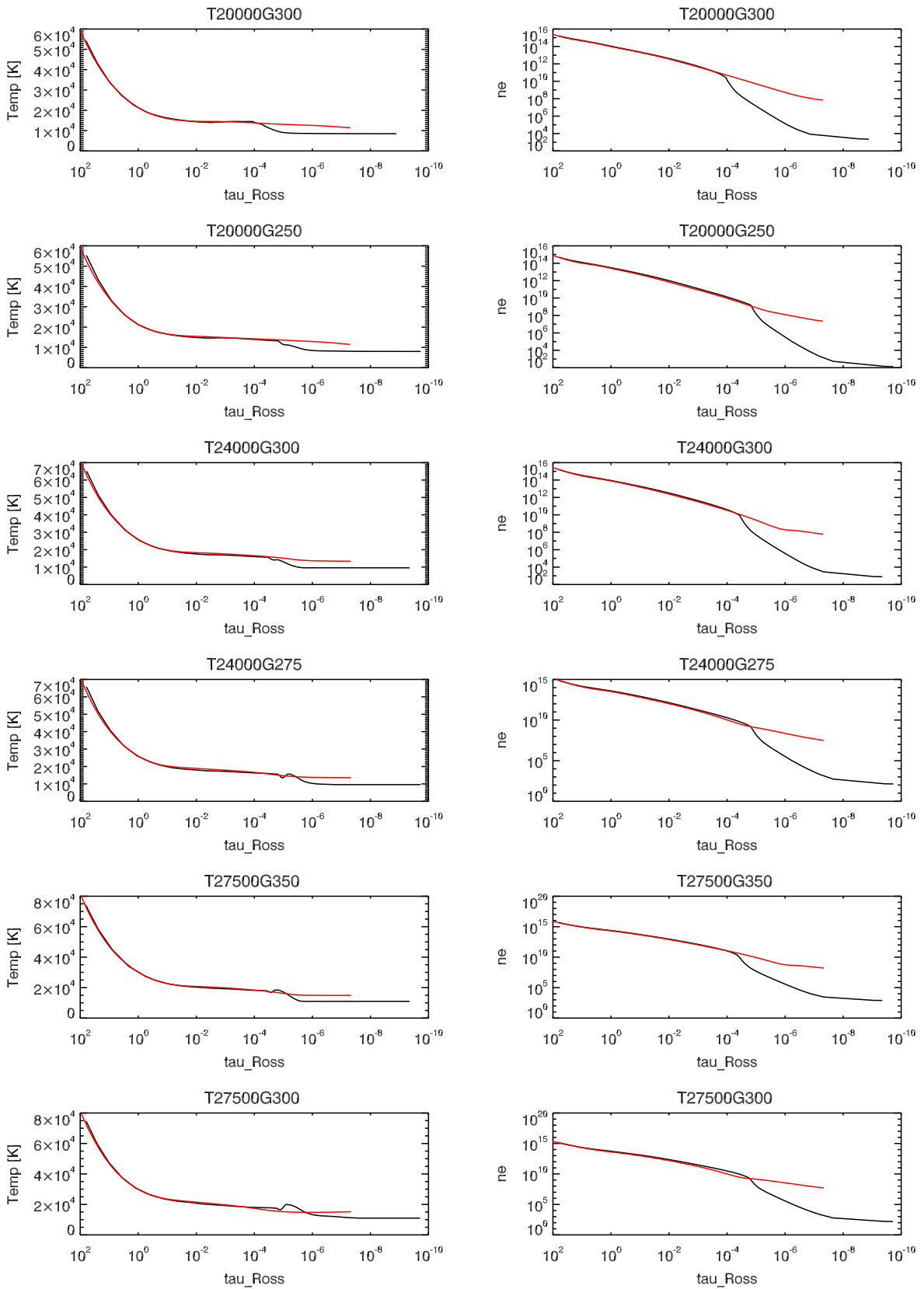}}
%\vspace{-.5cm}
\caption{ 
{\sc fastwind} (black) vs. {\sc tlusty} (red): comparison
of electron temperature and electron density as a function of Rosseland
optical depth, for the six models corresponding
to Figs.~\ref{comp-tlusty-2030} to \ref{comp-tlusty-2730}.} 
\label{mod-structure}
\end{figure*}

\begin{figure*}
\centerline{\includegraphics[width=11cm, angle=90]{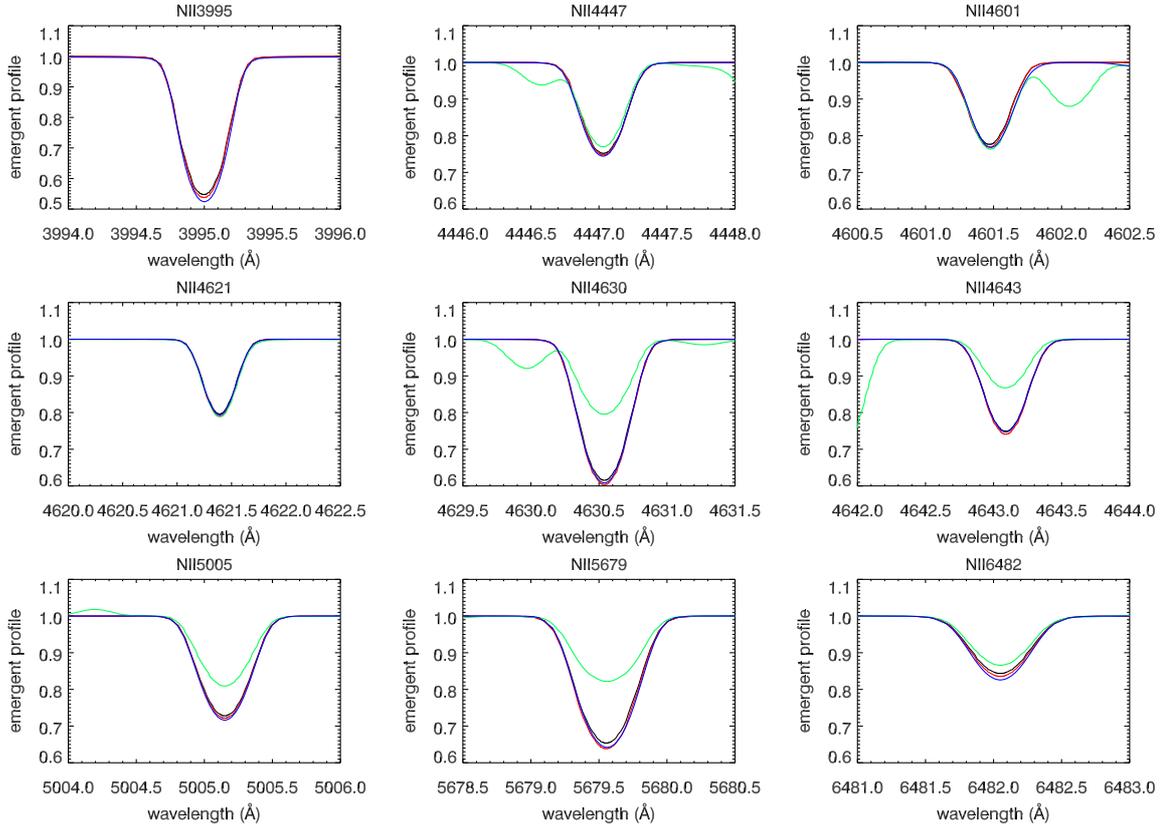}}
%\vspace{-.5cm}
\caption{Comparison of important optical \NII\ line profiles for a model
with \Teff\ = 20~kK and \logg\ = 3.0, for models -- see Table~\ref{nii_tests}
-- FW (black), FW2 (red), TL (green) and Prz (blue). Note that \NII\
$\lambda$ 3995 is not present in the BSTAR2006 grid.}
\label{comp-tlusty-2030}
\end{figure*}

\begin{figure*}
\centerline{\includegraphics[width=11cm, angle=90]{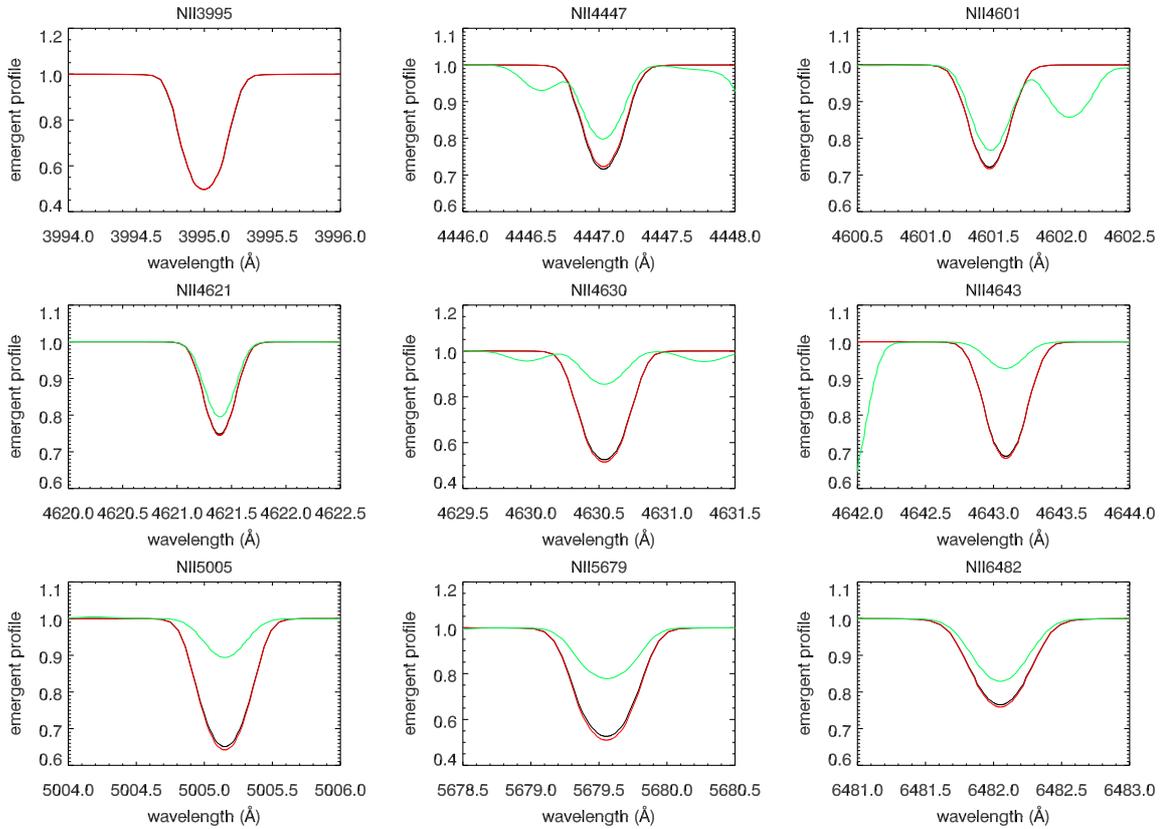}}
%\vspace{-.5cm}
\caption{As Fig.~\ref{comp-tlusty-2030}, but for \Teff\ = 20~kK and
\logg\ = 2.5, for models FW (black), FW2 (red), and TL(green).}
\label{comp-tlusty-2025}
\end{figure*}

\begin{figure*}
\centerline{\includegraphics[width=11cm, angle=90]{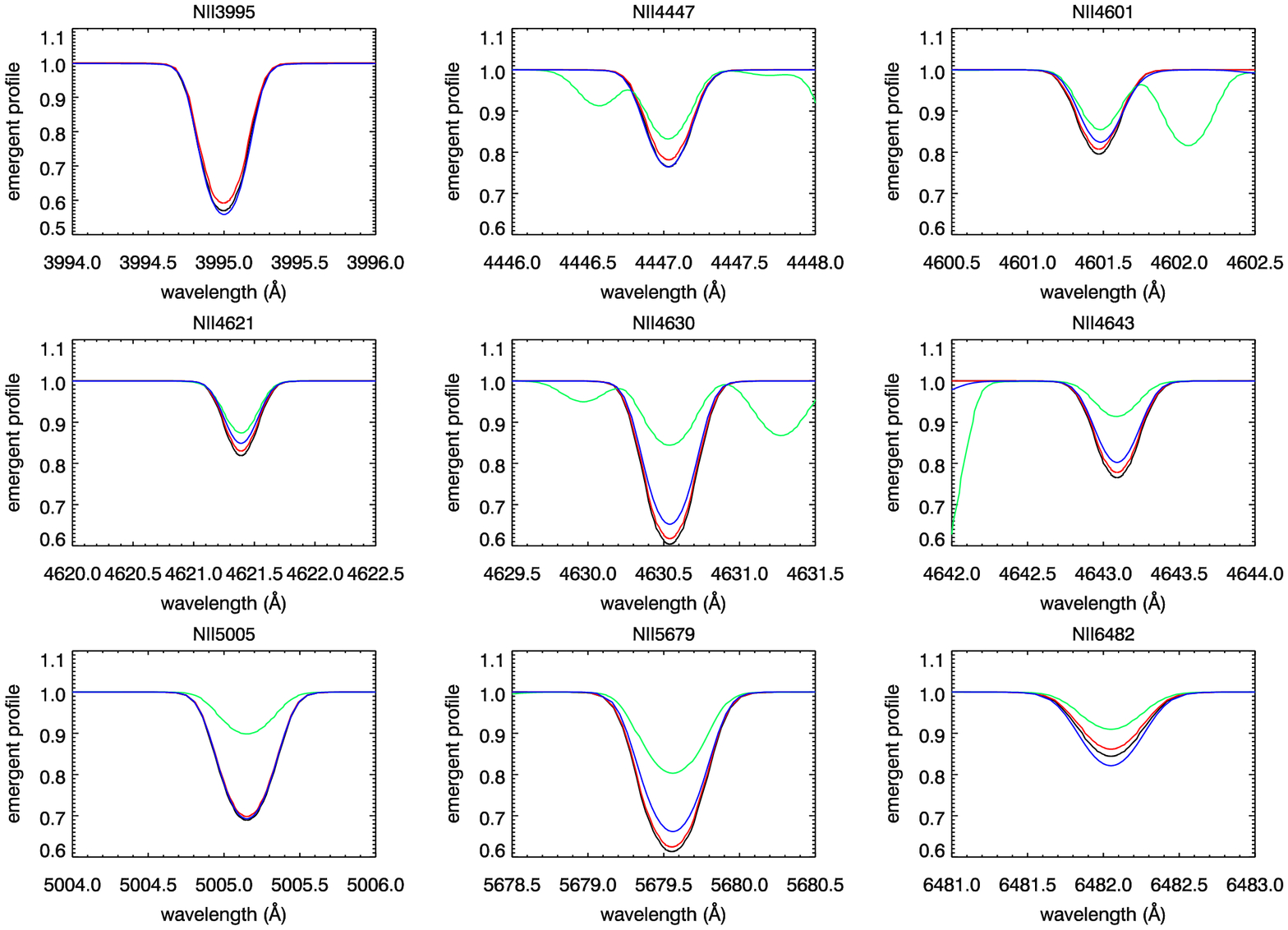}}
%\vspace{-.5cm}
\caption{As Fig.~\ref{comp-tlusty-2030}, but for \Teff\ = 24~kK and
\logg\ = 3.0.}
\label{comp-tlusty-2430}
\end{figure*}

\begin{figure*}
\centerline{\includegraphics[width=11cm, angle=90]{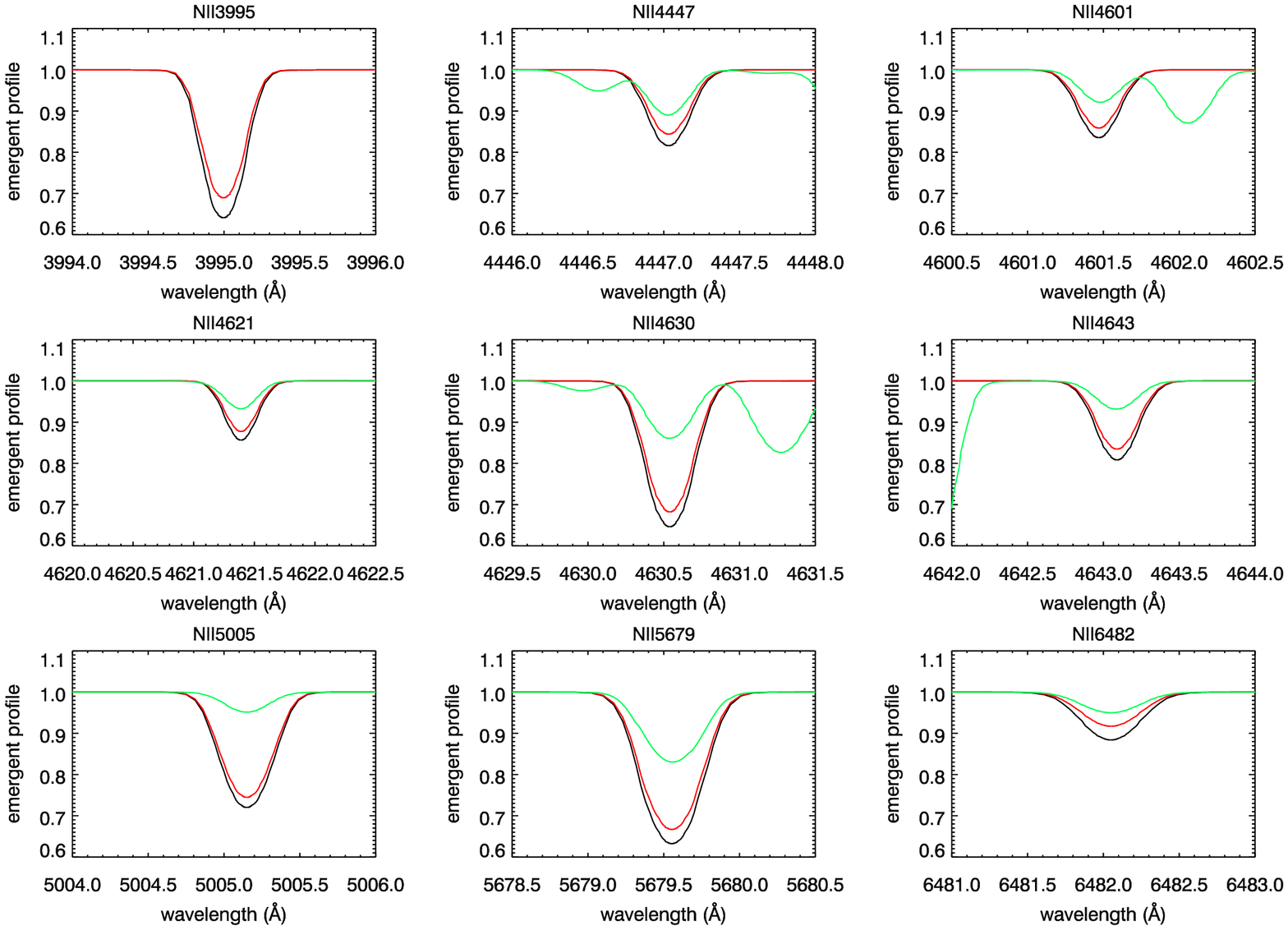}}
%\vspace{-.5cm}
\caption{As Fig.~\ref{comp-tlusty-2025}, but for \Teff\ = 24~kK and
\logg\ = 2.75.}
\label{comp-tlusty-2427}
\end{figure*}

\begin{figure*}
\centerline{\includegraphics[width=11cm, angle=90]{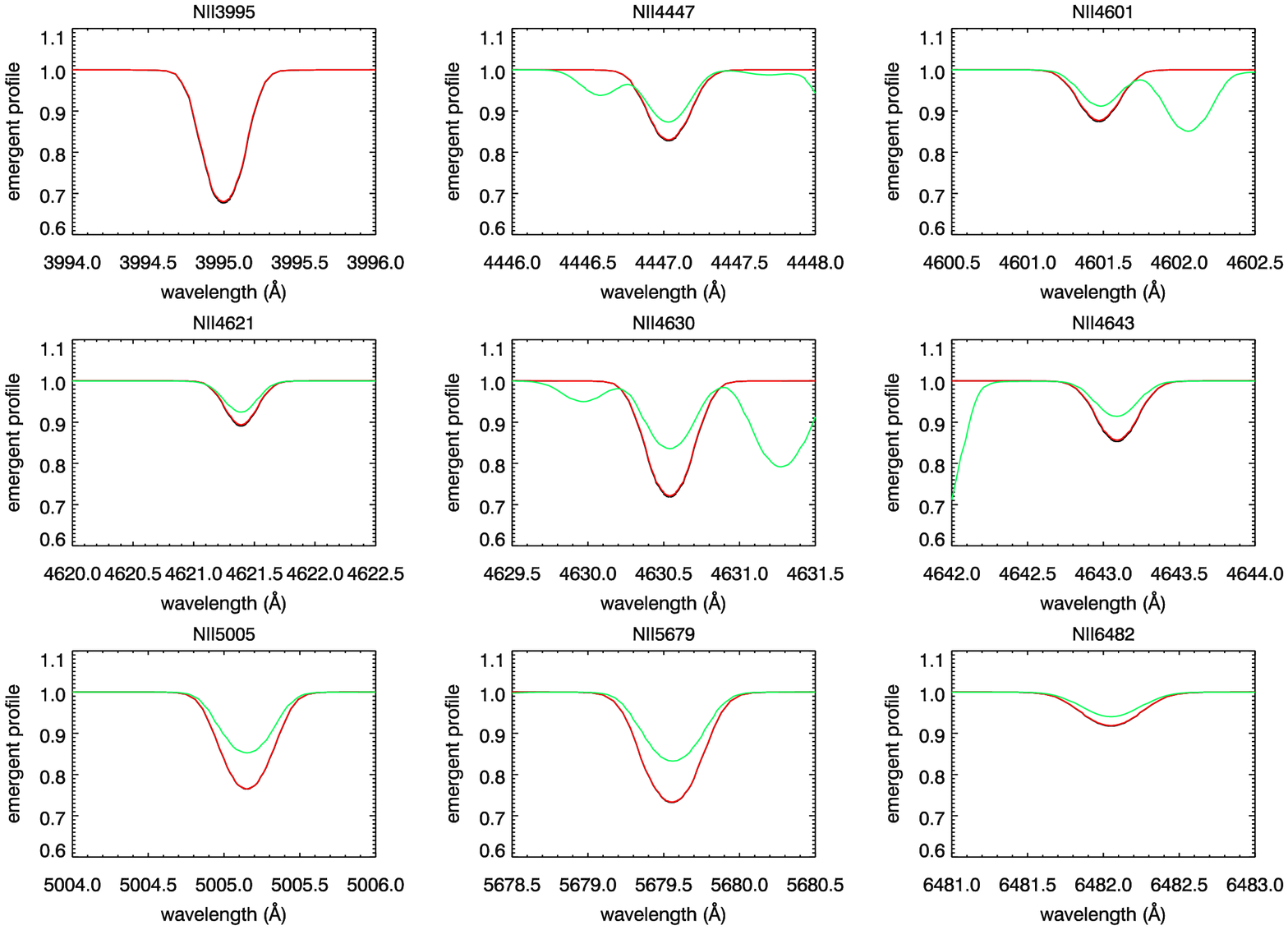}}
%\vspace{-.5cm}
\caption{As Fig.~\ref{comp-tlusty-2025}, but for \Teff\ = 27.5~kK and
\logg\ = 3.5.}
\label{comp-tlusty-2735}
\end{figure*}

\begin{figure*}
\centerline{\includegraphics[width=11cm, angle=90]{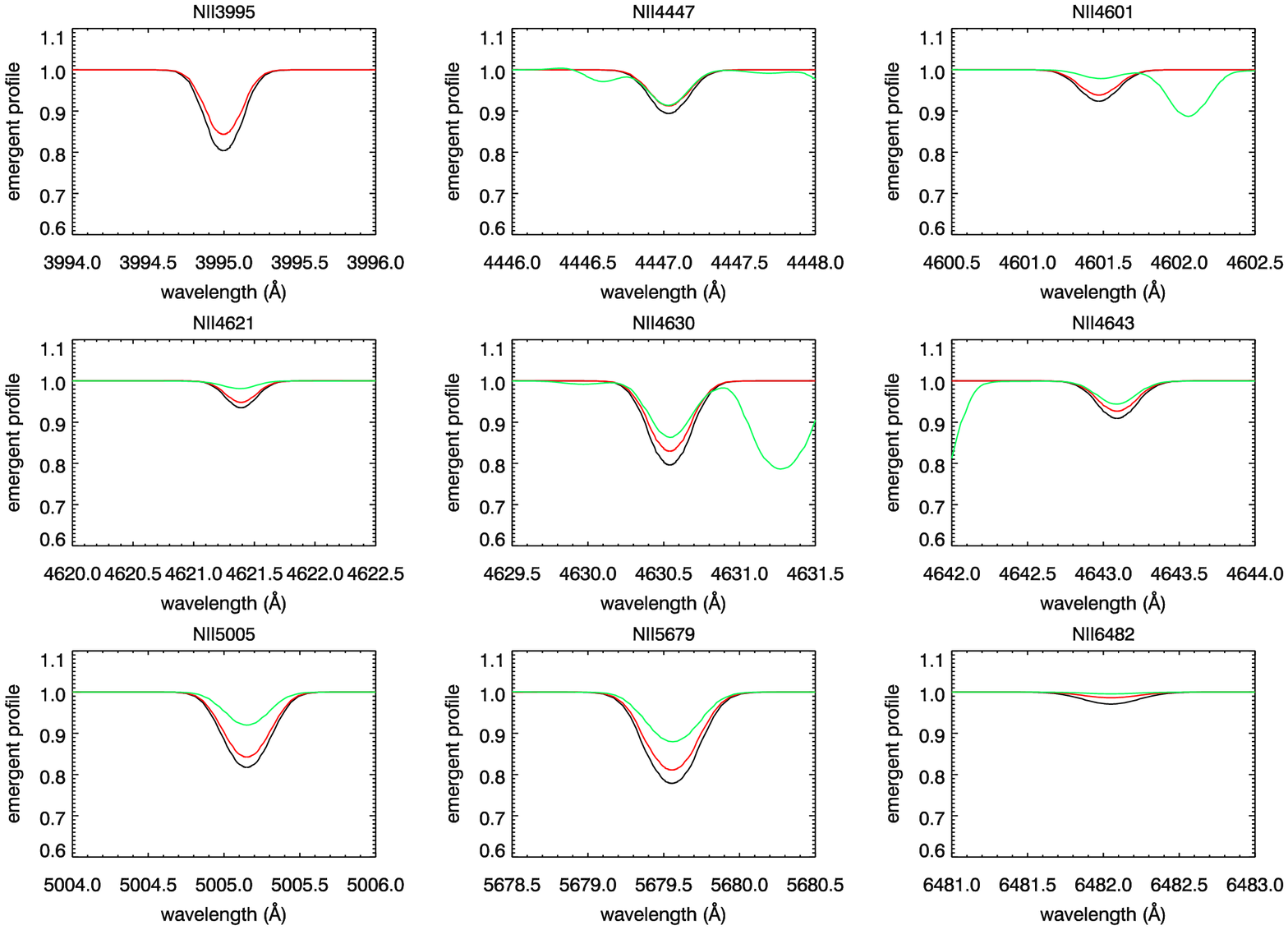}}
%\vspace{-.5cm}
\caption{As Fig.~\ref{comp-tlusty-2025}, but for \Teff\ = 27.5~kK and
\logg\ = 3.0.}
\label{comp-tlusty-2730}
\end{figure*}

Figures~\ref{mod-structure} to \ref{comp-tlusty-2730} refer to
tests of our \NII\ model ion, as described in Sect.~\ref{tests_nii}.
Figure~\ref{mod-structure} compares electron temperatures and densities
for B-star parameters calculated by {\sc fastwind} and {\sc
tlusty}, whilst Figs.~\ref{comp-tlusty-2030} to \ref{comp-tlusty-2730}
compare corresponding synthetic \NII\ line profiles from these two
codes and from calculations by Przybilla et al. (priv. comm.).


\section{Line fits for individual objects}
\label{fits_ap}

\begin{figure*}
\center
%\resizebox{\hsize}{!}
  {\includegraphics[totalheight=0.45\textheight]{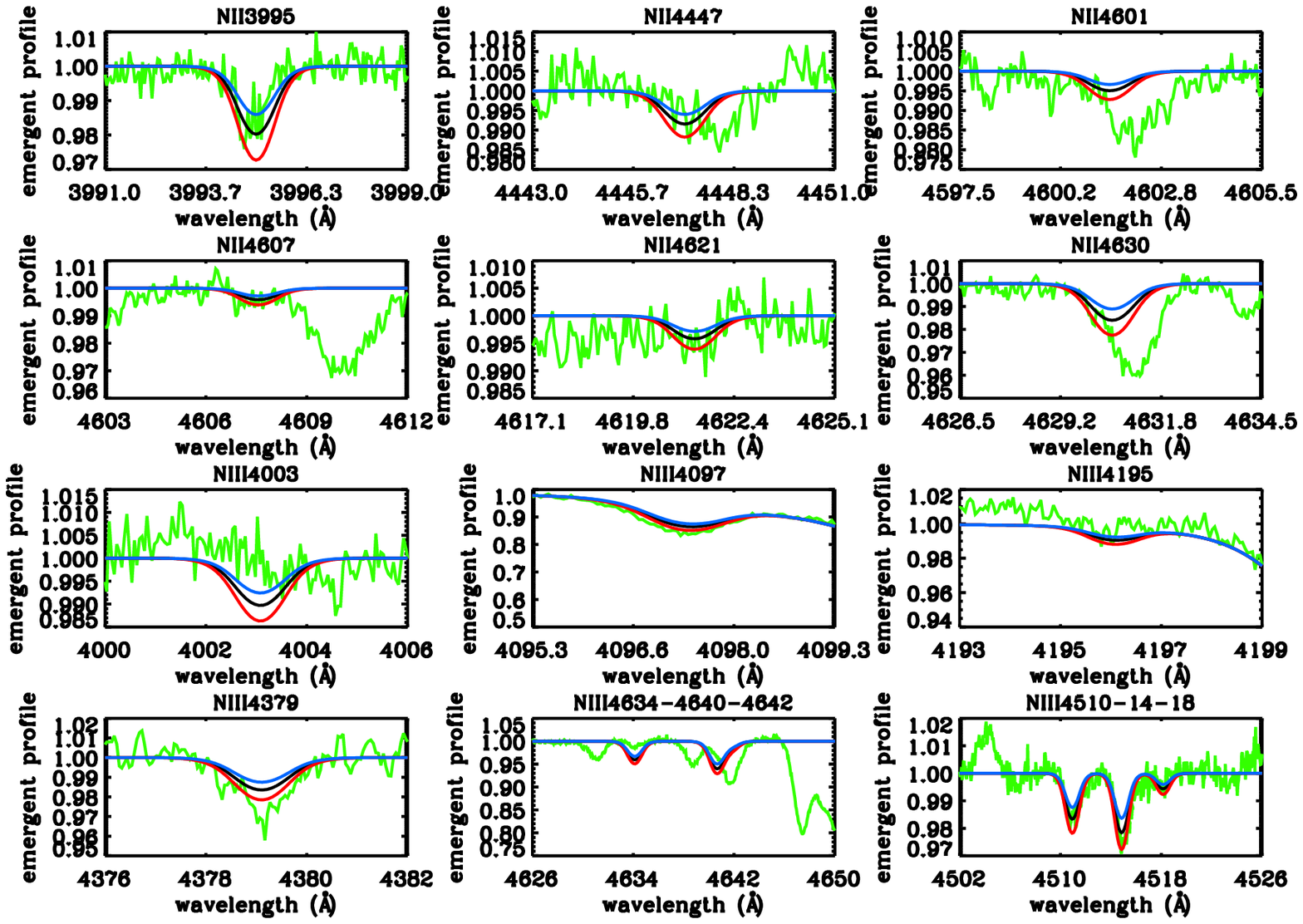}}
%\vspace{-0.5cm}
\caption{N11-029 -- O9.7 Ib. Observed (green) and best-fitting optical
nitrogen spectrum (black). Blue and red spectra correspond to
synthetic line profiles with [N] at the lower and upper limit,
respectively. For details, see Sect.~\ref{comments}.} 
\label{N11-029}
\end{figure*}

\begin{figure*}
\center
%\resizebox{\hsize}{!}
  {\includegraphics[totalheight=0.45\textheight]{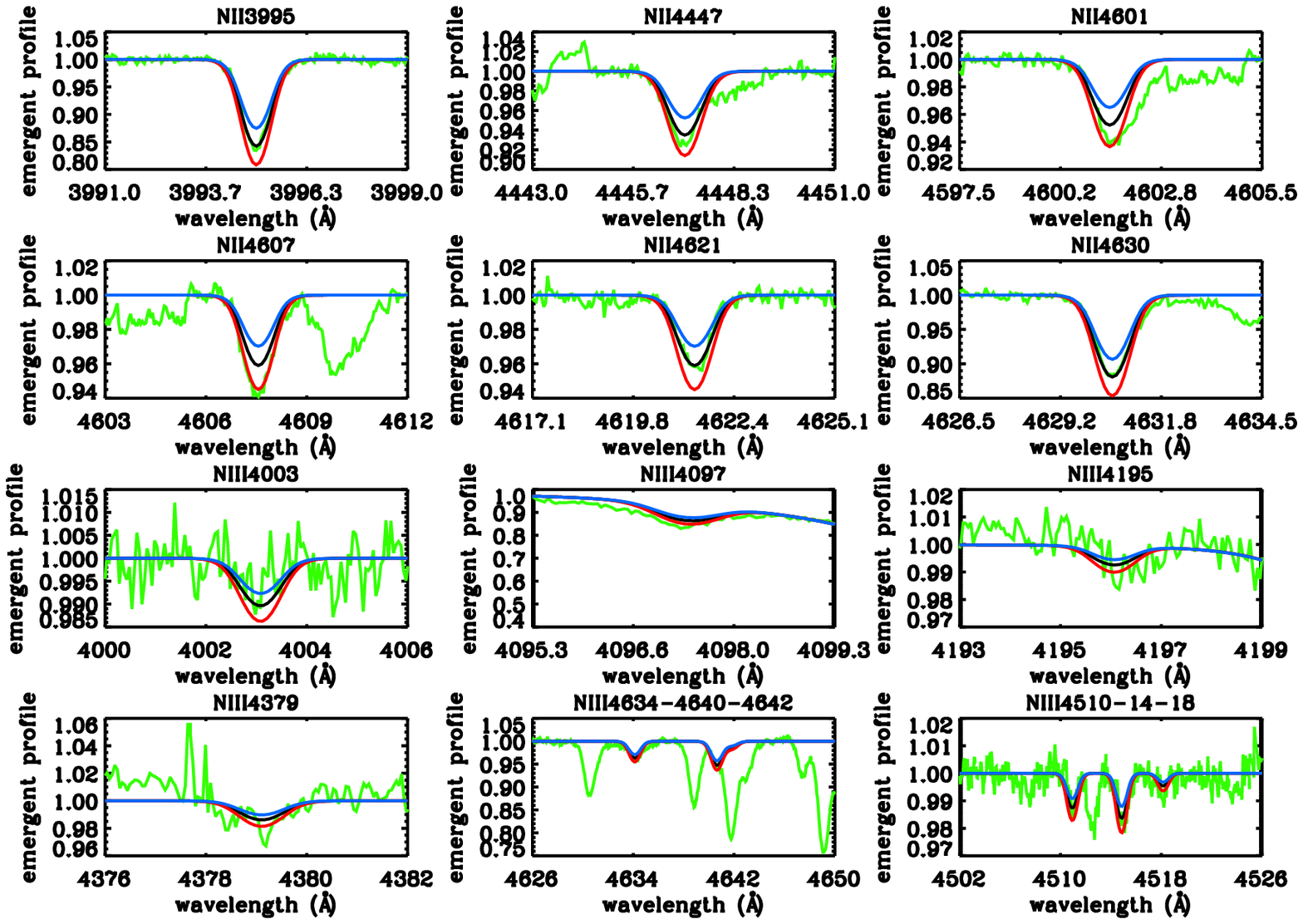}}
%\vspace{-0.5cm}
\caption{N11-036 -- B0.5 Ib} 
\label{N11-036}
\end{figure*}

\begin{figure*}
\center
%\resizebox{\hsize}{!}
{\includegraphics[totalheight=0.45\textheight]{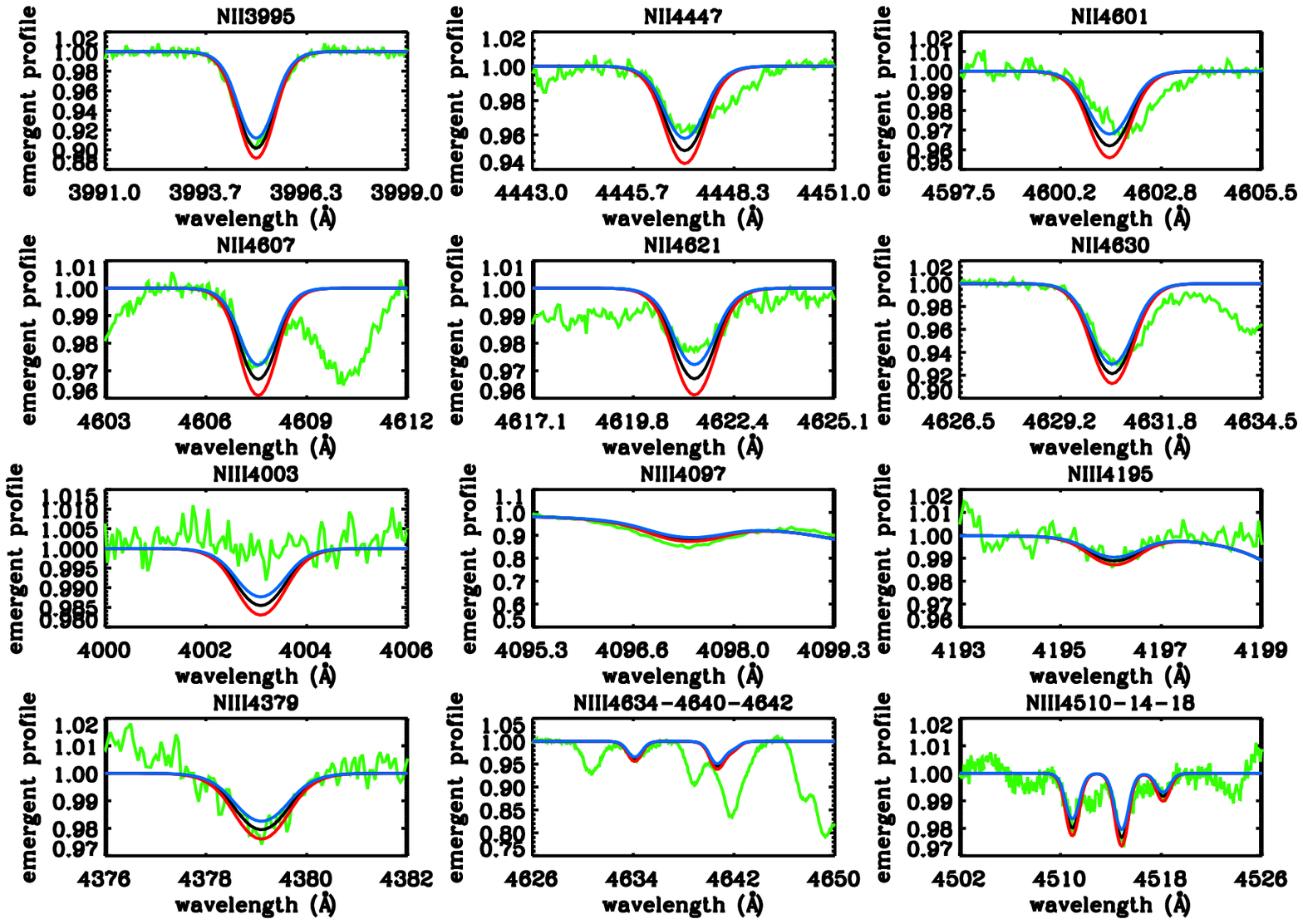}}
%\vspace{-0.5cm}
\caption{N11-008 -- B0.7 Ia} 
\label{N11-008}
\end{figure*}

\begin{figure*}
\center
%\resizebox{\hsize}{!}
  {\includegraphics[totalheight=0.45\textheight]{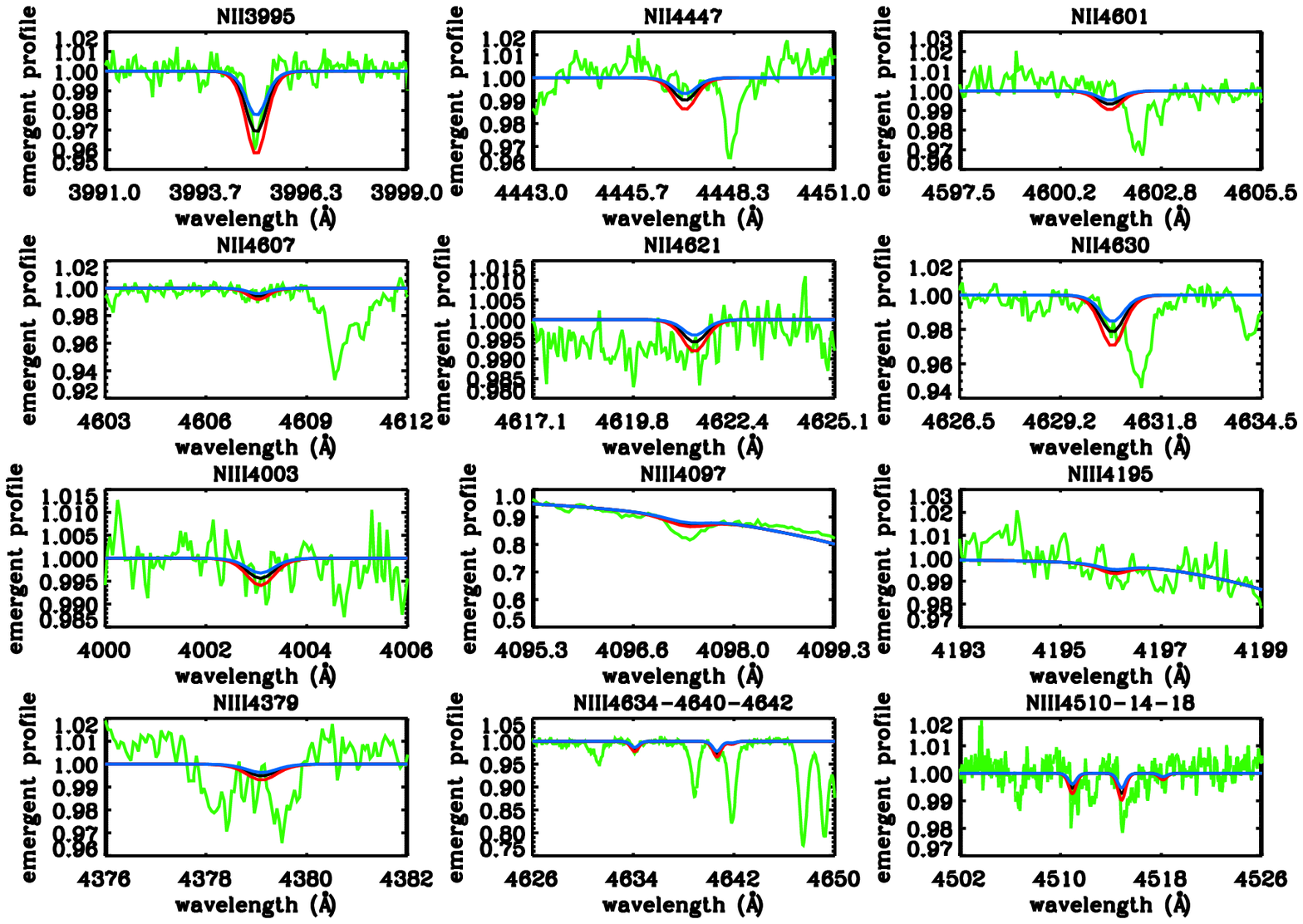}}
%\vspace{-0.5cm}
\caption{N11-042 -- B0 III} 
\label{N11-042}
\end{figure*}

\begin{figure*}
\center
%\resizebox{\hsize}{!}
  {\includegraphics[totalheight=0.45\textheight]{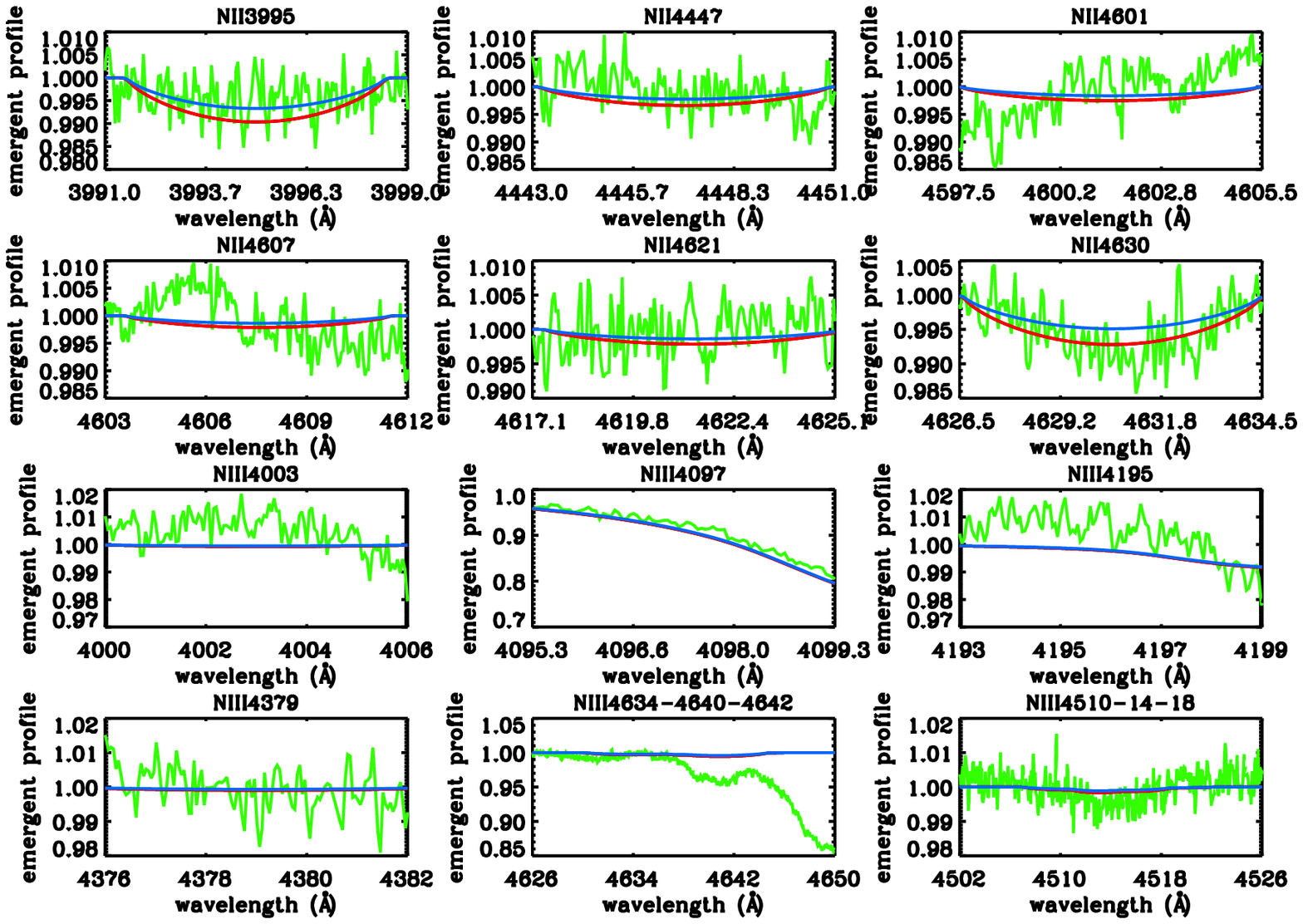}}
%\vspace{-0.5cm}
\caption{N11-033 -- B0 IIIn} 
\label{N11-033}
\end{figure*}

\begin{figure*}
\center
%\resizebox{\hsize}{!}
 {\includegraphics[totalheight=0.45\textheight]{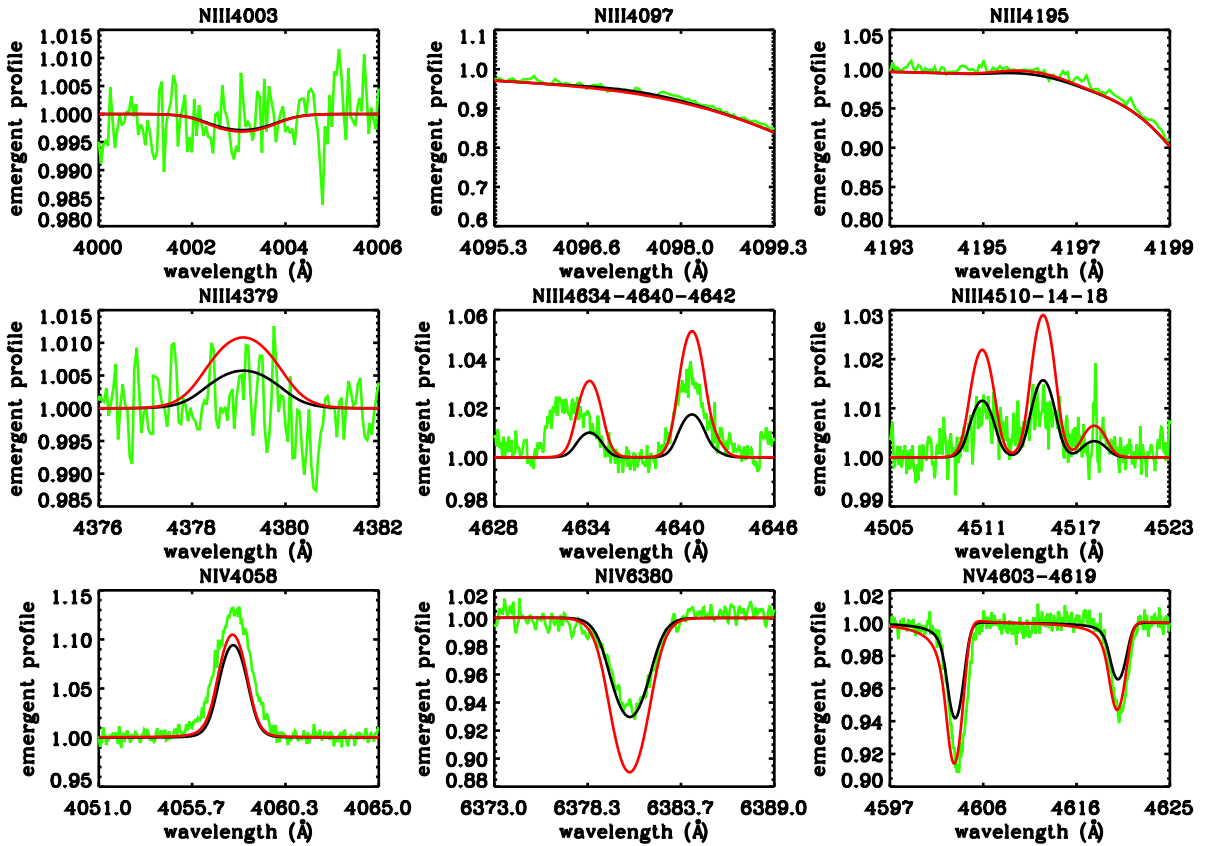}}
%\vspace{-0.5cm}
\caption{N11-026 -- O2 III(f$^*$)} 
\label{N11-026}
\end{figure*}

\begin{figure*}
\center
 %\resizebox{\hsize}{!}
 {\includegraphics[totalheight=0.45\textheight]{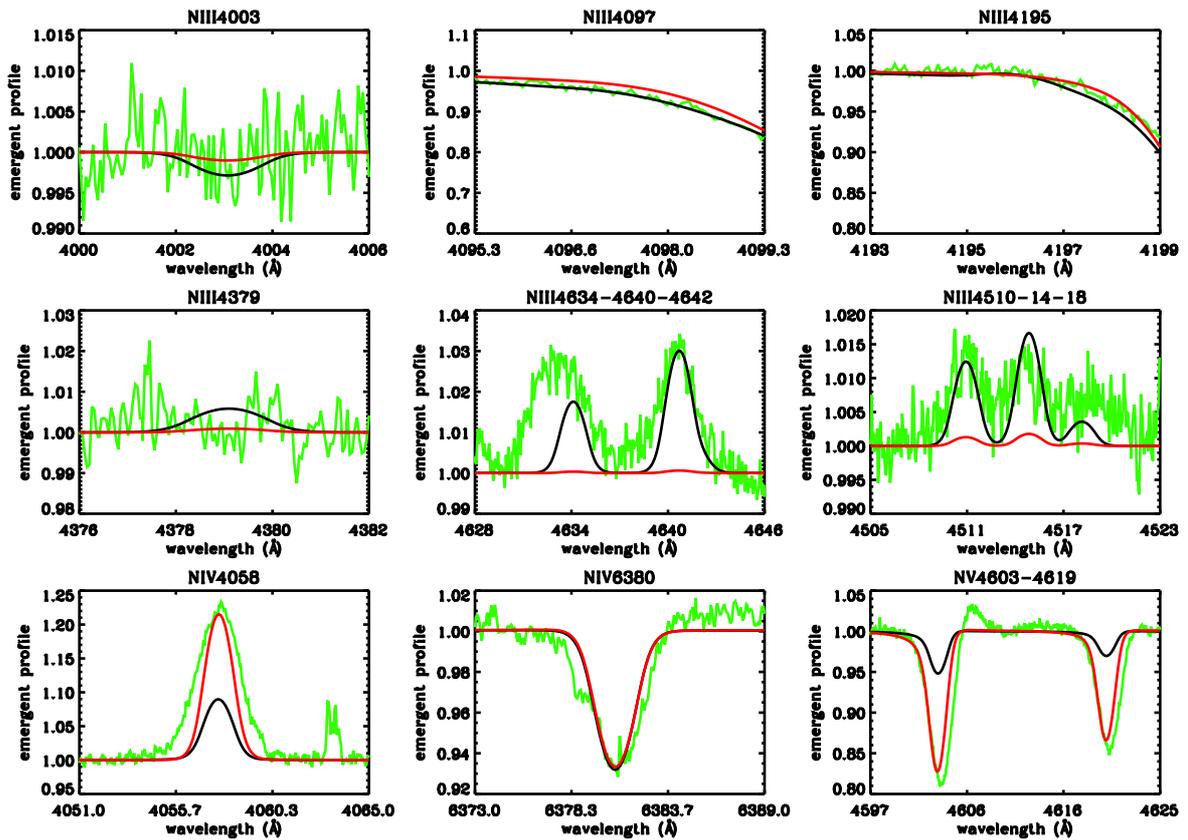}}
 %\vspace{-0.5cm}
\caption{N11-031 -- ON2 III(f$^*$). Black: cooler solution,
supported by \HeI$\lambda$4471, \NIII\ and \NIV\nivab.  Red: hotter
solution, supported by the \NIV/\NV\ lines (see Sect.~\ref{comments}).} 
\label{N11-031}
\end{figure*}

\begin{figure*}
\center
%\resizebox{\hsize}{!}
  {\includegraphics[totalheight=0.45\textheight]{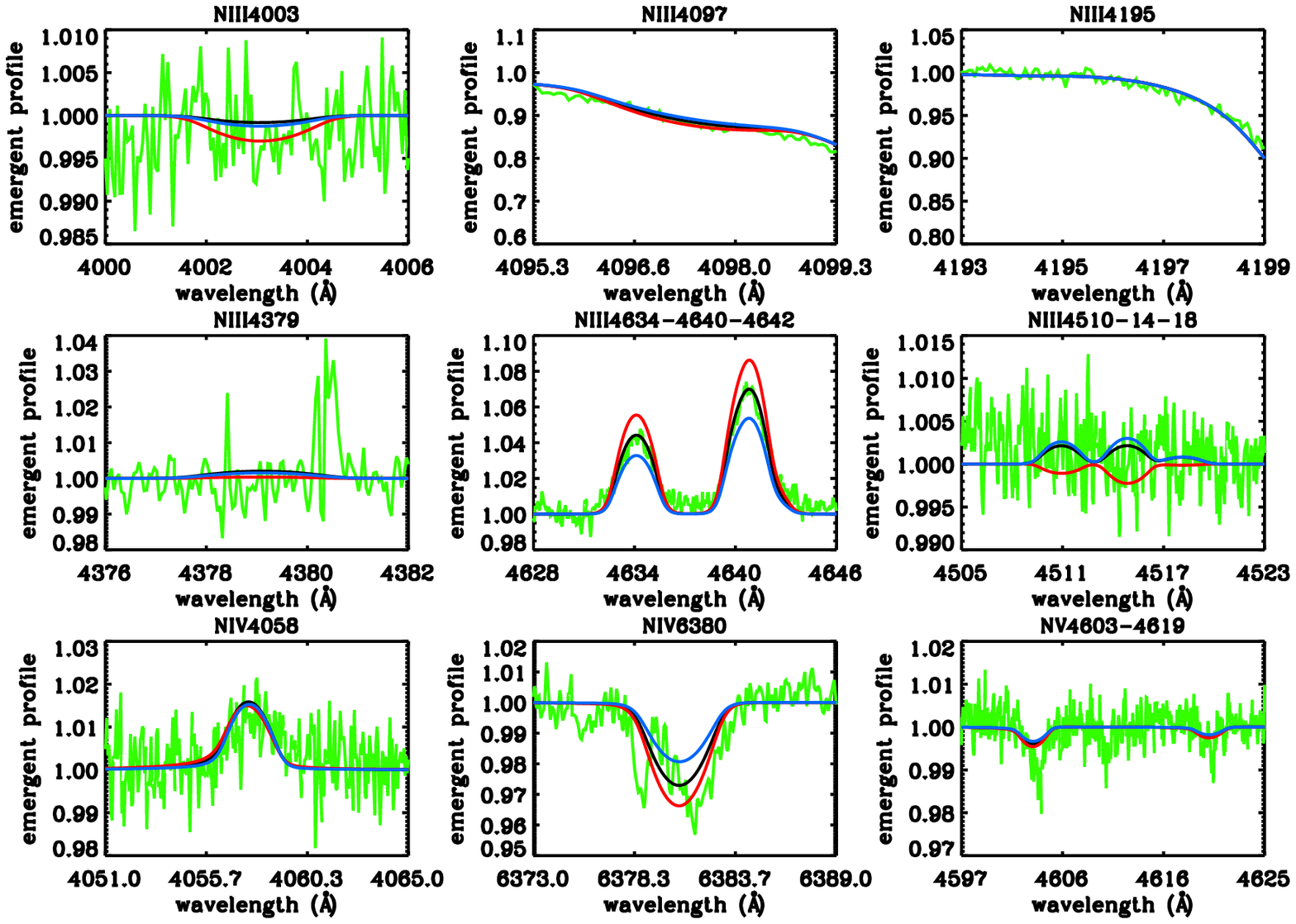}}
%\vspace{-0.5cm}
\caption{N11-038 -- O5 II(f$^+$)} 
\label{N11-038}
\end{figure*}

\begin{figure*}
\center
%\resizebox{\hsize}{!}
  {\includegraphics[totalheight=0.45\textheight]{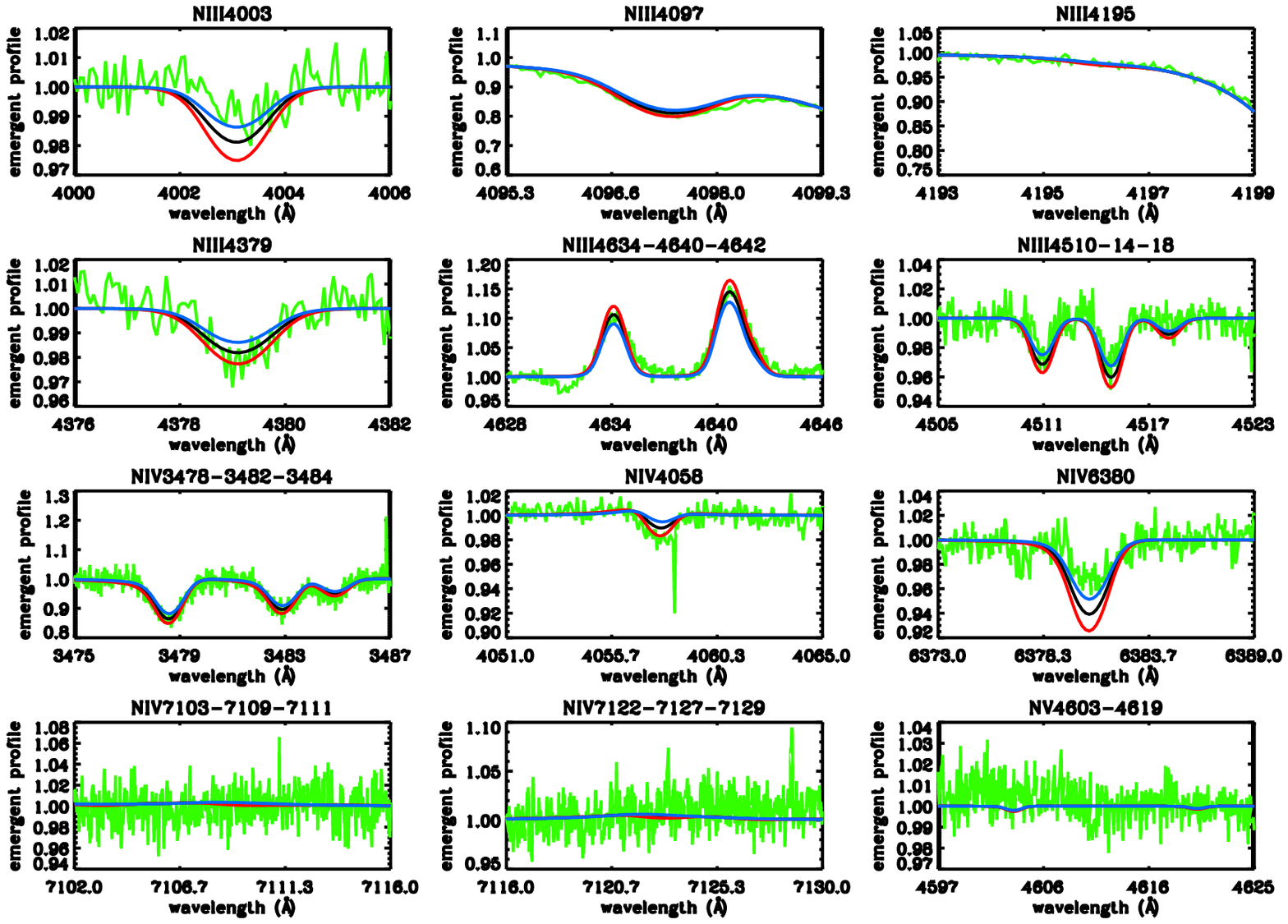}}
%\vspace{-0.5cm}
\caption{Sk--66$^{\circ}$ 100 -- O6 II(f)} 
\label{SK66100}
\end{figure*}

\begin{figure*}
\center
%\resizebox{\hsize}{!}
  {\includegraphics[totalheight=0.45\textheight]{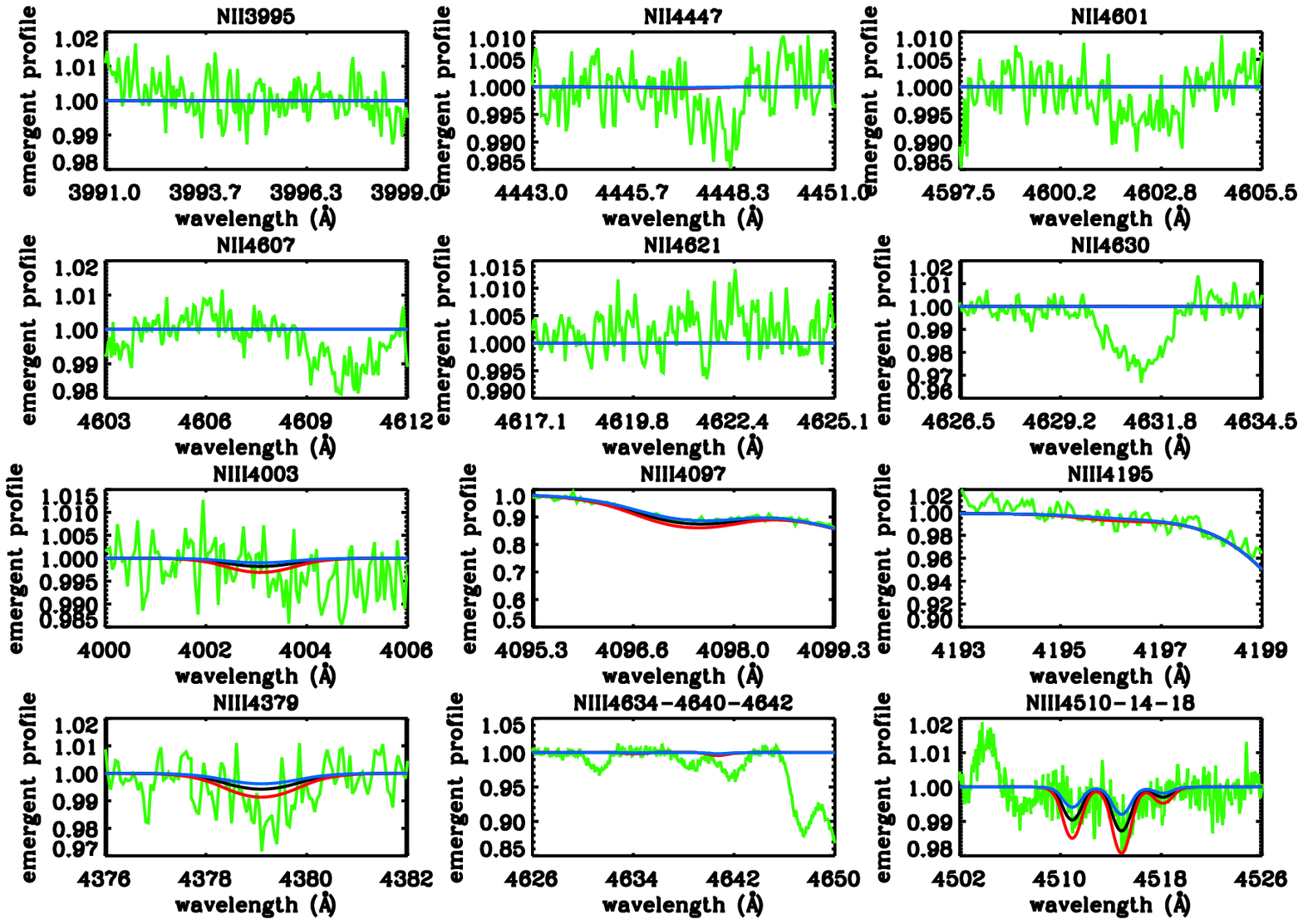}}
%\vspace{-0.5cm}
\caption{ N11-045 -- O9 III} 
\label{N11-045}
\end{figure*}

\begin{figure*}
\center
%\resizebox{\hsize}{!}
  {\includegraphics[totalheight=0.45\textheight]{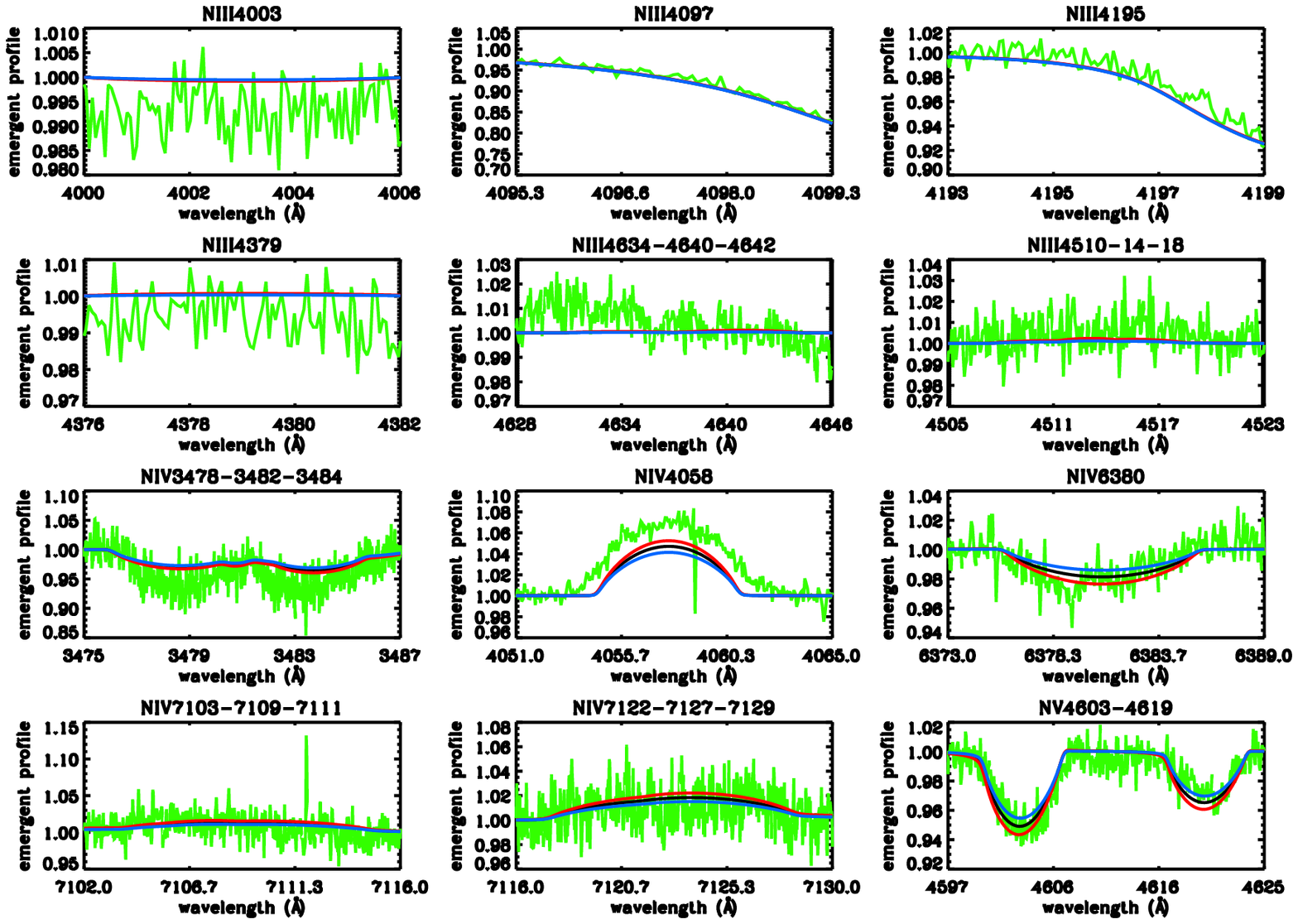}}
%\vspace{-0.5cm}
\caption{BI253 -- O2 V((f$^*$))} 
\label{BI253}
\end{figure*}

\begin{figure*}
\center
%\resizebox{\hsize}{!}
  {\includegraphics[totalheight=0.45\textheight]{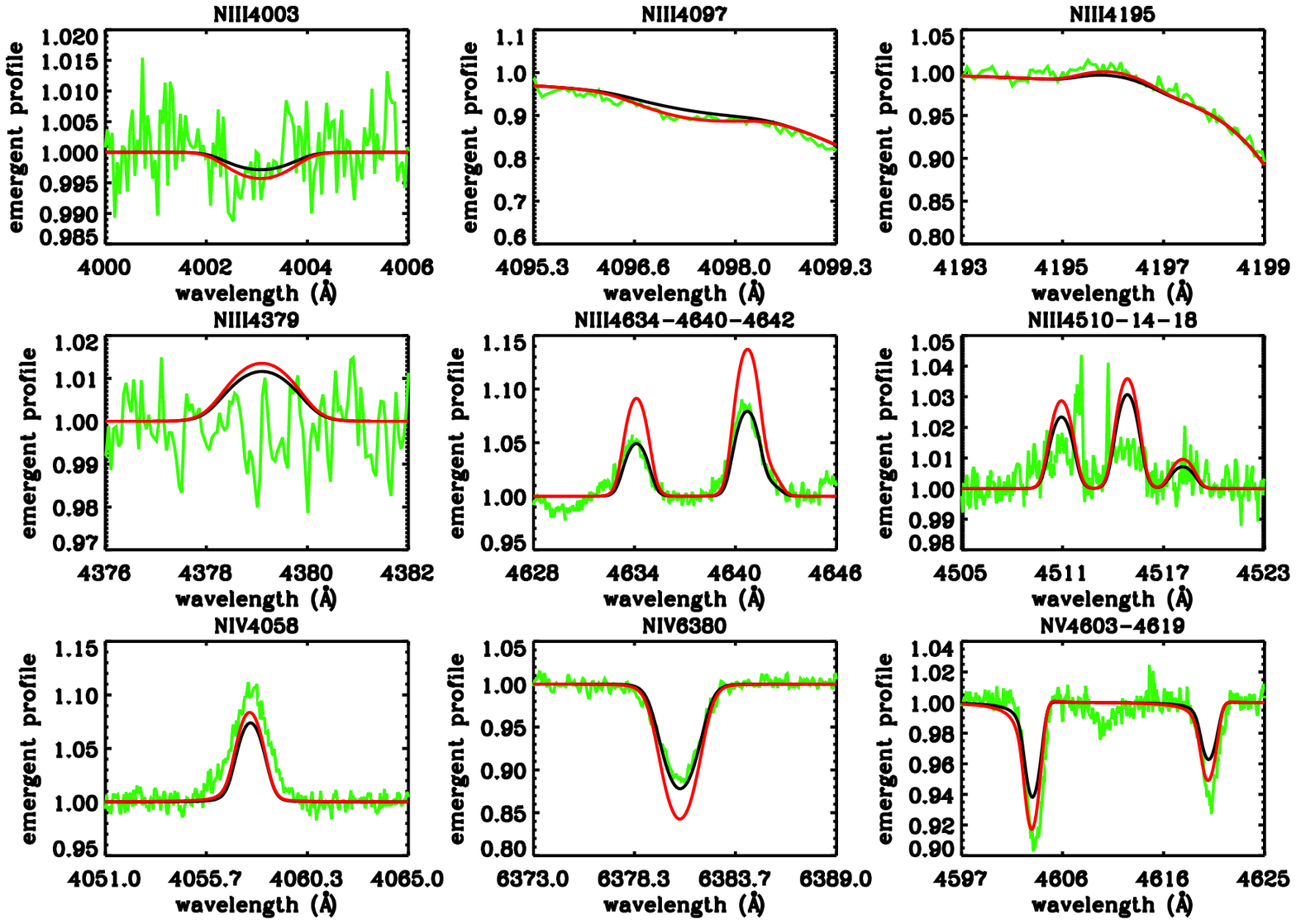}}
%\vspace{-0.5cm}
\caption{N11-060 -- O3 V((f$^*$))} 
\label{N11-060}
\end{figure*}

\begin{figure*}
\center
%\resizebox{\hsize}{!}
  {\includegraphics[totalheight=0.45\textheight]{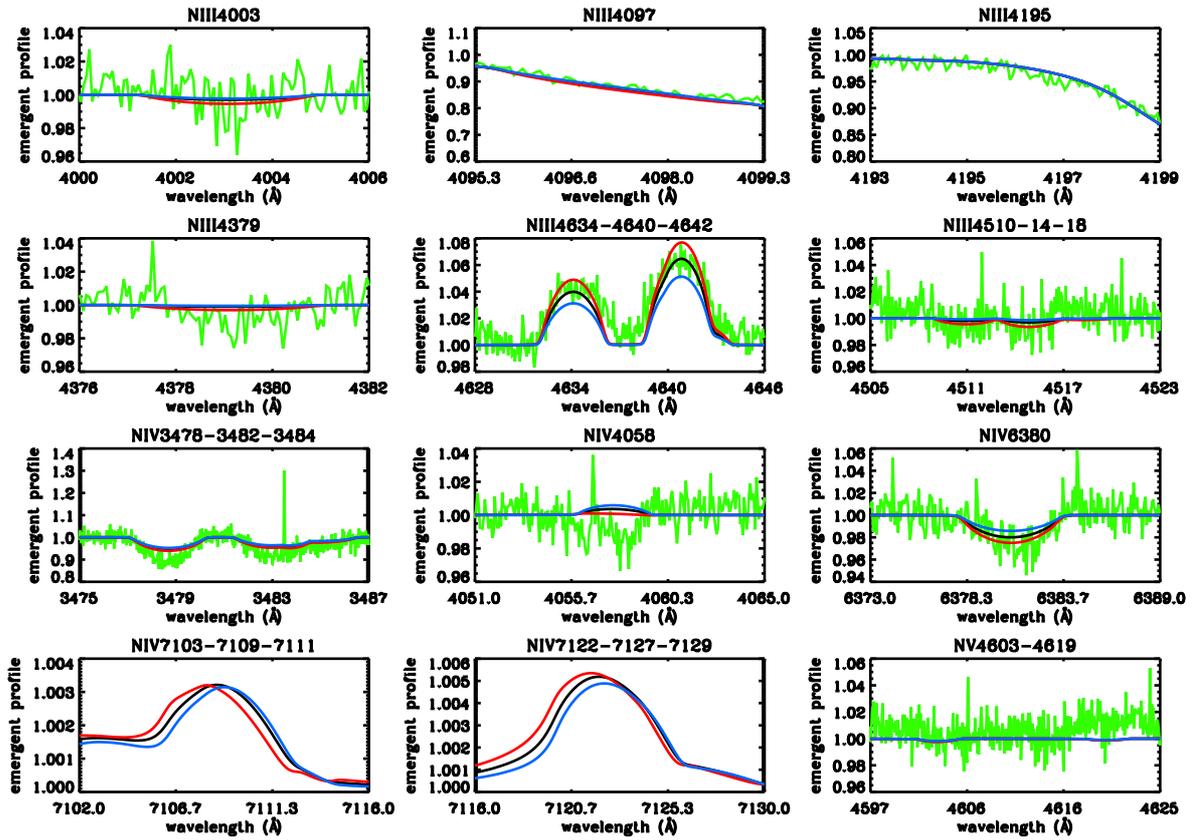}}
%\vspace{-0.5cm}
\caption{Sk--70$^{\circ}$ 69 -- O5 V((f)). For this star, the 
\NIV\ multiplet at 7103-7129~\AA\ has not been observed.}
\label{SK7069}
\end{figure*}

\begin{figure*}
\center
%\resizebox{\hsize}{!}
  {\includegraphics[totalheight=0.45\textheight]{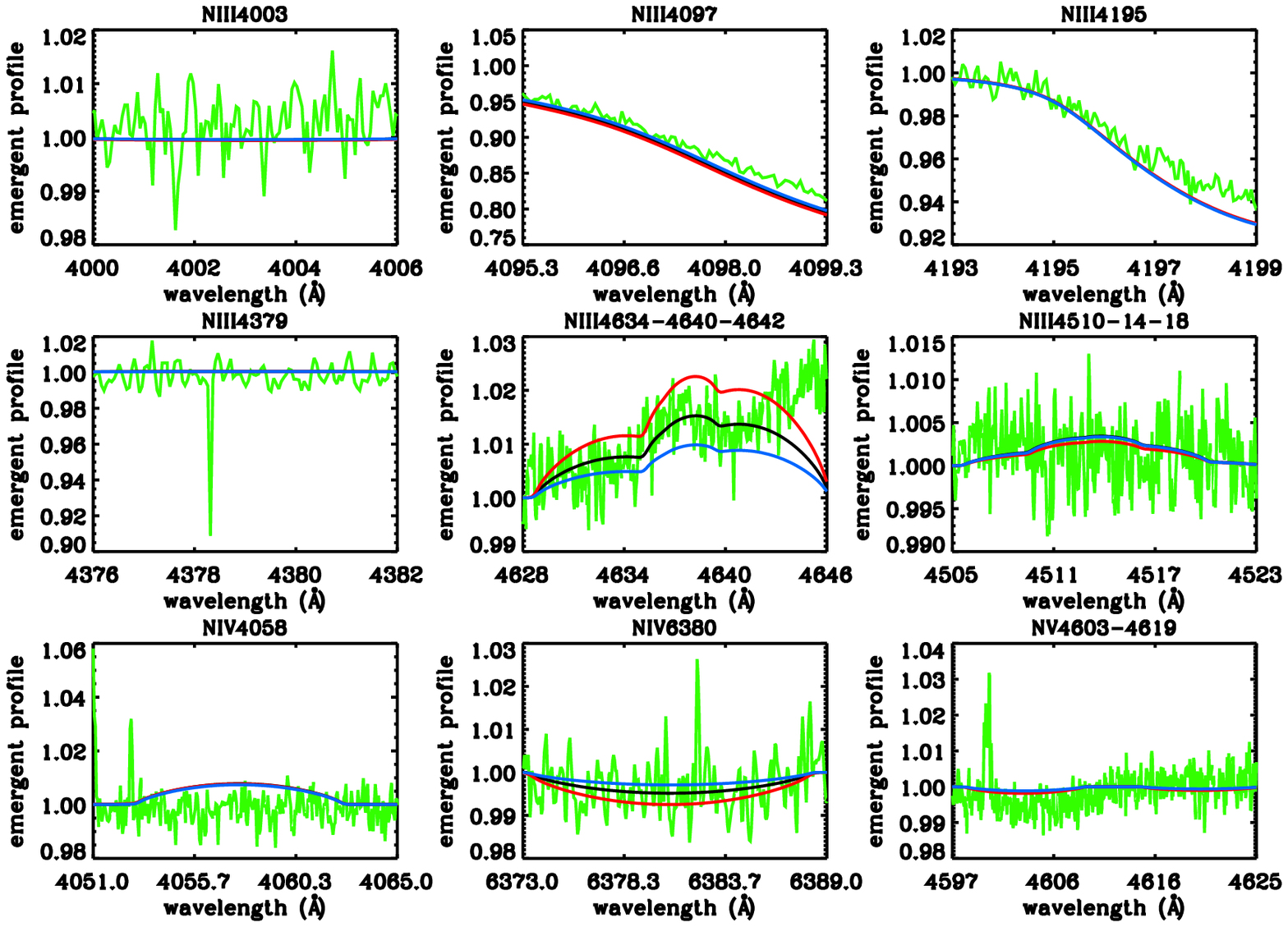}}
%\vspace{-0.5cm}
\caption{ N11-051-- O5 Vn((f))} 
\label{N11-051}
\end{figure*}

\begin{figure*}
\center
%\resizebox{\hsize}{!}
  {\includegraphics[totalheight=0.45\textheight]{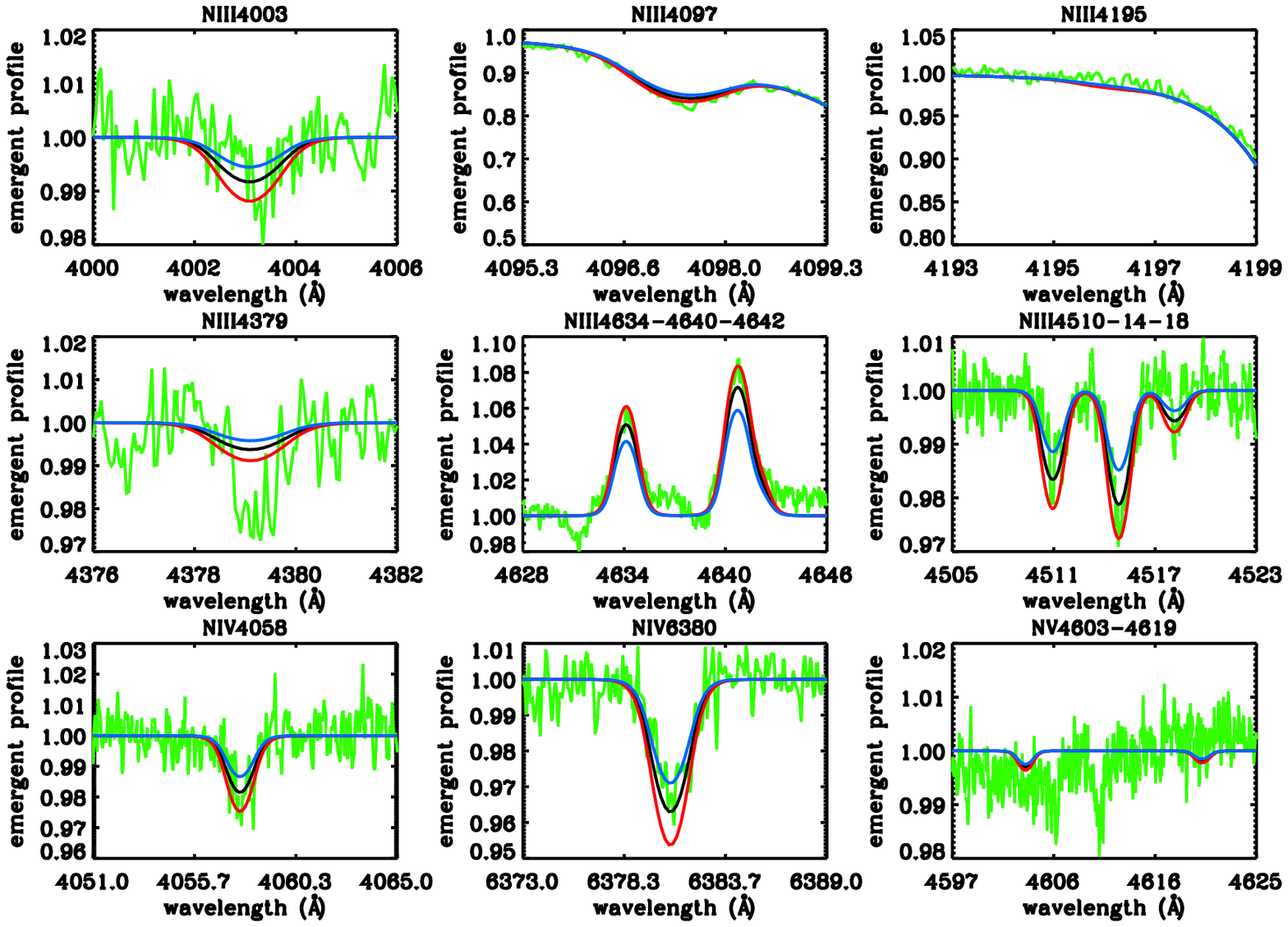}}
%\vspace{-0.5cm}
\caption{N11-058 -- O5.5 V((f))} 
\label{N11-058}
\end{figure*}

\begin{figure*}
\center
%\resizebox{\hsize}{!}
  {\includegraphics[totalheight=0.45\textheight]{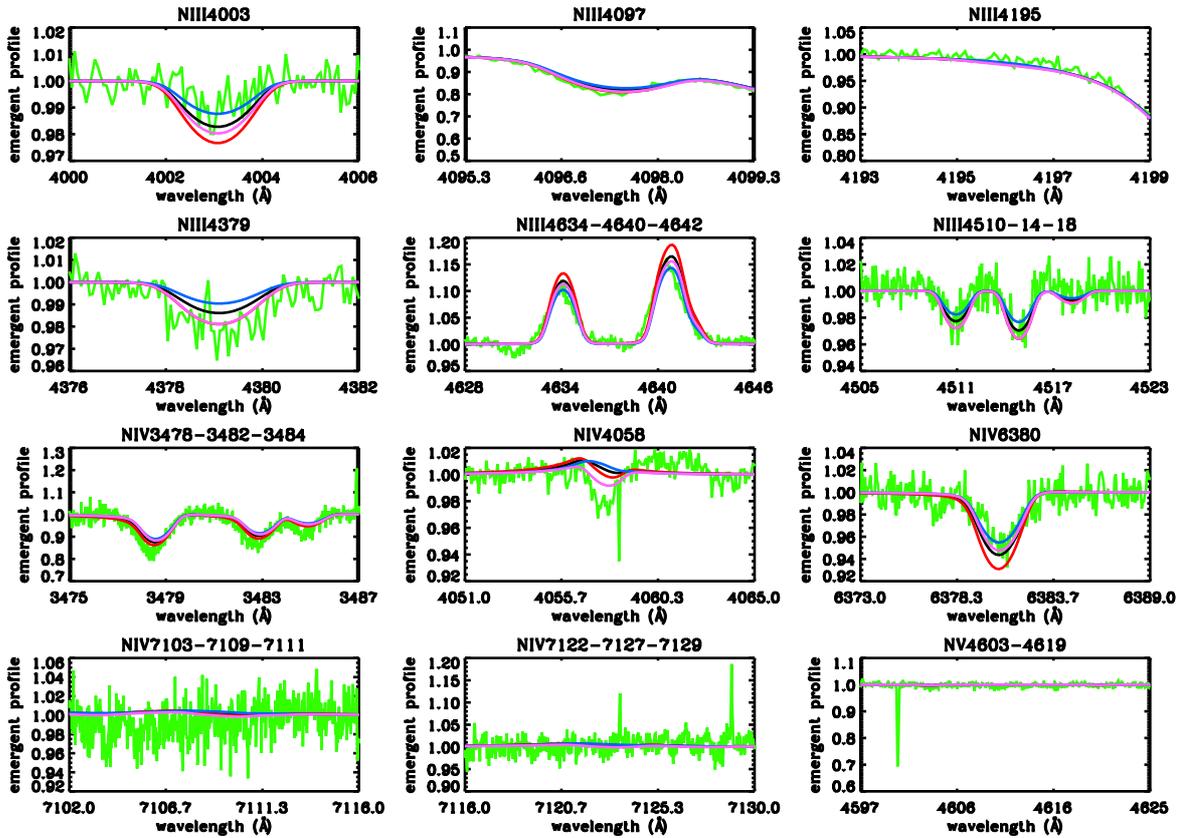}}
%\vspace{-0.5cm}
\caption{Sk--66$^{\circ}$ 18 -- O6 V((f)). Magenta spectra
correspond to a weakly clumped model. For details, see Sect.~\ref{comments}.} 
\label{SK6618}
\end{figure*}

\begin{figure*}
\center
%\resizebox{\hsize}{!}
  {\includegraphics[totalheight=0.45\textheight]{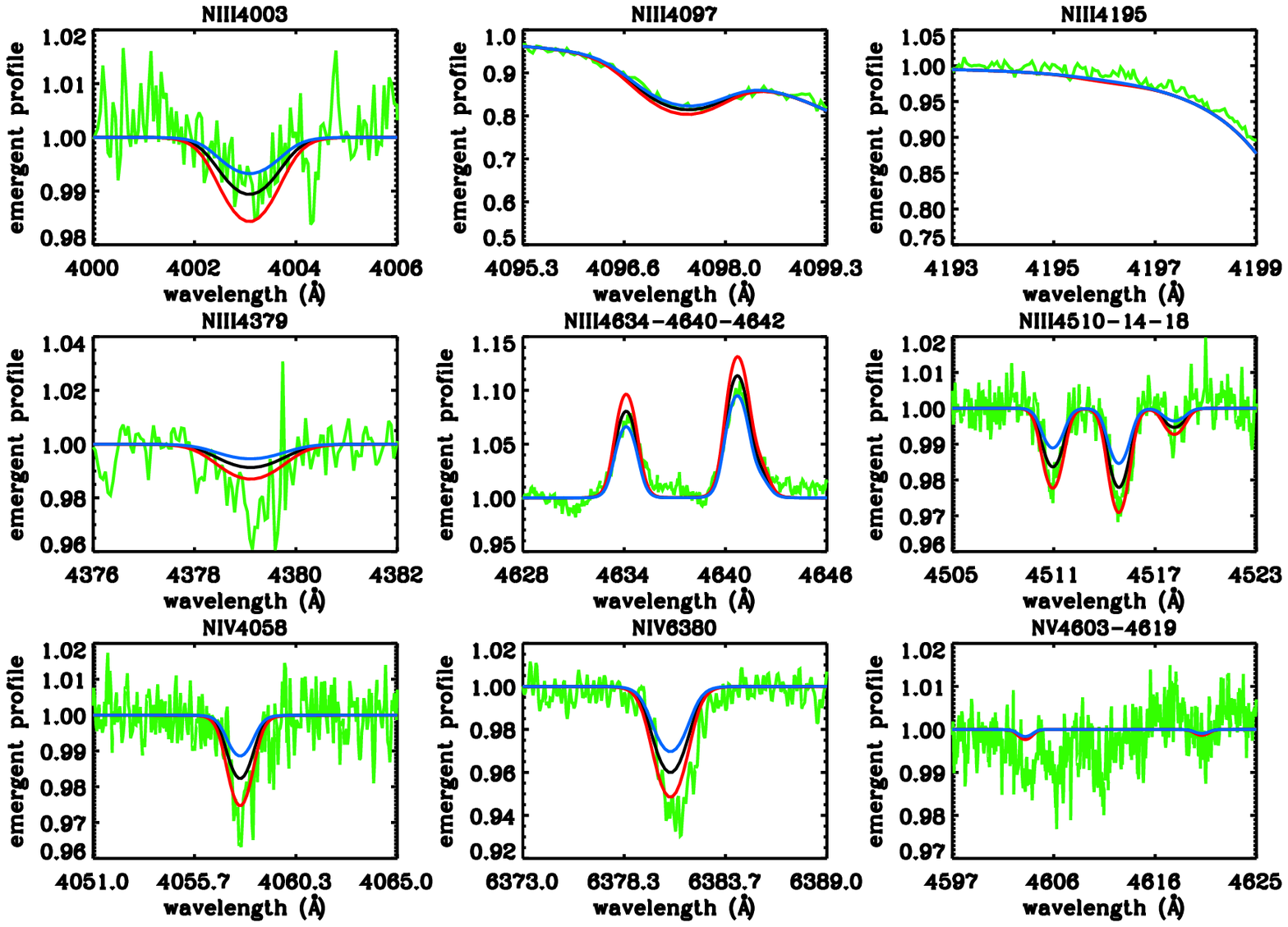}}
%\vspace{-0.5cm}
\caption{N11-065 -- O6.5 V((f))} 
\label{N11-065}
\end{figure*}

\begin{figure*}
\center
%\resizebox{\hsize}{!}
  {\includegraphics[totalheight=0.45\textheight]{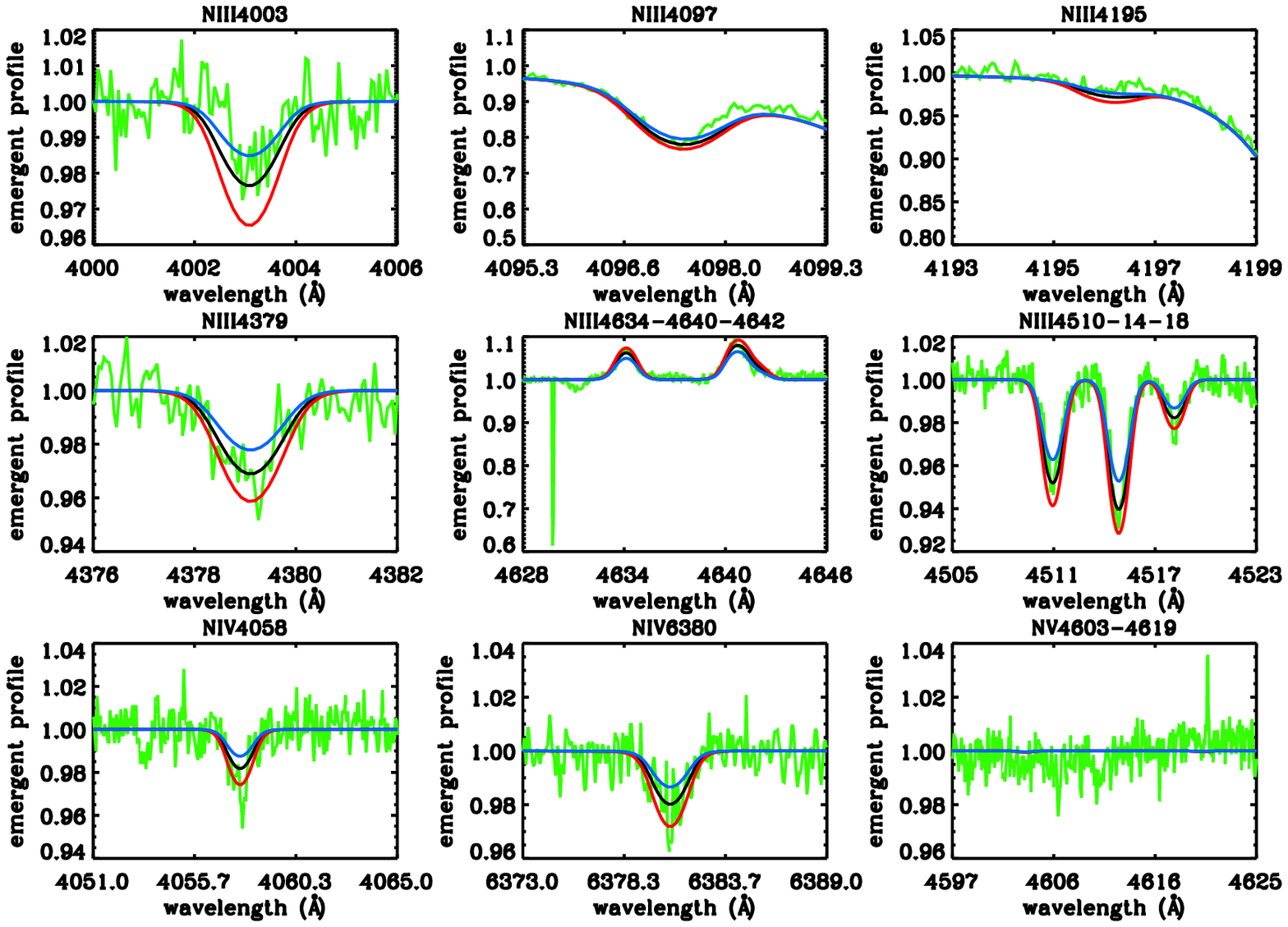}}
%\vspace{-0.5cm}
\caption{N11-066 -- O7 V((f))} 
\label{N11-066}
\end{figure*}

\begin{figure*}
\center
%\resizebox{\hsize}{!}
  {\includegraphics[totalheight=0.45\textheight]{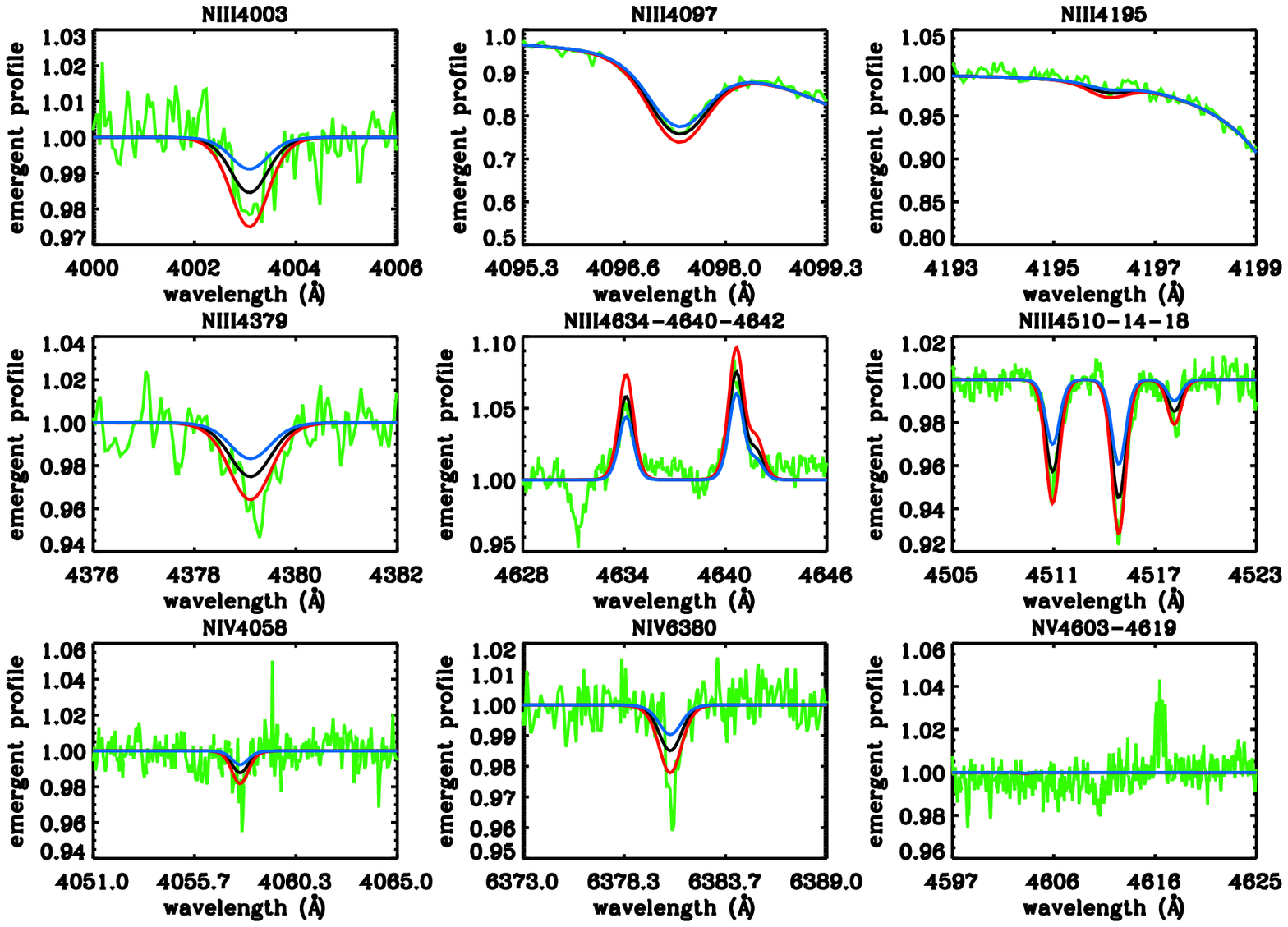}}
%\vspace{-0.5cm}
\caption{N11-068 -- O7 V((f))} 
\label{N11-068}
\end{figure*}

\begin{figure*}
\center
%\resizebox{\hsize}{!}
  {\includegraphics[totalheight=0.45\textheight]{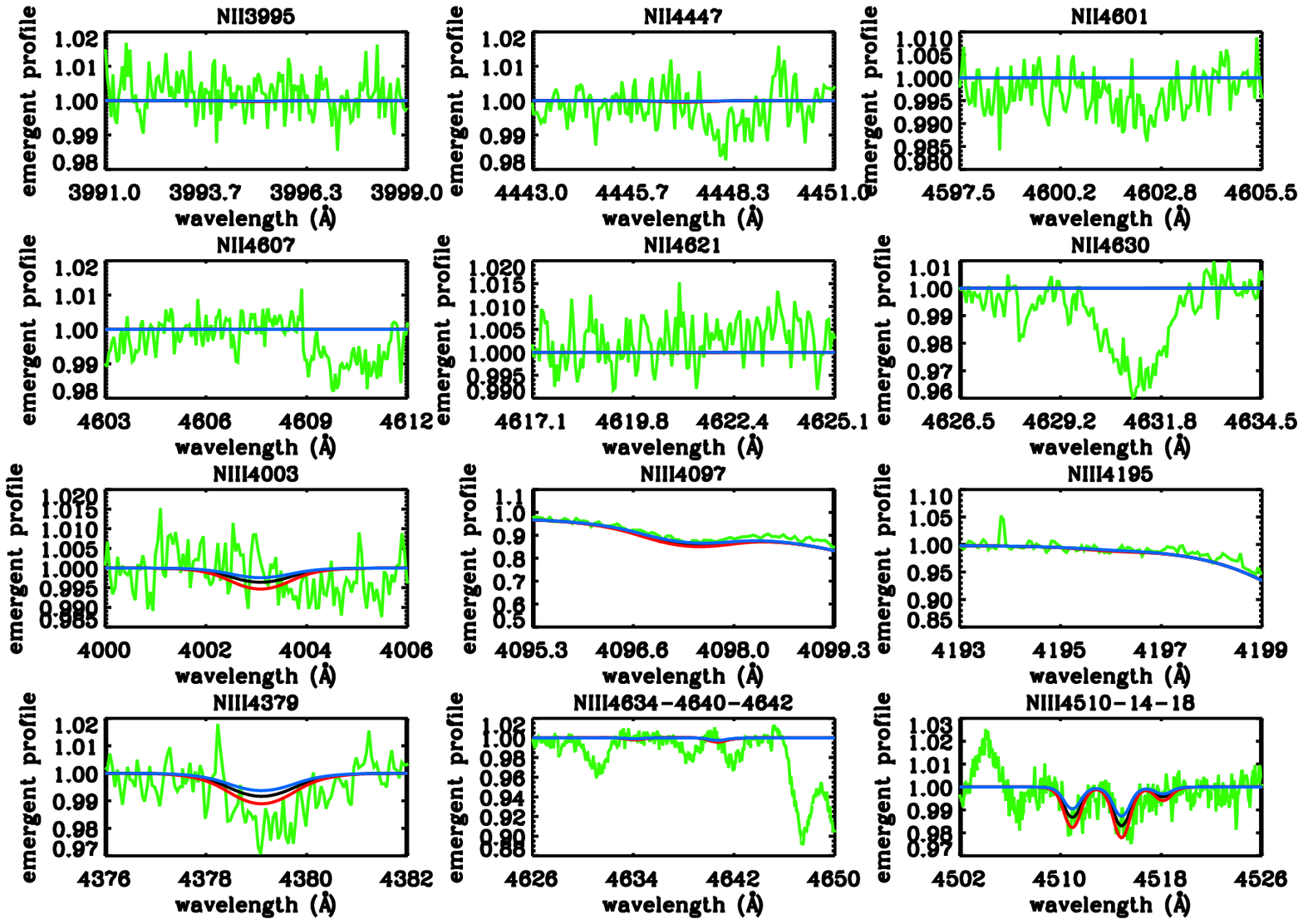}}
%\vspace{-0.5cm}
\caption{N11-061 -- O9 V} 
\label{N11-061}
\end{figure*}

\begin{figure*}
\center
%\resizebox{\hsize}{!}
  {\includegraphics[totalheight=0.45\textheight]{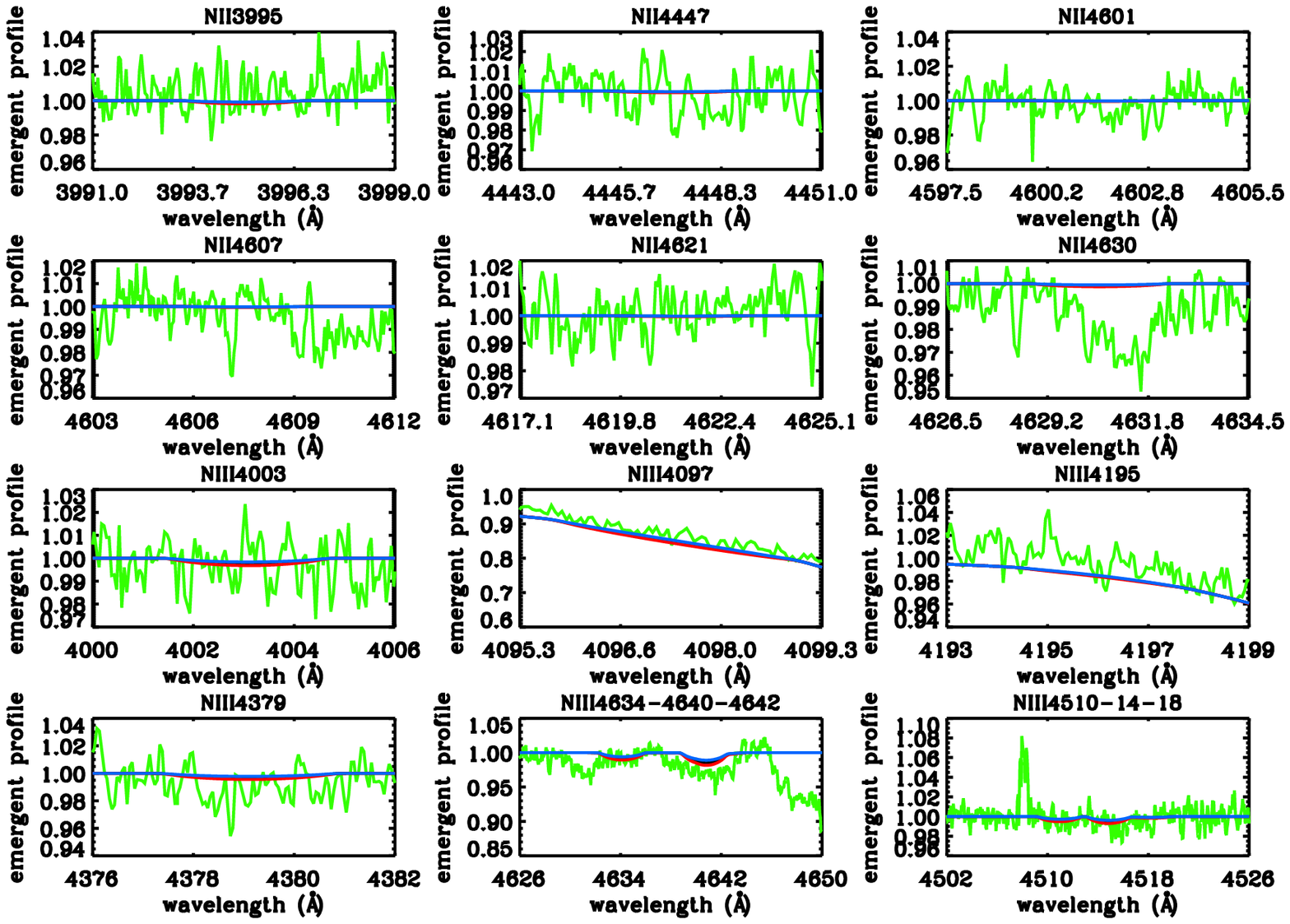}}
%\vspace{-0.5cm}
\caption{N11-123 -- O9.5 V} 
\label{N11-123}
\end{figure*}

\begin{figure*}
\center
%\resizebox{\hsize}{!}
  {\includegraphics[totalheight=0.45\textheight]{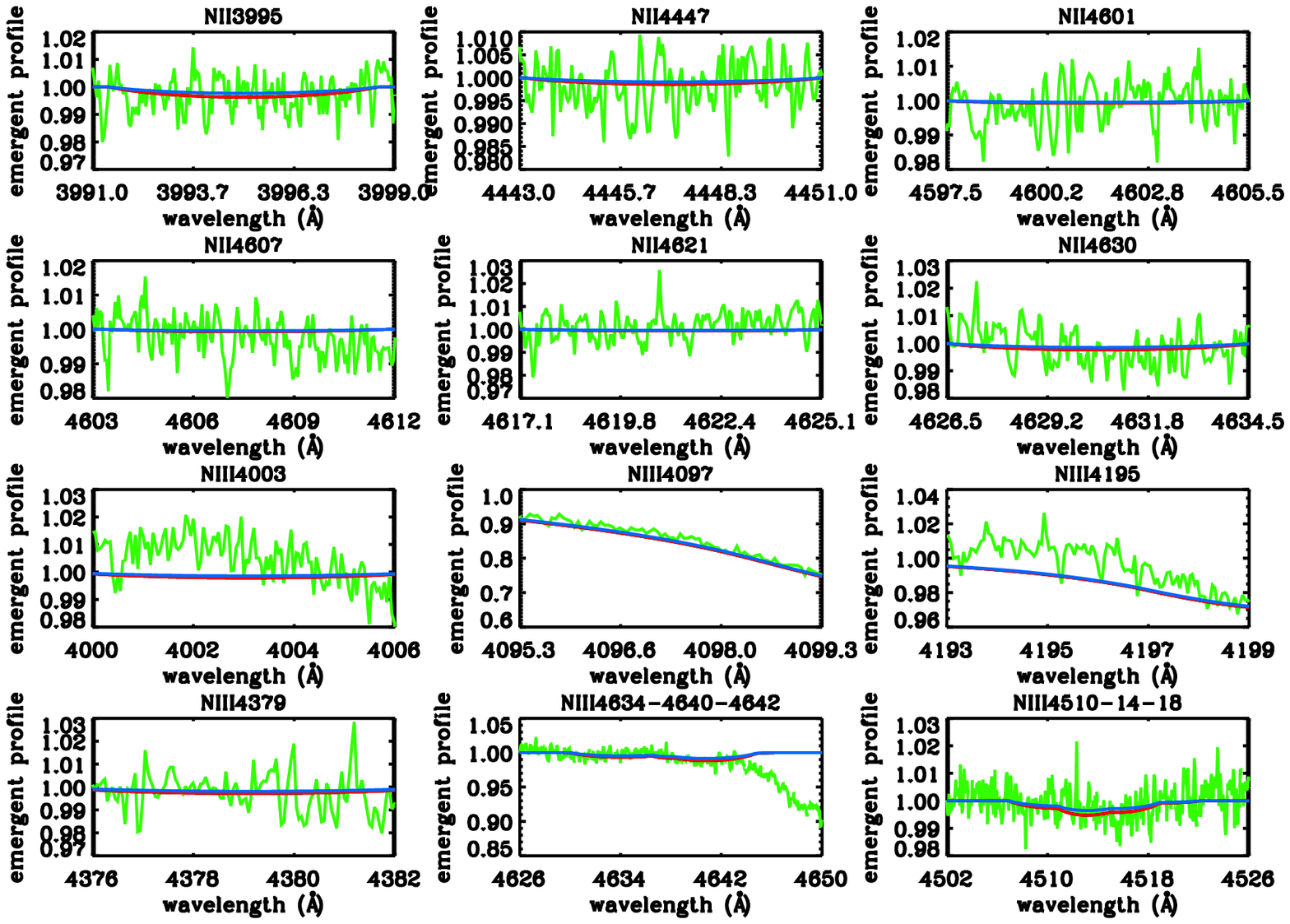}}
%\vspace{-0.5cm}
\caption{N11-087 -- O9.5 Vn} 
\label{N11-087}
\end{figure*}

Figures.~\ref{N11-029} to \ref{N11-087} display the observed (green) and
best-fitting optical nitrogen spectra (black) for all our objects,
except for N11-072, N11-032, and BI237 which are contained in the main
section. Blue and red spectra show corresponding synthetic line
profiles with [N] at the lower and upper limit, respectively. For N11-031
(Fig.~\ref{N11-031}), we show the fits corresponding to the two alternative
solutions for this star (see Sect.~\ref{comments}). For
details on the line fits, see Sect.~\ref{comments}, and for adopted
stellar parameters and derived nitrogen abundances inspect
Table~\ref{tab_param} and Table~\ref{tab_abun}, respectively. 
All fits are based on unclumped winds except explicitly stated.
\clearpage

%\begin{figure}
%\resizebox{\hsize}{!}
%  {\includegraphics[width=1.\textwidth]{plots/fits/N11_072.ps}}
%\vspace{-0.5cm}
%\caption{ N11-072} 
%\label{}
%\end{figure}

% \begin{figure*}
% \center
% %\resizebox{\hsize}{!}
%   {\includegraphics[totalheight=0.45\textheight]{plots/fits/N11_031.ps}}
% %\vspace{-0.5cm}
% \caption{N11-031 -- ON2 III(f$^*$)} 
% \label{N11-031}
% \end{figure*}

%\begin{figure}
%\resizebox{\hsize}{!}
%  {\includegraphics[width=1.\textwidth]{plots/fits/N11_032.ps}}
%\vspace{-0.5cm}
%\caption{ N11-032 -- O7II(f)} 
%\label{}
%\end{figure}

%\begin{figure*}
%\resizebox{\hsize}{!}
%  {\includegraphics[width=1.\textwidth]{plots/fits/BI237.ps}}
%\vspace{-0.5cm}
%\caption{BI237 -- O2V((f$^*$))} 
%\label{}
%\end{figure*}

\clearpage

\end{document}